\documentclass[12pt,reqno]{amsart}
\usepackage{graphicx,times,bbm,url}

\usepackage[margin=1.5in]{geometry}

\usepackage{float}
\usepackage{tikz}
\usetikzlibrary{arrows}
\tikzset{>=stealth'}
\usepackage{listings}
\usepackage{booktabs}
\usepackage{pgfplots}
\usepackage{placeins}

\pgfplotsset{
tick label style = {font=\tiny},
}

\usepackage[bookmarks=false,unicode,colorlinks,urlcolor=red!70!black,citecolor=red!70!black,linkcolor=blue]{hyperref}

\makeatletter
\newcommand{\refcheckize}[1]{%
  \expandafter\let\csname @@\string#1\endcsname#1%
  \expandafter\DeclareRobustCommand\csname relax\string#1\endcsname[1]{%
    \csname @@\string#1\endcsname{##1}\wrtusdrf{##1}}%
  \expandafter\let\expandafter#1\csname relax\string#1\endcsname
}
\makeatother
\usepackage{subfig}
\newcommand{\R}{\mathrm{R}}

\newcommand{\RR}{\mathrm{I\kern-0.20emR}}
\newcommand{\D}{\mathrm{d}\kern0.2pt}

\newcommand{\vp}{{\varphi}}
\newcommand{\pd}[2][]{\frac{\partial #1}{\partial #2}}

\newcommand{\mref}[1]{%
\href{http://www.ams.org/mathscinet-getitem?mr=#1}{#1}}

\title[Sloshing in axisymmetric containers]{On the shape of the fundamental sloshing mode in axisymmetric containers}
\author[T. Kulczycki]{Tadeusz Kulczycki}
\author[M. Kwa{\'s}nicki]{Mateusz Kwa{\'s}nicki}
\author[B. Siudeja]{Bart{\l}omiej Siudeja}
\address{T. Kulczycki and M. Kwa{\'s}nicki \\ Institute of Mathematics and Computer Science \\  Wroc{\l}aw University of Technology  \\ Wybrze{\.z}e Wyspia{\'n}skiego 27 \\ 50-370 Wroc{\l}aw, Poland.}
\address{B. Siudeja \\ Department of Mathematics \\ University of Oregon \\  Eugene, OR 97403. }
\email{Tadeusz.Kulczycki@pwr.wroc.pl}
\email{Mateusz.Kwasnicki@pwr.wroc.pl}
\email{siudeja@uoregon.edu}

\begin{document}

\begin{abstract}
In the paper we numerically study positions of high spots (extrema) of the fundamental sloshing mode of a liquid in an axisymmetric tank. Our approach is based on a linear model reducing the problem to appropriate Steklov eigenvalue problem. We propose a numerical scheme for calculating sloshing modes and a novel method of making images of oscillating fluid. We also describe the relation of the high spot problem to the celebrated hot spots conjecture.
\end{abstract}

\maketitle

\section{Introduction}

\begin{figure}[b]

\subfloat[A cup.\label{cupnglassa}]{
\includegraphics{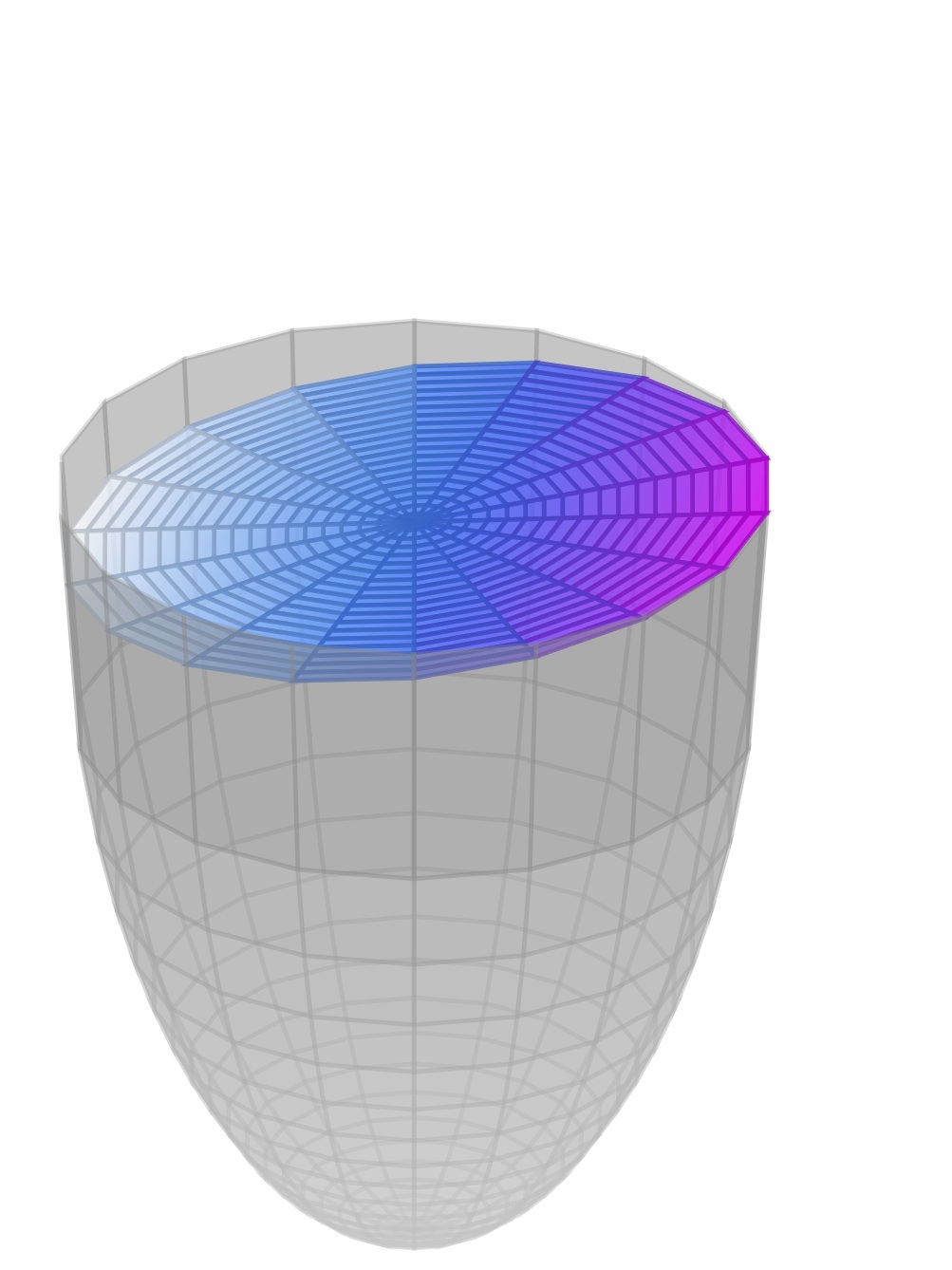}
}
\subfloat[A snifter.\label{cupnglassb}]{
\includegraphics{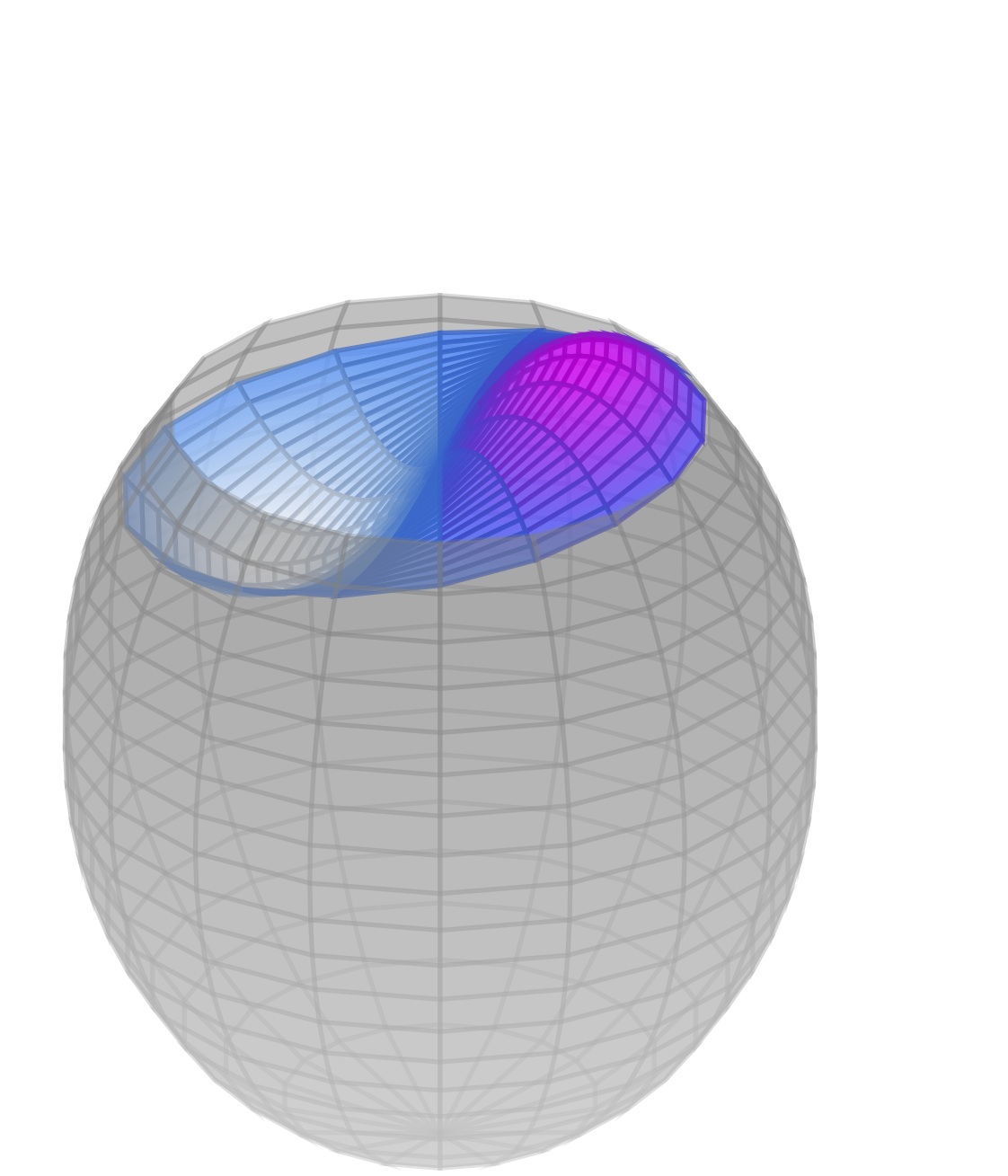}
}
\vspace{-6cm}

\hspace{0.04cm}
\includegraphics{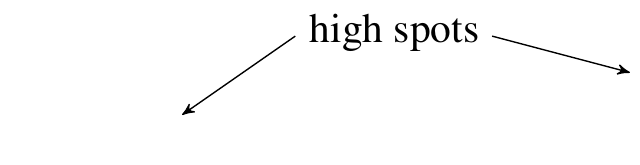}
\vspace{4cm}

\caption{High spots in a container with straight sides (a cup) compared to a bulbous one (a snifter).}
  \label{cupnglass}
\end{figure}

We consider the linearized sloshing problem in axisymmetric containers for a perfect fluid. The aim of the paper is to numerically study the positions of the {\it{high spots}}, the extremal points of the fundamental sloshing mode $\vp$. Their locations coincide with the extrema of the free surface elevation of liquid (see \autoref{cupnglass}), under the assumption that the liquid oscillates freely with a fundamental frequency according to $\vp$. We work with a simplified linear model (see \autoref{mathematical}), studied by many authors, including Faltinsen-Timokha \cite{FT2012,FT2014}, Fox-Kuttler \cite{FK83}, Kozlov-Kuznetsov \cite{KK2004} and McIver \cite{McI89}. Linear sloshing in simple moving containers was also considered by Wu-Chen \cite{WC09}, Ardakani-Bridges \cite{AB11} and Herczy\'nski-Weidmann \cite{HW12}. More accurate physical models can also be considered, but nonlinear phenomena force a severly restricted choice of containers (see a review of the numerical methods by Rebouillat and Liksonov \cite{RL10}). For general overview of both linear and nonlinear models of sloshing see monographs by Ibrahim \cite{Ibr}, Kopachevsky-Krein \cite{KK01}, Lamb \cite{L1932}, Moiseev \cite{M64} or Troesch \cite{T60}.

In a bulbous container high spots are located inside the container (see \autoref{cupnglassb}). On the other hand in a convex container, such that the angle between a wall of the container and a free surface is smaller or equal to $\pi/2$, high spots are located on the boundary of the container (see \autoref{cupnglassa}). This effect was proved rigorously by Kulczycki and Kwa{\'s}nicki in \cite{KK2012}. It has also been studied for spherical tanks by Faltinsen and Timokha \cite{FT2012}, using analytic approximations. 

In this paper we propose a new numerical scheme for calculating sloshing modes for solids of revolution. This scheme, applied to the fundamental mode, allows us to find positions of high spots with great accuracy. The same scheme can also be used to study higher sloshing modes. In \autoref{sec:numres} we present numerical results for the first few modes of various containers, we inspect convergence of the numerical scheme and compare to known results. 

We also describe an ingenious method of obtaining images of oscillating liquid showing positions of high spots. Our method is based on photographing a reflection of a dotted piece of paper on a very slightly disturbed surface of the liquid (see \autoref{photo}). We experimented with 3 different containers and positions of the high spots agree with our numerical results and/or known results. Our experiment was first briefly described in the Steklov memoirs published in Notices of AMS \cite{Notices2014}.

Finally, in \autoref{sec:relation} we describe the relation of the high spot problem to the celebrated hot spots conjecture.

\section{Mathematical model} \label{mathematical}
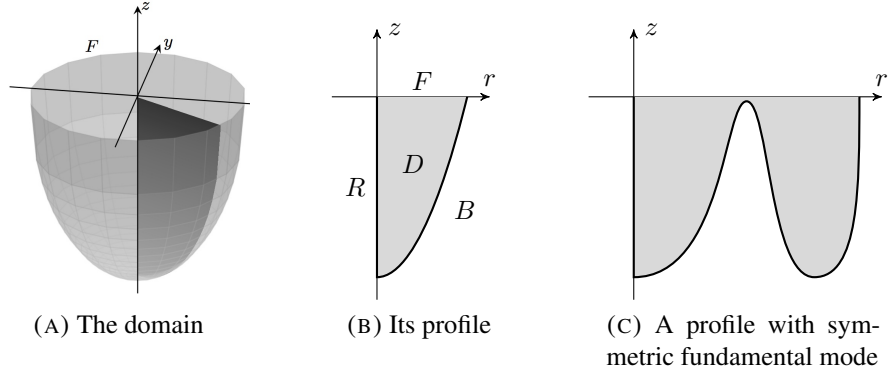
\begin{figure}[t]
\hspace{\fill}
\subfloat[The domain\label{fig8}]{
\includegraphics{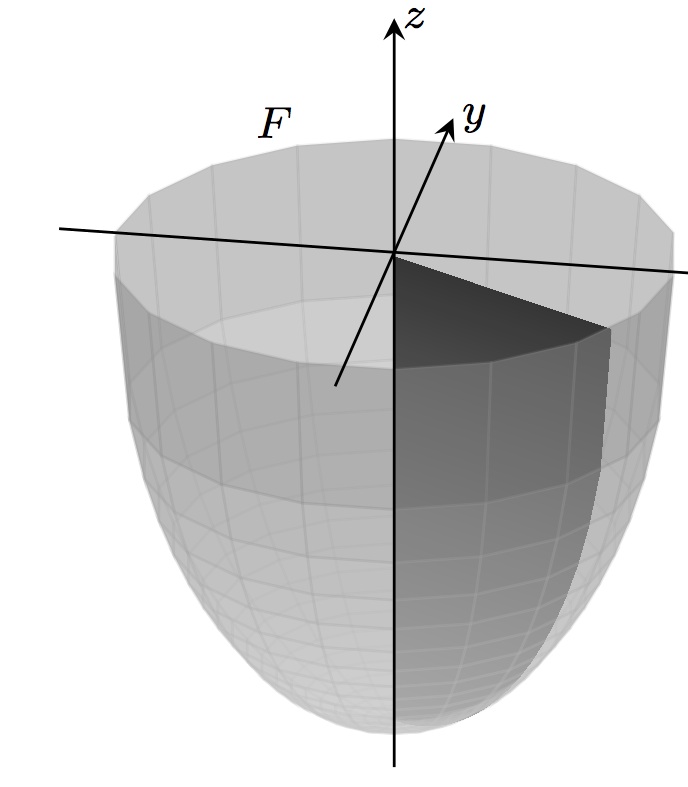}
}
\hspace{\fill}
\subfloat[Its profile\label{fig8b}]{
\begin{tikzpicture}[scale=0.6]
  \draw[->] (-1,0) -- (2.5,0) node[above] {\footnotesize $r$};
  \draw[->] (0,-4.5) -- (0,1.5) node[right] {\footnotesize $z$};
  \fill[fill=black!15!white,draw=black,thick]
    
  (0,0) -- (0,-4) node [pos=0.5,left] {\footnotesize $R$} parabola bend(0,-4) (2,0) ;
  \draw (1,-0.1) node[above] {\footnotesize $F$};
  \node at (0.8,-1.5) {\footnotesize $D$};
  \node at (1.95,-2.5) {\footnotesize $B$};
\end{tikzpicture}
}
\hspace{\fill}
  \subfloat[A profile with symmetric fundamental mode\label{fig8c}]{
  \begin{tikzpicture}[scale=0.6]
  \draw[->] (-1,0) -- (5.5,0) node[above] {\footnotesize $r$};
  \draw[->] (0,-4.5) -- (0,1.5) node[right] {\footnotesize $z$};
  \fill[fill=black!15!white,draw=black,thick]
   (0,0) -- (0,-4) .. controls (2,-4) and (2,-0.1) .. (2.5,-0.1) .. controls (3,-0.1) and (3,-4) .. (4,-4) .. controls (5,-4) and (5,-2) .. (5,0);
  \end{tikzpicture}
  }
\hspace{\fill}
  \caption{Solid of revolution and the generating profiles.}
  \label{solrev}
\end{figure}

We assume that the inviscid, incompressible, heavy liquid occupies a three-dimensional container bounded from above by a free surface, which in its mean position is a simply connected two-dimensional domain of finite diameter. Let Cartesian coordinates $(x,y,z)$ be chosen so that the mean free surface lies in the $(x,y)$-plane and the $z$-axis is directed upwards (see \autoref{fig8}). The surface tension is neglected on the free surface, and we assume the liquid motion to be irrotational and of small amplitude. These assumptions allow us to linearise the boundary conditions on the free surface and get the following boundary value problem for $\vp (x,y,z)$ (the velocity potential of the flow with a time-harmonic factor removed):

\begin{align}
 \Delta \vp & = 0  \text{ in $W$,} \label{slosh1} \\
 \pd[\vp]{z} & = \nu \vp  \text{ on $F$,} \label{slosh2} \\
 \frac{\partial \vp}{\partial \vec{n}} & = 0  \text{ on $B$,} \label{slosh3}
\end{align}
where $W$ is the part of the container filled with liquid (in its mean position), $B$ is the wetted part of its boundary, and $F$ is the free surface of the liquid (see \autoref{fig8}).  The outer normal derivative on $B$ is denoted by $\frac{\partial}{\partial \vec{n}}$. 

The zero eigenvalue obviously exists for the problem (\ref{slosh1}--\ref{slosh3}), but we exclude it with the help of the following orthogonality condition:
\begin{align}
 \int_F \vp & = 0, \label{slosh4}
\end{align}
This condition also ensures that the volume of the fluid is preserved throughout the sloshing process.

Let us reiterate that we are working with a linear model. We are not concerned with the interactions between the fluid and the environment. As such, the model is well suited for infinitesimal sloshing of a perfect liquid (cf. \cite{FK83,KK2004,FT2012,FT2014}). In other words, our approximations only apply to a nearly still water surface, as opposed to turbulent phenomena in strong sloshing (as in \cite{RL10}).

For sufficiently regular domains it is known that the mixed Steklov problem~(\ref{slosh1}--\ref{slosh4}) has a discrete sequence of eigenvalues
\[
  0 < \nu_1 \le \nu_2 \le \nu_3 \le \ldots \to \infty ,
\]
and the corresponding modes $\varphi_n \in H^1(W)$, $n= 1, 2, 3, \ldots$, restricted to the free surface $F$, form (together with a constant function elliminated by \eqref{slosh4}) a complete orthogonal set in $L^2(F)$. 

The following observation plays a key role in our study: if the liquid oscillates freely with the fundamental frequency $\nu_1$ then at every moment the free-surface elevation of the liquid is proportional to the fundamental eigenfunction $\varphi(x,y,0)$ (see e.g. Lamb \cite{L1932}).

Let us point out, that general containers are very hard to work with, even with the linearized model. Axisymmetric domains described below encompass many practical shapes and are somewhat easier to handle. Another simplification leading to manageable calculations involves two-dimensional containers, or long troughs with fluid sloshing only sideways. These were studied by Kulczycki-Kuznetsov \cite{KK2011} and Faltinsen-Timokha \cite{FT2014}, among others.
\subsection{Axisymmetric containers}
Now we turn to the problem of sloshing in axisymmetric containers. It is convenient to introduce the cylindrical coordinates $(r,\theta,z)$ so that
\begin{equation*}
x = r \cos \theta, \quad y = r \sin \theta. 
\end{equation*}
and to take the $z$-axis as the axis of symmetry for $W$. In this case $F$ is typically a disc on $xy$-plane (\autoref{solrev}). Moreover, we will consider $W$ as being obtained from the rotation of a domain $D$ adjacent to both axes in the $rz$-plane (\autoref{fig8b}). It is convenient to think of $D$ as the cross-sections of $W$ along the half-plane $\theta = 0$.  

It is clear that the ansatz
\begin{equation}
\vp = \psi_m(r,z) \cos (m \theta), \quad \text{or} \quad 
\vp = \psi_m(r,z) \sin (m \theta),
 \quad m = 0,1,2, \ldots , \label{rep}
\end{equation}
where $\psi$ is  bounded near $R=\{r=0\}\cap D$, reduces the eigenvalue problem (\ref{slosh1}--\ref{slosh3}) in $W$ to the following sequence of boundary value problems:
\begin{align}
\label{sloshD1}
  & (\psi_m)_{zz}+(\psi_m)_{rr}+\frac{1}{r}(\psi_m)_r-\frac{m^2}{r^2}\psi_m=0,\text{ on }D\\
  \label{sloshD2}
  &(\psi_m)_z=\lambda_m \psi_m,\text{ on }F\\
  \label{sloshD3}
  &\frac{\partial \psi_m}{\partial \vec{n}}=0,\text{ on } B
\end{align}
These relations must hold for all $m$, and  by~\eqref{slosh4}, for $m = 0$ we also need
\begin{equation}
\label{sloshD4}
\int_{F} \psi(r,0) r \, dr = 0 .
\end{equation}

The above reduction was applied by many authors (e.g. \cite{KK2012,GHLST,LBK}). Note that there is no boundary condition on $R$, but if $m>0$ we need continuity of the three dimensional solution, while for $m=0$ we need continuous derivative (three dimensional solutions are harmonic, hence smooth). Therefore we have either Dirichlet ($m>0$) or Neumann ($m=0$) condition implied on $R$. 

The variational methods (e.g. \cite{BB92}) guarantee that for every $m = 0,1,2, \ldots$, the spectral problem (\ref{sloshD1}--\ref{sloshD4}) has a sequence of eigenvalues 
\begin{equation}
0 < \lambda_{m,1} \le \lambda_{m,2} \le \lambda_{m,3} \le \ldots , \quad m = 0,1,2, \dots , \label{meigenvalues}
\end{equation}
and corresponding eigenfunctions $\psi_{m,k}$, $m \ge 0$, $k \ge 1$. Every eigenvalue in (\ref{meigenvalues}) has a finite multiplicity. Moreover, for every $m \geq 0$ we have $\lambda_{m,k} \to \infty$ as $\ k \to \infty$. 

It is clear that, the sequence of eigenvalues $\{\nu_n\}_{n = 1}^{\infty}$ of problem (\ref{slosh1})--(\ref{slosh4}) for $W$ coincides with the double sequence $\{\lambda_{m,k}\}_{m = 0, k = 1}^{\infty}$, with every number $\lambda_{m,k}$ repeated twice when $m \ge 1$. Thus the sequence of problems (\ref{sloshD1})--(\ref{sloshD4}) is equivalent to the original sloshing problem. 

\subsubsection{Known rigorous results}\label{rigor}
Usually, the modes corresponding to $\nu_1$ are antisymmetric (that is $\nu_1 = \lambda_{1,1}$). Nevertheless, some rather strange containers (profiles from \autoref{fig8c} and \autoref{fig:shapesym}) have rotationally symmetric fundamental eigenfunctions. For the latter we show numerically that $\nu_1\ne \lambda_{1,1}$, and $\nu_1=\lambda_{0,1}$ instead. 

Under the assumption $\nu_1 = \lambda_{1,1}$, there are two linearly independent antisymmetric fundamental sloshing modes corresponding to the least eigenvalue $\nu_1$ ($=\nu_2$, has double multiplicity). In cylindrical coordinates they have the form \eqref{rep} with $m=1$.

By~\cite[Theorems~1.1 and 1.2]{KK2012}, if $W$ is a convex solid of revolution such that the angle between $B$ and $F$ is smaller or equal to $\pi/2$, then indeed $\nu_1 = \nu_2 = \lambda_{1,1}$ correspond to antisymmetric eigenfunctions $\varphi_1$ and $\varphi_2$, and the high spots of the modes described by $\varphi_1$ and $\varphi_2$ are located on the boundary, e.g. containers from \autoref{cupnglassa} and \autoref{solrev}. On the other hand, \cite[Proposition~1.3]{KK2012} (see also \cite{FT2012}) asserts that if $B$ and $F$ form an obtuse angle then $\varphi_1(x,y,0)$, $\varphi_2(x,y,0)$ attain their maxima inside $F$, as shown on \autoref{cupnglassb}. We check our numerical methods described below against these theoretical results in \autoref{sec:numres}.

\section{Numerical approximations}
In this section we design a two dimensional numerical finite element scheme (FEM) for calculating the sloshing eigenvalues (and eigenfunctions) of the solids of revolution described in \autoref{mathematical}. Three dimensional FEM can also be used (see e.g. \cite{Notices2014}), however one should be able to improve accuracy by exploiting decomposition \eqref{rep} (see also \autoref{solrev}). Unfortunately, this leads to a singular weak formulation, due to the singular form of the gradient in polar coordinates. We introduce two modifications, leading to nonsingular schemes. The first is more natural, but leads to badly conditioned discrete problems for higher modes $m$, due to fast decaying weights near $r=0$. The second modification is designed to keep a balance between singularity removal and decay of the weights. We use the second approach to recover, essentially exactly, the results for partially filled tanks obtained by McIver \cite{McI89}, for modes $m=0,1,2,3$. The same approach can be applied to arbitrary axisymmetric containers.

Our derivations are based on the classical weak formulations of eigenvalue problems (see e.g. Babu{\v{s}}ka-Osborn \cite{BO91} and Blanchard-Br{\"u}ning \cite{BB92}). One can also consult Bernardi, Dauge and Maday \cite{BDM99} for axisymmetric domains.

\subsection{Modified decompositions}

\subsubsection{Harmonic polynomials}\label{sec:harmonic}
We introduce the first modified, but equivalent decomposition by putting $\psi_m(r,z)=r^m \phi_m(r,z)$. The solutions of the original three dimensional problem take the form (cf. \eqref{rep})
\begin{align}\label{harmonicpoly}
  \varphi_m = \phi_m(r,z)r^m\cos(m\theta),
  \qquad\text{ or }\qquad
  \varphi_m = \phi_m(r,z)r^m\sin(m\theta).
\end{align}
Note that $c_m=r^m\cos(m\theta)$ and $s_m=r^m\sin(m\theta)$ are polar forms of the harmonic polynomials, e.g. $c_1=x$, $c_2=x^2-y^2$, $s_2=2xy$. Indeed, if we interpret $(x,y)=(r,\theta)$ as a complex number $w=x+iy$, then $c_m=\Re w^m$ and $s_m=\Im w^m$. 

Furthermore, note that $\varphi_m$ is analytic inside $W$ (it is in fact harmonic). In particular, the first $m$ $r$-derivatives must be continuous, forcing $\varphi_m$ to decay at least as fast as $r^m$, hence our modified decomposition can be used to solve the sloshing problem. We will show that this new decomposition leads to a nonsingular weak formulation. Note that the Neumann boundary condition on $B$ must be imposed on $r^m \phi_m$. This is however irrelevant, since Neumann boundary conditions are naturally incorporated into weak formulations.

Troesch \cite{T60} used a similar approach (family of harmonic polynomials, solid spherical harmonics) to find the shape of the domain on which a polynomial is one of the eigenfunctions for the sloshing problem. He found that the lowest $m=1$ mode for a cone with aperture angle $\pi/2$ has eigenfunction with $\psi_1=r+rz/2$. Our modification gives $u=1+z/2$, and even the simplest linear numerical approximation provides the exact solution. In fact for any $m>0$, Troesch found a cone such that $r^m(1+c z)$ is the lowest mode.

\subsubsection{Linear aproach}\label{sec:linear}
The above approach gives the best possible accuracy near $r=0$, effectively removing the decay at $r=0$ from all eigenfunctions. However, in order to remove the singularity of the angular derivative, it is enough to introduce $\psi_m(r,z)=r u_m(r,z)$ for $m>0$. This means that low degree approximations (linear or quadratic) cannot recover the exact decay around $r=0$ for $m$ larger than $2$. However, this limitation does not influence overall performance of the numerical method, especially that the high spots for the fundamental mode occur close to the boundary (far from $r=0$). All the numerical results described in \autoref{sec:numres} were obtained using this approach. 

\subsection{General weak formulations.}

We will derive appropriate weak formulations based on weighted Sobolev spaces described in \cite{K80} and \cite{BDM99}. Let
\begin{align*}
  H^1_\alpha(D)=\{u:D\to\R:\; \int_D |\nabla u|^2 r^\alpha drdz + \int_D u^2 r^\alpha drdz <\infty\},\\
  V^1_\alpha(D)=\{u:D\to\R:\; \int_D |\nabla u|^2 r^\alpha drdz+\int_D u^2 r^{\alpha-2} drdz<\infty\}.
\end{align*}
The former is the classical weighted Sobolve space, while latter is more appropriate for $m>0$, as we show below.

Start with the differential equation \eqref{slosh1} written in a divergence form
\begin{align}\label{pde}
  L\psi := \nabla\cdot(r\nabla \psi)-\frac{m^2}{r}\psi=0.
\end{align}

First we find the weak formulation for the classical decomposition (\ref{rep}) and the Steklov problem (\ref{sloshD1}--\ref{sloshD4}).
We want to construct a symmetric bilinear form using the $L^2$ inner product of $L\psi$ and smooth test functions. Then we can formally close the form in appropriate Sobolev space to get a valid weak formulation. For smooth functions, the equation $\langle L\psi,v\rangle=0$ is equivalent to
\begin{align}
  \int_D\frac{m^2}{r} \psi v\,drdz&=\int_D[\nabla\cdot (r\nabla \psi)] v\,drdz=-\int_D \nabla \psi\cdot \nabla v\,rdrdz+\int_{\partial D} r \frac{\partial \psi}{\partial n}v\,dS\nonumber
  \\&=-\int_D r\nabla \psi\cdot \nabla v\,drdz+\int_F \lambda_m \psi v\,rdr.\label{weak1}
\end{align}
In the last equality used the fact that the boundary integral equals $0$ on $B$ (due to Neumann BC for $\psi$) and on $R$ (since $r=0$). Note that we did not need to know the boundary condition for $R$, as long as $\psi$ is bounded there. Note that the above weak formulation is consistent with the reduction of the integral $\langle \Delta \phi,v\rangle$ from the three dimensional axisymmetric container to its two dimensional profile. In particular we get $r dr$, the polar Jacobian. Therefore, it is natural to consider the reduced problem on $H^1_1(D)$.

When $m=0$ (axisymmetric eigenfunctions), formulation \eqref{weak1} reduces to 
\begin{align}\label{weakm0}
  \int_D   \nabla \psi\cdot \nabla v\,rdrdy = \lambda_0\int_F \psi v\,rdr.
\end{align}
The left side naturally extends to $H^1_1(D)$, and the space of all axisymmetric functions $H^1(W)$ (on the 3D container $W$) is isomorphic to $H^1_1(D)$ \cite[Section II]{BDM99}. Furthermore, traces of such functions on $F$ are well defined, and belong to $H^{1/2}_1(F)$. We get a compact embedding of $H^1_1(D)$ into $L^2_1(F)$ and the standard spectral argument (e.g. \cite[Section 6.3]{BB92}) leads to discreteness of the spectrum of the quadratic form \eqref{weakm0}.

\subsection{Non-singular weak formulations.}
When $m>0$, the weak formulation \eqref{weak1} contains singular $L^2_{-1}$ inner product. Therefore it is natural to consider the space $V^1_1(D)$ as its domain. Functions in this space have null trace on $R$ (see \cite{MR82}), and the space is isomorphic to $H^1(W)$ with zero condition on the axis of symmetry \cite[Section II]{BDM99}. This last condition exactly characterizes the sloshing modes $m>0$. However, the presence of the singular $L^2_{-1}$ norm hinders applicability of that weak formulation. 

The first modified decomposition \eqref{harmonicpoly} solves this issue, by introducing $\psi(r,\theta)=r^m\phi(r,\theta)$, effectively eliminating the singularity. Replacing $\psi$ with $r^m \phi$ and the test function $v$ with $r^m v$ (to keep symmetry of the formulation) leads to

\begin{align*}
  \lambda_m &\int_F \phi v r^{2m+1}\,dr = \int_D   \nabla (r^m \phi)\cdot \nabla (r^m v)\,rdrdy+m^2\int_D \phi v\,r^{2m-1}drdy 
  \\&=\int_D  r^{2m-2}(m \phi+r \phi_r, r \phi_z)\cdot (m v+rv_r,r v_z)\,rdrdy+\int_D m^2\phi v\,r^{2m-1}drdy
  \\&=\int_D \nabla \phi\cdot\nabla v\,r^{2m+1}dr+2m^2\int_D \phi v\, r^{2m-1}dr+m\int_D (v\phi_r+\phi v_r)\,r^{2m}dr.
\end{align*}
The last expression defines a symmetric quadratic form that can be extended to the Sobolev space $V^1_{2m+1}(D)$. Again, one gets appropriate compact embeddings into $L^2_{2m+1}(F)$, and
\begin{align}
  \lambda_m \int_F uv r^{2m+1}\,dr
  &=\int_D \nabla u\cdot\nabla v\,r^{2m+1}dr+2m^2\int_D uv\, r^{2m-1}dr\nonumber
  \\&\qquad\qquad+m\int_D (vu_r+uv_r)\,r^{2m}dr.\label{weakm1}
\end{align}
can be used to numerically approximate the eigenvalues and the eigenfunctions for $m>0$. 

Unfortunately, the presence of $r^{2m+1}$ in the weak formulation requires high degree quadratures for accurate numerical integration near $r=0$ when $m$ is large. Furthermore, fast decay of $r^{2m+1}$ leads to discrete matrix eigenvalue problems with nearly singular matrices, for very fine meshes. This degrades the convergence of the numerical method, or even leads to numerically singular matrices. 

These problems are mitigated by a linear approach $\psi(r,\theta)=r u(r,\theta)$ (described in \autoref{sec:linear}). We get
\begin{align}
  \lambda_m \int_F &u v r^{3}\,dr = \int_D   \nabla (ru)\cdot \nabla (r v)\,rdrdy+m^2 \int_D u v\,r drdy\nonumber 
  \\&=\int_D  (u+r u_r, r u_z)\cdot (v+rv_r,r v_z)\,rdrdy+\int_D m^2\phi v\,r drdy\nonumber
  \\&=\int_D \nabla u\cdot\nabla v\,r^3 dr+(m^2+1)\int_D u v\, r dr+ \int_D (vu_r+u v_r)\,r^2dr.\label{weak2}
\end{align}
Note that when $m=1$ we recover the first modification, while for $m>1$ we still use the Sobolev space $V^1_3(D)$, hence avoiding high powers of $r$, high degree quadratures in numerical integration, and nearly singular matrices in the discrete approximation.

\subsection{Numerical method}
The simplified weak formulations \eqref{weakm1} and \eqref{weak2} allow us to use commonly available numerical method suitable for Laplace eigenvalue problems. However, we only present the latter approach due to numerical instability of \eqref{weakm1} we experienced for $m>1$. Note that formulations \eqref{weakm1} and \eqref{weak2} are the same for $m=1$, usually containing the fundamental mode. 

We triangulate the domain and use conforming $P1$ (first-order) or $P2$ (second-order) finite elements to discretize the weak formulation \eqref{weak2}. Resulting piecewise linear (or quadratic) functions give approximate shapes of the eigenfunctions, in particular the sloshing profiles on $F$, showing high-spots.

We also design an adaptive mesh refining strategy to decrease errors in approximations, and fit the possibly curved boundary. We use classical local residual error a posteriori estimators (see e.g. Ainsworth and Oden \cite{AO00}), that were recently studied for pure Steklov problems by Armentano-Padra \cite{AP08}, Garau-Morin \cite{GM11}. We modify their approach to fit the weighted weak formulation and variable boundary conditions. In each step we choose a small fraction of triangles to be refined, based on the largest local error estimators \cite[Definitions 2.5 and 2.7]{GM11}). Even though we are using \eqref{weak2} to find approximation $\widetilde u$, we evaluate the error estimators on $\widetilde\psi=r \widetilde u$ and the original operator $L$ given by \eqref{pde}. More precisely, for any triangle $T$ and its edge $E$ in the triangulation of the domain we find
\begin{align}
  R(T) &= \int_T \left(\nabla\cdot \left(r\nabla \widetilde \psi\right) -\frac{m^2}{r}\widetilde \psi\right)^2 dr dz,\quad\text{ (element residual)}\label{error1}\\
  S(T,E) &= \int_E \left( \widetilde \psi_z - \widetilde\lambda\widetilde\psi \right)^2 r dr\qquad\text{ if }E\subset F,\quad\text{ (boundary residual)}\\
  J(T,E) &= \int_E \left( \frac{\partial_n \widetilde \psi+\partial_{\bar n} \widetilde \psi}{2} \right)^2 r ds\qquad\text{ if }E\subset D,\quad\text{ (jump residual)}\label{error3}
\end{align}
where $\partial_n$ is the inner, and $\partial_{\bar n}$ is the outer normal derivative for the edge $E$. Note that $\partial_{\bar n}$ is evaluated outside of $T$, on an adjacent piece of the triangulation. The element residual measures deviation of the approximation from harmonicity. The jump residual is responsible for the discontinuity of the derivative of the approximation across triangles. Finally, the boundary residual measures deviation from the Steklov eigenvalue condition on $F$. 

We add all errors corresponding to the same triangle and refine the least accurate $10\%$ of the triangles. Finally, we add all errors and take square root to find a global residual norm, used in the numerical results \autoref{sec:numres} to control convergence. It turns out that this norm correlates almost exactly with the eigenvalue error in known cases. In cases where the eigenvalues are not known, the plots of the squared norm and the differences to the best approximation are also nearly identical. This indicates that we have a reliable error estimator. 

In order to get more than just the first eigenvalue, we considered the following strategies. Execute the adaptive loop just for the first eigenvalue and find all eigenvalues based on fully refined mesh. This leads to rather poor approximations for higher eigenvalues. It is better to add residual errors from all eigenfunctions and refine worst triangles based on the accumulated estimators. This gives much better results for higher eigenvalues, but slightly degrades the accuracy of the lowest eigenvalue estimation. Finally, we tried to run the adaptive algorithm for each eigenvalue separately. This is obviously optimal, but one must first know fair approximations for these eigenvalues in order to target the specific one. In practice, the optimal separate calculations must be preceded by a collective adaptive calculation yielding good starting points for targeted searches.

\subsection{Implementation}

We used FEniCS \cite{LMW} to implement the above numerical scheme. It provides intuitive mathematical interface for defining weak forms and solving differential equations using Finite Element Methods. Adaptive methods are also easy to implement, including residual errors and snapping mesh boundary to a given theoretical domain.

\subsubsection{Eigenvalue problem}

Assuming that variable $mesh$ contains an already created triangulation, we define appropriate function space (quadratic Continuous-Galerkin method), test and trial functions

\lstset{
	language=Python,
	showstringspaces=false,
        breaklines=true,
        frame=tb
	xleftmargin=\parindent,
        basicstyle=\tiny\ttfamily,
        keywordstyle=\bfseries\color{green!40!black},
        commentstyle=\itshape\color{purple!80!black},
        identifierstyle=\color{blue!80!black},
	stringstyle=\itshape\color{orange!90!black},	
	escapeinside={*@}{@*}
}
\begin{lstlisting}
V = FunctionSpace(mesh, "CG", 2) # second order method
u = TrialFunction(V)
v = TestFunction(V)
\end{lstlisting}

We do not need to specify any boundary conditions, but we need to create subdomain $F$, since we integrate over it later. We assume that $F\subset \{z=0\}$ and the mesh is contained in $\{z\le 0\}$. We can use variable $s$ to limit $F$ to $r\le s$. We use this facility to find the eigenvalues of the ice-fishing problem \cite{KK2004}.

\begin{lstlisting}
# x[0]=z, x[1]=r
s = 1 # F contained r<=s
class Steklov(SubDomain): # subdomain F
    def __init__(self,s = 1):
        self.size = s
        super(Steklov,self).__init__()
    def inside(self, x, on_boundary):
        return on_boundary and near(x[1],0) and x[0]<=self.size
steklov=Steklov(s)
boundary = MeshFunction("size_t", mesh,1) # 1D elements
boundary.set_all(0)
steklov.mark(boundary, 1) # mark F for later integration
\end{lstlisting}
Next we create both sides of the weak formulation \eqref{weak2}
\begin{lstlisting}
ds = Measure("ds")[boundary]
r = Expression('x[0]',cell=triangle,degree=1)
# right side
a = ( r**3*inner(grad(u),grad(v)) + r**2*(u*Dx(v,0)+v*Dx(u,0))
    + Constant(m**2+1)*r*u*v )*dx
# left side
b= r**3*u*v*ds(1) # integration over surface (F marked with 1)
A = PETScMatrix()
B = PETScMatrix() 
assemble(a, tensor=A)
assemble(b,tensor=B)
\end{lstlisting}

Finally, we configure and run the eigenvalue solver

\begin{lstlisting}
eigensolver = SLEPcEigenSolver(A,B)
eigensolver.parameters["spectral_transform"] = "shift-and-invert"
eigensolver.parameters["problem_type"] = "gen_hermitian"
eigensolver.parameters["spectrum"] = "smallest real"
eigensolver.parameters["spectral_shift"] = 1E-10
eigensolver.solve(3) # find 3 eigenvalues
eig = eigensolver.get_eigenpair(0) # get the first
\end{lstlisting}
The variable $eig$ contains a tuple of four elements: $\Re \lambda$, $\Im \lambda$, $\Re f$, $\Im f$. Imaginary parts are obviously zero in our case, while $\Re f$ can be turned into the eigenfunction, but it needs to be multiplied by $r$ (if $m>0$)
\begin{lstlisting}  
u = Function(V)
u.vector()[:] = eig[2]
u=project(u*r,V)
\end{lstlisting}
Now function $u$ can be evaluated at any point in the domain using standard Python evaluation, e.g. $u(r,0)$ will give a value at point $(r,0)$ in $F$. We used this to generate plots of the expected heights of the sloshing fluid.

Note that one can find an eigenvalue near a target number $T$ by changing two solver parameters 
\begin{lstlisting}
eigensolver.parameters["spectrum"] = "target real"
eigensolver.parameters["spectral_shift"] = T
\end{lstlisting}
This allows us to optimize the adaptive loop described below for a specific eigenvalue.

\subsubsection{Adaptive mesh refinement}
We choose a small fraction of cells, e.g. 10\% to be refined. The refinement procedure is built into FEniCS, we only need to find the local errors. To acheive this we find indicators that are constant on triangles and equals to the local error for each cell. Then largest 10\% gets marked and refined.

We declare the space of discontinuous, piecewise constant functions, a test function and few helper functions.
\begin{lstlisting}
C = FunctionSpace(mesh,"DG",0) # piecewise constant function
w = TestFunction(C)
h=CellSize(mesh) 
n=FacetNormal(mesh) # normal vectors to edges
r = Expression('x[0]',cell=triangle,degree=1)
ds = Measure("ds")[boundary] 
\end{lstlisting}
Here $boundary$ is the object marked for integration over $F$ in the eigenvalue solving code. Assuming $u$ and $e$ are the just found eigenfunction and eigenvalue we calculate local residual error indicators (\ref{error1}--\ref{error3})
\begin{lstlisting}  
# three kinds of error
# dS inner egdes, ds(1) Steklov part
errform = ( h**2*(div(r*grad(u))-Constant(m**2)*u)**2*w*dx 
            + h*(Dx(u,1)-Constant(e)*u)**2*r*w*ds(1) 
            + avg(h)*jump(grad(u),n)**2*avg(w)*r*dS )
indicators = assemble(errform) # errors for each cell
\end{lstlisting}
Finally we find the cut-off value of the error for the 10\% of worst cells, mark those cells and refine the mesh.
\begin{lstlisting}
fraction=0.1
cutoff = sorted(indicators, reverse=True)[int(len(indicators)*fraction)-1]
marker = CellFunction("bool",mesh)
marker.array()[:] = indicators>cutoff # mark worst errors
mesh = refine(mesh,marker)
\end{lstlisting}

We can perform the refinement based on a single eigenfunction, or add indicators for all eigenfunctions to get the collective refinement. We use both methods in \autoref{sec:numres}. The collective method is faster, but less accurate. While the separate refinements give better results, but we still need to perform a collective search to determine the targets for separate searches. 

\subsubsection{Perimeter refinement and mesh snapping}\label{snapping}

Assuming that the left boundary $R$ is inside $\{r=0\}$, we can select all boundary cells except those touching $\{r=0\}$ and refine them. This increases number of mesh vertices on $B$ (potentially curved boundaty) as well as on the Steklov part. The refinement on the Steklov part $F$ of the boundary is beneficial, since we need to integrate over $F$. While the refinement on curved part $B$ allows us to snap new boundary vertices to the theoretical boundary $B$, getting better shape approximation.

First we extract all triangles with an edge on the boundary.
\begin{lstlisting}
perimeter = [c for c in cells(mesh) if any([f.exterior() for f in facets(c)])]
perimeter = [c for c in perimeter if min([v.x(0) for v in vertices(c)])>0]
\end{lstlisting}
Next we find 25\% triangles with largest diameter and refine those.
\begin{lstlisting}
marker = CellFunction('bool', mesh, False)
diameters = [c.diameter() for c in perimeter]
cutoff = sorted(diameters, reverse=True)[int(len(diameters)*0.25)-1]
for c in perimeter:
    marker[c] = c.diameter()>cutoff
mesh = refine(mesh,marker)
\end{lstlisting}
Whenever we have new boundary vertices, either via adaptive eigenvalue refinement, or perimeter refinement, we can use boundary snapping to achieve better boundary approximation on $B$. The appropriate algorithms are already available in FEniCS, we only need to specify the function describing curved boundary $B$. Assuming $r=f(z)$ describes $B$, we define
\begin{lstlisting}
class Curved(SubDomain):
    def snap(self, x):
        # snap only on B, not on R or F
        if not near(x[0]*x[1],0): 
            x[0]=f(x[1])
\end{lstlisting}
Using this class we can snap all boundary vertices (not satisfying $r=0$ or $z=0$) to the prescribed curve, by simply redefining their $r$ coordinates using $f(z)$. And we apply the snapping function to existing $mesh$. 
\begin{lstlisting}
curve = Curved()
mesh.snap_boundary(curve)
\end{lstlisting}
Before we start solving for eigenvalues on a curved domain (e.g. spherical containers), we perform a few loops containing perimeter refinement and boundary snapping, to improve the initial shape of the domain. Later, perimeter refinement and boundary snapping are also incorporated into each adaptive step. 

\subsubsection{Mesh generation}
FEniCS has many ready to use meshes to choose from, but it is also possible to build a mesh from vertices of a polygon, as long as those vertices are given in a counterclockwise order. This can be used to build a mesh with arbitrary ``curved'' boundary $B$. Say $B$ is described by $r=f(z)$. Then we can take a mesh with vertices $(0,0)$, $(0,-d)$ (where $d$ is the depth of the tank), and a number $n$ of vertices of the form $(f(dk/n),-dk/n)$, with $k=n,\cdots,0$. One can choose a any parametrization, but this one is consistent with the boundary snapping described in the previous section. Having all the points, we create the mesh. 
\begin{lstlisting}
# points = list of tuples (points)
# size = initial mesh size parameter
polygon = Polygon([Point(*p) for p in points])
mesh = Mesh(polygon, size)  
\end{lstlisting}

We can also add a few extra points on the top of the profile $\{z=0\}$ in order to ensure there are enough boundary intervals in $F$ to perform integration over $F$. This is however not strictly necessary, as the mesh created by FEniCS is well balanced, and further refinements will be applied to that part of the  boundary.
\section{Numerical results}\label{sec:numres}
In this section we apply our numerical scheme to various shapes, including spherical and cylindrical tanks. We chose to limit the calculations to around 100 000 triangles with the quadratic method, and 400 000 triangles with the linear method (about 0.5GB memory usage for both methods). 

We recover the exact solutions for cylindrical tanks with great accuracy. We also get nearly the same results as McIver \cite{McI89} in the numerically challenging spherical case (curved boundary). Finally we study quite general domains end use residual norms to check stability and convergence of our approximations.

\subsection{Cylindrical tanks - comparisons to exact solutions. }

\begin{table}
  \centering
  \begin{tabular}{crrrrr}
    \toprule        
    		    & exact        & P2 separate & P2 collective   & P1 separate & P1 collective\\
    \midrule
    $m=0$ & 3.8281081396 & {\tiny3.8281081}405 & {\tiny3.8281081}418 & {\tiny3.8281}387227 & {\tiny3.8281}483323\\
          & 7.0155753606 & {\tiny7.01557536}57 & {\tiny7.01557536}69 & {\tiny7.015}6844394 & {\tiny7.015}6809670\\
          & 10.173468105 & {\tiny10.1734681}24 & {\tiny10.1734681}26 & {\tiny10.173}695372 & {\tiny10.173}706920\\
    \midrule
    $m=1$ & 1.7507975745 & {\tiny1.750797574}6 & {\tiny1.750797574}6 & {\tiny1.75079}89922 & {\tiny1.750}8023457\\
          & 5.3311932955 & {\tiny5.33119329}70 & {\tiny5.33119329}72 & {\tiny5.331}2426129 & {\tiny5.331}2472275\\
          & 8.5363157091 & {\tiny8.5363157}164 & {\tiny8.5363157}170 & {\tiny8.536}4660207 & {\tiny8.536}4704653\\
    \midrule
    $m=2$ & 3.0406821799 & {\tiny3.0406821}800 & {\tiny3.0406821}802 & {\tiny3.04068}64963 & {\tiny3.0406}919358\\
          & 6.7061131204 & {\tiny6.70611312}23 & {\tiny6.70611312}29 & {\tiny6.7061}691529 & {\tiny6.7061}730312\\
          & 9.9694677794 & {\tiny9.9694677}892 & {\tiny9.9694677}900 & {\tiny9.969}6239668 & {\tiny9.969}6306520\\
    \bottomrule
  \end{tabular}
  \caption{Numerical approximations of the three smallest eigenvalues for modes $m=0,1,2$, for a cylinder with radius $1$ and height $1$ compared to true eigenvalues, using either linear $P1$ (400k triangles) or quadratic $P2$ elements (100k triangles). Finally we use a collective adaptive loop, or each eigenvalue separately. }
  \label{tab:cyl}
\end{table}

\pgfplotsset{
large/.style={
xmin=10000, xmax=400000,
axis on top,
ytick={1E-10,1E-8,1E-6,1E-4,1E-2,1},
xtick={10000,20000,40000,80000,160000,320000},
xticklabels={10000,20000,40000,80000,160000,320000},
width=0.45\textwidth,
height=0.45\textwidth,
tick label style = {font=\tiny},
},
smally/.style={
xmin=10000, xmax=400000,
axis on top,
xtick={10000,20000,40000,80000,160000,320000},
xticklabels={},
width=0.33\textwidth,
height=0.35\textwidth,
},
small/.style={smally,
yticklabels={},
},
residuals/.style={ymin=1e-11, ymax=1E0},
diffs/.style={ymin=1e-11, ymax=1E0}
}

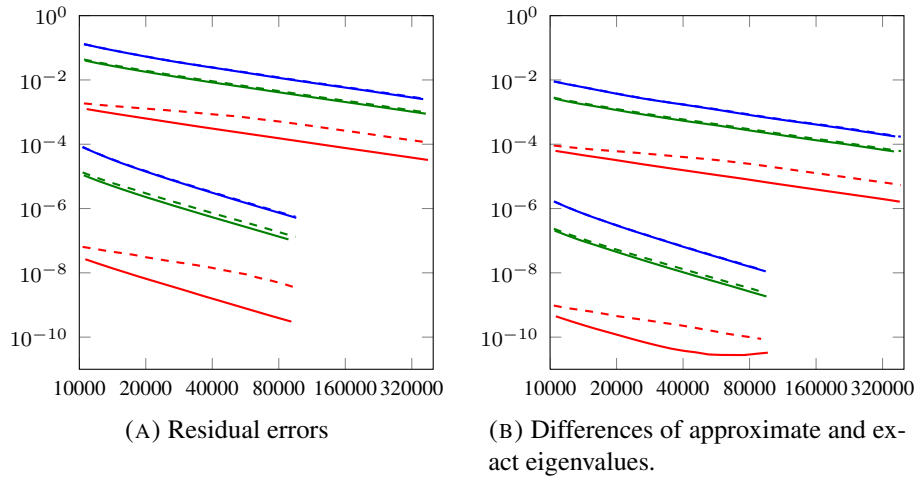
\begin{figure}[t]
  \begin{center}
    \subfloat[Residual errors]{
   
\begin{tikzpicture}
\begin{loglogaxis}[large,residuals]
\addplot [thick, red, dashed]
coordinates {(10352,6.40843655285518e-08)(11808,5.55013783009042e-08)(13460,4.73663094940073e-08)(15367,4.16988811488189e-08)(17542,3.65752966451118e-08)(20054,3.05292622506664e-08)(22860,2.64341071367319e-08)(26097,2.26760183453762e-08)(29801,1.99006863646908e-08)(34016,1.7258060390253e-08)(38719,1.48751495043766e-08)(44069,1.24660344891416e-08)(50140,1.05501906567234e-08)(57113,8.92718770823427e-09)(64972,6.97951127898409e-09)(73779,5.74115112063808e-09)(83796,4.56245532020967e-09)(95125,3.46884009168772e-09)};
\addplot [thick, red]
coordinates {(10649,2.64492883548379e-08)(12205,1.92655158251587e-08)(13994,1.41306630474896e-08)(16054,1.05009387325328e-08)(18437,7.77657141197885e-09)(21143,5.85808854843446e-09)(24235,4.43344950817791e-09)(27750,3.36545141620009e-09)(31744,2.54622356864394e-09)(36265,1.92364585716964e-09)(41468,1.46803468936558e-09)(47365,1.12245472311426e-09)(54059,8.59228576444668e-10)(61656,6.58255793808829e-10)(70297,5.05247649433901e-10)(80153,3.9030321281643e-10)(91224,3.0375426879135e-10)};
\addplot [thick, green!50.0!black, dashed]
coordinates {(10352,1.32741341406789e-05)(11808,9.45503948087724e-06)(13460,6.91620285024758e-06)(15367,5.18899536989871e-06)(17542,3.95707021405633e-06)(20054,2.96294574083435e-06)(22860,2.24061762330893e-06)(26097,1.70708988715624e-06)(29801,1.31314920674399e-06)(34016,1.00338380599857e-06)(38719,7.73711323563825e-07)(44069,5.97395567132152e-07)(50140,4.66667089016076e-07)(57113,3.63664839253617e-07)(64972,2.82187578348235e-07)(73779,2.18563910062994e-07)(83796,1.69132708388205e-07)(95125,1.3189631592949e-07)};
\addplot [thick, green!50.0!black]
coordinates {(10477,1.08226520937592e-05)(12031,7.46458308123074e-06)(13717,5.40404162102856e-06)(15722,3.91756598244221e-06)(17987,2.8844568344197e-06)(20556,2.14301787329448e-06)(23546,1.61113750949731e-06)(26888,1.21079005504371e-06)(30783,9.18313703582948e-07)(35275,6.93252616678571e-07)(40268,5.29141143436747e-07)(46006,4.05792843736854e-07)(52517,3.11080432536426e-07)(59904,2.39100658680894e-07)(68301,1.84133701777514e-07)(77689,1.42089780344777e-07)(88496,1.09665099809878e-07)};
\addplot [thick, blue, dashed]
coordinates {(10352,8.26052647310202e-05)(11808,5.49923127294278e-05)(13460,3.84272848847926e-05)(15367,2.71422572368733e-05)(17542,1.97335216610124e-05)(20054,1.45760023994963e-05)(22860,1.07776082604816e-05)(26097,8.0305926162082e-06)(29801,6.01652355643956e-06)(34016,4.54878928172008e-06)(38719,3.46244084288487e-06)(44069,2.64534553621735e-06)(50140,2.02023828697482e-06)(57113,1.54787001839427e-06)(64972,1.19212427768049e-06)(73779,9.22231475961766e-07)(83796,7.16723497781638e-07)(95125,5.55536563901238e-07)};
\addplot [thick, blue]
coordinates {(10343,8.10297106053958e-05)(11836,5.43256984092677e-05)(13556,3.70033352906121e-05)(15478,2.60503782248575e-05)(17676,1.88609865800777e-05)(20139,1.39101983064989e-05)(23003,1.02679558092993e-05)(26290,7.61505888405444e-06)(30016,5.7198150704999e-06)(34295,4.28199128485418e-06)(39122,3.24194808429304e-06)(44557,2.4631724533643e-06)(50719,1.87463178755969e-06)(57684,1.43999045367889e-06)(65564,1.10931364556791e-06)(74449,8.58678525848352e-07)(84627,6.64073852132253e-07)(96017,5.12672491102616e-07)};
\addplot [thick, red, dashed]
coordinates {(10500,0.00186515525092731)(12117,0.00169453911118167)(13973,0.00155390811212843)(16051,0.00144723128329062)(18449,0.00134164637680764)(21250,0.00124127171735472)(24461,0.00115527892340223)(28186,0.00106570754167663)(32535,0.000975589878815144)(37552,0.000889493734815965)(43248,0.000816373792286289)(49684,0.000753694722307436)(56991,0.000680813226384606)(65375,0.000602426826163852)(74822,0.000540219481191825)(85666,0.000482983126354293)(98175,0.00042304878107274)(112522,0.000371914815621737)(128870,0.000324307732791901)(147499,0.000284116381465634)(168375,0.000251277085579128)(191813,0.000221326100652783)(218177,0.000195131388794947)(247874,0.000171068609000875)(281460,0.000150521038485882)(319533,0.000133362838134475)(362749,0.000118351359043066)};
\addplot [thick, red]
coordinates {(10800,0.00124237529004315)(12464,0.00105366890330098)(14408,0.000896760811806023)(16611,0.000770227603589867)(19222,0.000659351412417382)(22195,0.000567591649419634)(25585,0.000488006864852902)(29444,0.000420451596013452)(33904,0.000363289265213539)(39090,0.000314502065504524)(45071,0.000272392318606196)(51906,0.000236267866272063)(59667,0.000205619487315162)(68566,0.000179077647200293)(78640,0.000156207003690261)(90057,0.000136347936092571)(103034,0.000119030021394965)(117778,0.000103913825920393)(134613,9.07420811864792e-05)(153769,7.93370057490138e-05)(175528,6.95751536699621e-05)(200170,6.10882638740462e-05)(227893,5.37526001782249e-05)(259326,4.7304546562619e-05)(294816,4.16999623247865e-05)(334325,3.68006938112004e-05)(378661,3.24894941946393e-05)};
\addplot [thick, green!50.0!black, dashed]
coordinates {(10500,0.0432325146796026)(12117,0.0352638899833639)(13973,0.029487296481928)(16051,0.0249489064287814)(18449,0.0211839967295041)(21250,0.0180711423502177)(24461,0.0154123938560841)(28186,0.0131319456662427)(32535,0.0112786286855002)(37552,0.00970820890378517)(43248,0.00845196269245923)(49684,0.00734320213380118)(56991,0.00637206550747862)(65375,0.00552809893201969)(74822,0.00482587242424508)(85666,0.00421064878931413)(98175,0.00365267659813809)(112522,0.00317754042314317)(128870,0.0027802271455488)(147499,0.00243991037491267)(168375,0.00215120454249357)(191813,0.00189309563236861)(218177,0.00166207951315945)(247874,0.00146244617154675)(281460,0.00128835807699694)(319533,0.00113475205365031)(362749,0.000997962970678414)};
\addplot [thick, green!50.0!black]
coordinates {(10603,0.0401113805796391)(12232,0.0327436469280561)(14156,0.0271392299772928)(16293,0.022818644640676)(18802,0.0192406766177155)(21641,0.0163625750420084)(24911,0.0139558340561652)(28683,0.0119412437513916)(33061,0.0102440557560393)(38040,0.00882129873849453)(43898,0.00762352890780474)(50501,0.0066128532998904)(58082,0.00574262896307152)(66665,0.00498654153897975)(76342,0.00433454818654837)(87396,0.00378140477400174)(100108,0.00329618778601704)(114483,0.00287399167839393)(130867,0.00250803664252391)(149678,0.00219652922271172)(170927,0.00192919997399959)(195021,0.00169553703716188)(222228,0.00148948679671822)(252975,0.00130809181433842)(287516,0.00114974882465711)(326326,0.00101278933003285)(370540,0.000892284001365071)};
\addplot [thick, blue, dashed]
coordinates {(10500,0.130154913997735)(12117,0.10439211231093)(13973,0.086001011661808)(16051,0.0713062891632803)(18449,0.0597056096142226)(21250,0.0499342353697337)(24461,0.0422116963943666)(28186,0.0358164366043235)(32535,0.0305957947429702)(37552,0.0263526334401398)(43248,0.0227426044872345)(49684,0.0197084626936914)(56991,0.0171142277916608)(65375,0.0148543275592808)(74822,0.0128579422965628)(85666,0.0111443135854371)(98175,0.00969808165165508)(112522,0.0084479473680721)(128870,0.00739114336588026)(147499,0.00648441498055609)(168375,0.00569978260901434)(191813,0.0050151931014656)(218177,0.0044131058481732)(247874,0.00387551329456055)(281460,0.00340014638546919)(319533,0.00298086253944574)(362749,0.00261466600897461)};
\addplot [thick, blue]
coordinates {(10502,0.129944752091882)(12122,0.104201884085292)(13953,0.0855418808059959)(16061,0.0708396794258001)(18486,0.0589729301574595)(21289,0.0493110266075116)(24524,0.0413958143693688)(28286,0.0351399809162014)(32635,0.0300688161126113)(37604,0.0258583051943237)(43229,0.022383836963392)(49584,0.0193413279024676)(56832,0.0167410305407727)(65060,0.0144630111921156)(74545,0.0125037849308898)(85370,0.0108273385642955)(97862,0.00941327446396989)(112013,0.00821682511616125)(128114,0.00719013348699457)(146259,0.00631742818784463)(166686,0.00555639304653339)(189788,0.00488406482658155)(215922,0.00428717204625906)(245534,0.00375251629778172)(278932,0.00328387543615854)(317016,0.0028767972374285)(360474,0.00252404668641106)};
\end{loglogaxis}

\end{tikzpicture}

    }
    \subfloat[Differences of approximate and exact eigenvalues. ]{
\begin{tikzpicture}
\begin{loglogaxis}[large,diffs]
\addplot [thick, red, dashed]
coordinates {(10403,9.58880530532724e-10)(11886,8.04263100562252e-10)(13508,7.12144565540029e-10)(15374,6.22241813630353e-10)(17490,5.35520516820043e-10)(19936,4.53239668019023e-10)(22694,4.00659505572776e-10)(25874,3.5269609455213e-10)(29454,3.10194092634219e-10)(33570,2.70084399289772e-10)(38129,2.38542741115566e-10)(43223,2.06747285957931e-10)(48959,1.77085457409021e-10)(55421,1.52486911986216e-10)(62669,1.29158239658977e-10)(70827,1.13255627098852e-10)(79951,1.00604191644038e-10)(90237,8.81383854789419e-11)};
\addplot [thick, red]
coordinates {(10621,4.43663106253211e-10)(12210,3.24978932653153e-10)(13974,2.42453390697506e-10)(15970,1.84477766396185e-10)(18312,1.4196022135593e-10)(20982,1.09466880005016e-10)(23971,8.46092085282635e-11)(27392,6.64146515561015e-11)(31168,5.33006971892291e-11)(35518,4.35642633078714e-11)(40440,3.73479025483903e-11)(45955,3.34743344154731e-11)(52064,2.89517299023601e-11)(58954,2.79762879529244e-11)(66957,2.7814417435934e-11)(75823,2.77529110803698e-11)(85638,3.03175262672539e-11)(96779,3.2873481714546e-11)};
\addplot [thick, green!50.0!black, dashed]
coordinates {(10403,2.34583746205885e-07)(11886,1.66117749778039e-07)(13508,1.23326626244591e-07)(15374,9.17832325697532e-08)(17490,7.00319349178358e-08)(19936,5.30664268083569e-08)(22694,4.08317406552783e-08)(25874,3.11766754634846e-08)(29454,2.42818050111282e-08)(33570,1.85014270570605e-08)(38129,1.43997738177859e-08)(43223,1.12227089843486e-08)(48959,8.74534844541586e-09)(55421,6.80223255500323e-09)(62669,5.35658717382148e-09)(70827,4.20892742880596e-09)(79951,3.36815730861417e-09)(90237,2.59779753264411e-09)};
\addplot [thick, green!50.0!black]
coordinates {(10476,2.0633417108229e-07)(11999,1.43845698019618e-07)(13707,1.03196031275843e-07)(15714,7.62936949172399e-08)(17967,5.67864466560764e-08)(20506,4.23544515015806e-08)(23399,3.18799413534521e-08)(26750,2.39287656356169e-08)(30529,1.81602937132652e-08)(34832,1.37871225547315e-08)(39594,1.06723740955772e-08)(44978,8.2189748340511e-09)(51124,6.38234798344683e-09)(58057,4.97510743713292e-09)(65853,3.89290999436298e-09)(74612,3.0150175689414e-09)(84322,2.35417019212036e-09)(95366,1.84295068095253e-09)};
\addplot [thick, blue, dashed]
coordinates {(10403,1.67057826416794e-06)(11886,1.12622736025969e-06)(13508,7.83170113649589e-07)(15374,5.65626645965267e-07)(17490,4.13363483886542e-07)(19936,3.05378382137178e-07)(22694,2.26329357033705e-07)(25874,1.6965853077977e-07)(29454,1.28236580465568e-07)(33570,9.67340376689663e-08)(38129,7.34070493280115e-08)(43223,5.65162068255631e-08)(48959,4.37281268972356e-08)(55421,3.38926327003719e-08)(62669,2.64299089280939e-08)(70827,2.07035988353255e-08)(79951,1.60705990737142e-08)(90237,1.2720427378099e-08)};
\addplot [thick, blue]
coordinates {(10467,1.63842191014396e-06)(11984,1.08962958655923e-06)(13668,7.59695568675056e-07)(15583,5.42608095344121e-07)(17818,3.94506137979533e-07)(20343,2.86672387161957e-07)(23266,2.11141644612667e-07)(26595,1.57401279565761e-07)(30341,1.1784437425888e-07)(34652,8.8685847998704e-08)(39372,6.70817517089972e-08)(44688,5.14066620382891e-08)(50688,3.95461405844344e-08)(57556,3.05843457226729e-08)(65258,2.36806574349657e-08)(73900,1.8240148946802e-08)(83725,1.4225886246777e-08)(94712,1.10745617121211e-08)};

\addplot [thick, red, dashed]
coordinates {(10387,9.11601821629304e-05)(11857,8.2204641264072e-05)(13551,7.54967919220917e-05)(15413,6.92275753300109e-05)(17593,6.46765593785847e-05)(20025,6.04800377472792e-05)(22829,5.66254788485487e-05)(26013,5.24418188392328e-05)(29623,4.84176681383452e-05)(33733,4.47129445897421e-05)(38431,4.10689936636288e-05)(43790,3.81852067634636e-05)(49620,3.51533168396756e-05)(56293,3.23895048806033e-05)(63796,2.94491792414142e-05)(72242,2.66215167479267e-05)(81758,2.41403679028807e-05)(92492,2.14355103658193e-05)(104685,1.90082545350467e-05)(118319,1.6913371187588e-05)(133645,1.49997891021325e-05)(150998,1.33167736793283e-05)(170386,1.18770276475999e-05)(191777,1.06482234520922e-05)(215886,9.56465515633553e-06)(242942,8.53261808519967e-06)(273220,7.58027222480528e-06)(306878,6.83641617982289e-06)(344949,6.09155095632197e-06)(387335,5.41521211960827e-06)};
\addplot [thick, red]
coordinates {(10580,6.22287356499829e-05)(12142,5.37480468671969e-05)(13871,4.6386097619866e-05)(15830,4.04874694681645e-05)(18102,3.54518612533106e-05)(20669,3.08350183511052e-05)(23561,2.67901514838798e-05)(26801,2.3335636547106e-05)(30530,2.04766063585105e-05)(34768,1.79311810366567e-05)(39488,1.58293445147173e-05)(44879,1.39154736333857e-05)(50949,1.22595371603307e-05)(57737,1.08460733965199e-05)(65454,9.60737352939667e-06)(74065,8.51874000051822e-06)(83830,7.50098931012744e-06)(94562,6.59872379360493e-06)(106675,5.84115052060241e-06)(120334,5.1836944245931e-06)(135532,4.59549888653932e-06)(152455,4.09926606370625e-06)(171622,3.64219183257575e-06)(192985,3.24592417633163e-06)(216716,2.89720521684522e-06)(243236,2.58637860683919e-06)(272694,2.31866296473981e-06)(305720,2.06839048444429e-06)(342427,1.84437321171238e-06)(383005,1.6363704402611e-06)};
\addplot [thick, green!50.0!black, dashed]
coordinates {(10387,0.00280946263022042)(11857,0.00231406497741204)(13551,0.00196392368356779)(15413,0.00168562204757539)(17593,0.00145407023374133)(20025,0.00124287931763423)(22829,0.00107173628800172)(26013,0.000928160409333856)(29623,0.000805007149239145)(33733,0.00070216609264584)(38431,0.000614340961877957)(43790,0.00054273089764667)(49620,0.00047937849238977)(56293,0.000423884199200586)(63796,0.000372433220191759)(72242,0.000325556641235458)(81758,0.000287227474618845)(92492,0.000254024413608001)(104685,0.000223782243270243)(118319,0.000197412357312743)(133645,0.000174722415336781)(150998,0.00015511361072118)(170386,0.000138693675102886)(191777,0.00012379911008864)(215886,0.000110570716835845)(242942,9.82561821452066e-05)(273220,8.69772575917693e-05)(306878,7.69219482830152e-05)(344949,6.84953966665347e-05)(387335,6.10784538546127e-05)};
\addplot [thick, green!50.0!black]
coordinates {(10520,0.00260418025729514)(12051,0.00213296476899671)(13800,0.00178530595254589)(15801,0.00153456177415023)(18039,0.0013138745674226)(20608,0.00111855269159289)(23510,0.000958017892109808)(26867,0.000827461540310104)(30617,0.000716143642966571)(34903,0.000621645218133793)(39749,0.000543687504718449)(45193,0.000478991033463494)(51422,0.000424436996173227)(58349,0.0003761732060239)(66169,0.000328590428981812)(75045,0.000287794287943122)(84959,0.000252004322494059)(96236,0.000222596631488159)(108961,0.000196112549767591)(123181,0.000172707569928932)(139267,0.000152829241530483)(157229,0.000135540821754709)(177310,0.000121010130095911)(199843,0.000108680668708061)(225164,9.71411360994878e-05)(253449,8.59194848343847e-05)(285604,7.59012824609329e-05)(321247,6.71085779293534e-05)(360519,5.97295609354731e-05)};
\addplot [thick, blue, dashed]
coordinates {(10387,0.00909464436266916)(11857,0.00748505438960478)(13551,0.00629165121313946)(15413,0.00532102644822174)(17593,0.0044861949057271)(20025,0.00376761760393229)(22829,0.00320690436629611)(26013,0.00272634015269801)(29623,0.002356351611839)(33733,0.00205616915692275)(38431,0.00180496455832468)(43790,0.00159637420674308)(49620,0.00140445525630817)(56293,0.00123262938769031)(63796,0.00108142348329743)(72242,0.000938406361166599)(81758,0.000823134033456796)(92492,0.000718435837464781)(104685,0.000633789107912008)(118319,0.000561450496647353)(133645,0.000498507965570383)(150998,0.000447108548783604)(170386,0.00040212134209483)(191777,0.000357812358167564)(215886,0.00031806079494956)(242942,0.000282926516971926)(273220,0.000248179862301967)(306878,0.000219236378868004)(344949,0.000193876318737907)(387335,0.000171335351975443)};
\addplot [thick, blue]
coordinates {(10428,0.00897157824691419)(11916,0.00738750915979125)(13639,0.00621137409216388)(15588,0.00518479724175158)(17784,0.00435654131357488)(20301,0.00365224814842691)(23205,0.00308221190869773)(26524,0.0026234044868545)(30233,0.00227794122413272)(34572,0.00197820734209486)(39391,0.00174298536227546)(44764,0.00152672647955043)(50925,0.00133179870336875)(57909,0.00116537946075645)(65655,0.00100893426236404)(74474,0.000880607796485577)(84552,0.0007643443831995)(95916,0.000670544270445106)(108632,0.000592336709406283)(123124,0.000523933243950836)(139513,0.000467208838406563)(157671,0.000417222904372849)(177904,0.000369811987333435)(200937,0.000327624398686055)(226595,0.000290936436153189)(255360,0.000255052006357559)(287731,0.000225188039047808)(324337,0.000197628509919667)(365363,0.000174791235322047)};

\end{loglogaxis}

\end{tikzpicture}    
    }
  \end{center}
  \caption{Covergence of the adaptive method for $m=1$. Dashed lines indicate one adaptive loop for all eigenvalues, while solid lines show results for each eigenvalue separately optimized. Longer lines show the first order, piecewise linear approximations, shorter the quadratic method. Finally red, green and blue stand for the first, second and third eigenvalues. }
  \label{fig:cyl_conv}
\end{figure}

The sloshing eigenvalues for the cylindrical tank (coffee cup) of radius $1$ and depth $d$ can be calculated using 
\begin{align*}
  \lambda_{m,k} = j_{m,k}' \tanh(d\,j_{m,k}'), 
\end{align*}
where $j_{m,k}'$ is the $k$-th zero of the derivative of the Bessel $J_m$ function. The exact formulas allow us to test our numerical method.

We use a square mesh, with side $1$ to get the results from \autoref{tab:cyl}. The solid of revolution obtained from this profile is a coffee cup with radius $1$ and height $1$. In each approximation we either apply the adaptive procedure to all eigenvalues together, or each one separately. \autoref{tab:cyl} shows almost no difference these strategies for very fine meshes, but the individual approach gives better results for the lowest eigenvalues, as evidenced by \autoref{fig:cyl_conv}. The slopes of the lines on that figure clearly differentiate the orders of the used methods, while being virtually the same in the scope of each method. We get consistent rates of convergence, obviously faster for the quadratic method. Also, note that the quadratic method seems to reach the limits of the floating point arithmetic for the lowest eigenvalue. Finally, squared residual error norm obtained from (\ref{error1}--\ref{error3}) correlates nearly perfectly with the actual approximation error (even the scales on the plots are identical).

Let us also point out that one could simply refine the mesh uniformly to around 100 000 triangles for quadratic method, without using any adaptive strategy. The results one would get are the same as starting with 5000 triangles and performing a few adaptive steps, until just 10 000 triangles are reached. Therefore adaptive approach saves a lot of memory, allowing for a more accurate results. Also, the advantage of the adaptive method should increase with complexity of the domain (e.g. corners). The adaptive approach is even more important for domains that cannot be exactly triangulated, e.g. spherical tanks (see \autoref{sec:sphres}), due to the necessity of the boundary snapping algorithm (improving perimeter approximation).

It is clear that the linear method performs poorly compared to the quadratic. The latter gives nearly exact answers for the lowest eigenvalues of each mode, and only slightly worse approximations for higher eigenvalues. However, one should expect great performance in this case, due to a very basic shape of the container. In the next section we study convergence for more irregular shapes, with no known eigenvalues. There, we will use residual errors to check convergence.

\subsubsection{Ice-fishing problem.}\label{sec:ice}
Before we proceed, we can use the same mesh to find a numerical approximation of the lowest eigenvalue for the ice-fishing problem. Formally, we should consider an infinitely wide and deep ocean covered with ice, with a small round fishing hole. Sloshing in such container was studied by McIver \cite{McI89} and Kozlov-Kuznetsov \cite{KK2004}. 

We will approximate this infinite case using a cylindrical tank with small round hole on top. The only difference from the cylindrical case discussed above is that we need to impose the Steklov condition on a part of the top boundary. More precisely we can take $F=\{(r,z): z=0\text{ and }r<\varepsilon\}$ to get a good approximation of the ice-fishing problem. Now we have a flat top part of the boundary with changing boundary condition, leading to a slower convergence. Nevertheless, for a fixed $\varepsilon$ the approximations stabilize rather quickly, and the domain approximation seems the main source of errors (see \autoref{tab:ice} for a comparison with the results due to McIver \cite{McI89}). Interestingly, only the $m=1$ mode seems hard to estimate, with $m=0$ and $m=2$ very quickly approaching McIver's values, even for relatively large $\varepsilon$. 

The profile of the fundamental mode of the sloshing liquid is shown on \autoref{fig:sphprof}, together with the profiles for spherical tanks filled to various levels. We chose to postpone the plot, since McIver \cite{McI89} treats the ice-fishing problem as a limiting case of almost full spherical tanks.

\begin{table}
  \centering
  \begin{tabular}{cllll}
    \toprule        
       &McIver   &$\varepsilon=1/64$&$\varepsilon=1/16$&$\varepsilon=1/4$ \\
    \midrule
   $m=1$  &  2.75475 &2.754750&2.754465&2.736291 \\
    &  5.89215 &5.892146&5.892041 &5.885388 \\
    &  9.03285  &9.032852 &9.032790 &9.028844 \\
    &  12.1741  &12.17412 &12.17407 &12.17130 \\
    \bottomrule

  \end{tabular}
  \caption{Quadratic approximations of the eigenvalues of the approximate ice-fishing problem: cylinder with height and radius one, and small hole of radius $\varepsilon$. Note that for modes $m=0$ and $m=2$ we obtained the same values as McIver (even for $\varepsilon=1/16$) hence we do not list the results.}
  \label{tab:ice}
\end{table}

\FloatBarrier

\subsection{Other polygonal shapes, and adaptive loop convergence. }
\begin{figure}[t]
  \begin{center}
    \subfloat[Silo\label{fig:silo}]{
    \includegraphics{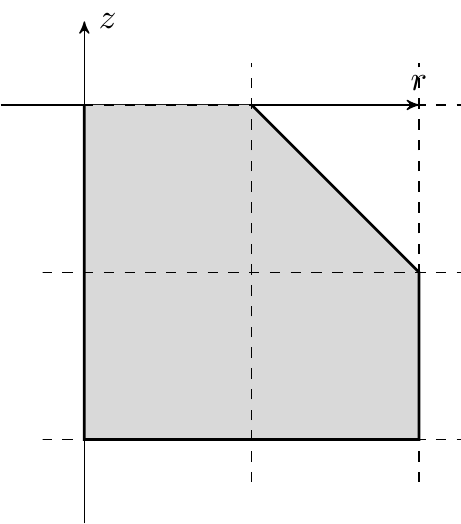}}
\subfloat[Vase\label{fig:vase}]{
\includegraphics{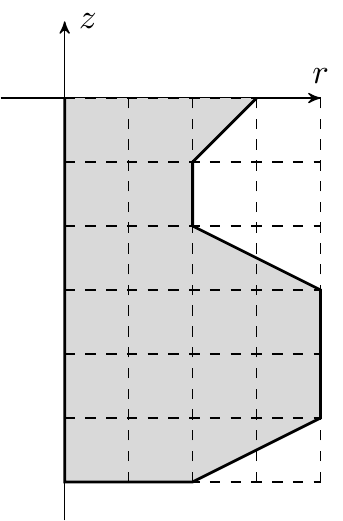}}
\subfloat[Shape with axisymmetric fundamental mode\label{fig:shapesym}]{
\includegraphics{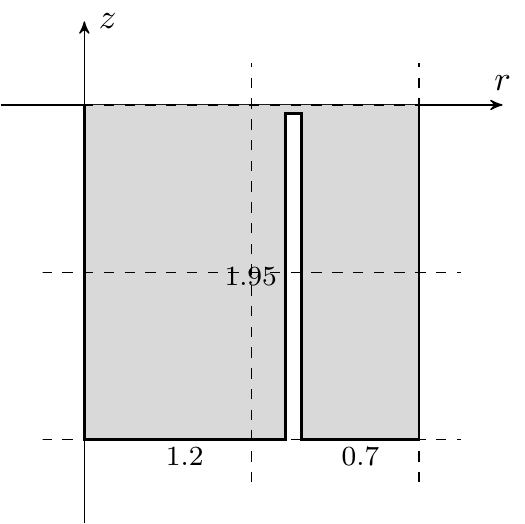}}
  \end{center}
  \caption{Profiles of the tanks.}
  \label{fig:shapes}
\end{figure}
For further testing, we choose a few polygonal shapes from \autoref{fig:shapes}, so that the boundary can be exactly represented using triangulations. Exact eigenvalues are not known, but we can still check stability and convergence of the residual errors. We will use only the optimal, separate adaptive loops for each eigenvalue.

\subsubsection{The silo and the vase.}
\begin{figure}[t]
  \begin{center}
\subfloat[Silo: residuals]{
\hspace{-0.5cm}
\begin{tikzpicture}

\begin{loglogaxis}[smally,residuals]
\addplot [thick, red]
coordinates {(10401,7.2584048178787e-06)(11901,3.81808228079795e-06)(13645,2.09455989349077e-06)(15625,1.2998463543857e-06)(17998,8.44993874752541e-07)(20651,5.87842094145288e-07)(23621,4.24320582462207e-07)(27052,3.14736020495153e-07)(31034,2.37317711130842e-07)(35513,1.79799615429596e-07)(40653,1.37072324046812e-07)(46409,1.05308156887111e-07)(53018,8.04221842579874e-08)(60514,6.15147732336099e-08)(69076,4.71040724921325e-08)(78760,3.60970861866112e-08)(89771,2.76848970386015e-08)};
\addplot [thick, green!50.0!black]
coordinates {(10273,0.000102545836775112)(11738,5.52170200404507e-05)(13381,3.08124929292367e-05)(15231,1.9261446444485e-05)(17382,1.27153270296142e-05)(19817,8.88425651219256e-06)(22585,6.44770477415731e-06)(25742,4.70805449951513e-06)(29248,3.50119456533527e-06)(33298,2.61624358668649e-06)(37991,1.97900515747021e-06)(43238,1.5102010227637e-06)(49162,1.15178643438172e-06)(55798,8.86392832970879e-07)(63455,6.84261162682715e-07)(72056,5.30644776642794e-07)(81778,4.11505396887121e-07)(92690,3.20250833649005e-07)};
\addplot [thick, blue]
coordinates {(10092,0.000414809305471183)(11482,0.000227805568834518)(13025,0.000130420242119142)(14786,8.22287966008462e-05)(16784,5.40383770859276e-05)(19100,3.73113536524611e-05)(21711,2.67196947452453e-05)(24674,1.9429366711791e-05)(28073,1.43538728708133e-05)(31839,1.07950221026668e-05)(36160,8.1670002784093e-06)(41046,6.21327826160769e-06)(46598,4.70685603903721e-06)(52900,3.58215532694617e-06)(59914,2.75761805309843e-06)(67878,2.13175477074092e-06)(76868,1.65019413567219e-06)(86922,1.28418406899598e-06)(98447,9.97095667451803e-07)};
\addplot [thick, red]
coordinates {(10569,0.00990283010034747)(12135,0.00829726786799437)(13951,0.00700698192223459)(16040,0.00595533694750566)(18474,0.00507047931501162)(21254,0.00435421624640352)(24499,0.00375322510951238)(28207,0.00325773081802891)(32445,0.00283754639362838)(37297,0.00246936238223643)(42750,0.00214939006413035)(49057,0.00187095422236952)(56163,0.00162815338258152)(64326,0.00141728224390177)(73660,0.00123440884220254)(84205,0.00107820713613091)(96219,0.000944051371084186)(109855,0.000828382957973208)(125298,0.000728107925761419)(142757,0.000639976848365701)(162282,0.000562874567666685)(184449,0.000495278658553925)(209332,0.000435824232641326)(237393,0.000383682750250097)(269075,0.000338313488006276)(304618,0.00029796734205276)(344698,0.000263194173691433)(390092,0.000232942195733754)};
\addplot [thick, green!50.0!black]
coordinates {(10375,0.0713380968938574)(11904,0.0570158222009143)(13643,0.0464214795157014)(15657,0.0385454927144039)(17945,0.0323070238779511)(20570,0.027290879108537)(23557,0.0232066644691231)(26993,0.0197897481127725)(30954,0.0169089493916387)(35424,0.0145021229658018)(40534,0.0125454537339897)(46477,0.010877276150924)(53142,0.00944587298120583)(60622,0.00823325754505156)(69159,0.0071841576407125)(78781,0.00628861924970505)(89700,0.00551121458834867)(102146,0.00482019091724796)(116290,0.0042234780952661)(132162,0.00369752303288919)(150236,0.00325257541818003)(170695,0.00286660484357008)(193636,0.00252978675392663)(219489,0.00223425704699302)(248716,0.00197204235183924)(281580,0.00174170102552145)(318452,0.00153971424428763)(360058,0.00136125944125888)};
\addplot [thick, blue]
coordinates {(10190,0.19057826486876)(11655,0.150943964022487)(13317,0.122919884048764)(15235,0.100719053950964)(17435,0.0834519725093102)(20025,0.0696420342986464)(22914,0.0586471138040224)(26281,0.0499077648387267)(30103,0.0427024579991596)(34399,0.0366706630230066)(39219,0.0317646247451805)(44655,0.0274776824182064)(50955,0.0237701526068194)(58077,0.0206221114815803)(66162,0.0179053914052807)(75466,0.0155734128347567)(85955,0.0136092538873557)(97679,0.0119513298937799)(110975,0.0105142098783173)(126042,0.00925251404547492)(142834,0.0081488885222704)(161689,0.00717438688127512)(183055,0.00631597410457201)(207158,0.00556390530336521)(234535,0.00489721438193364)(265337,0.00431276940057571)(300165,0.00381187096354275)(339620,0.00337043755229293)(383457,0.00299460730126209)};
\end{loglogaxis}
\end{tikzpicture} \hspace{-0.2cm}}   
  \subfloat[Silo: differences]{
\begin{tikzpicture}

\begin{loglogaxis}[small,diffs]
\addplot [thick, red]
coordinates {(10484,1.61601397152822e-07)(12007,6.45589501857557e-08)(13755,3.81889546652303e-08)(15767,1.95494629373627e-08)(18041,1.3173476531847e-08)(20656,8.4456890370177e-09)(23610,5.97739591157165e-09)(26987,4.29748991948031e-09)(30825,3.22229931626339e-09)(35202,2.39029018800352e-09)(40234,1.78934600469915e-09)(45890,1.2969127993756e-09)(52224,9.34390342877123e-10)(59352,6.53189502486384e-10)(67355,4.53984405623942e-10)(76418,2.94797963817928e-10)(86530,1.66147540170414e-10)(98091,6.9727335016978e-11)};
\addplot [thick, green!50.0!black]
coordinates {(10329,2.36365011208051e-06)(11819,1.0424770646722e-06)(13487,6.29229131021702e-07)(15390,3.51664101394533e-07)(17525,2.34264326159916e-07)(19954,1.57650596577241e-07)(22736,1.13542741431161e-07)(25835,8.21122387861806e-08)(29474,6.06176113748802e-08)(33626,4.45797434522888e-08)(38258,3.25067608386576e-08)(43467,2.31951871043634e-08)(49362,1.66416596059094e-08)(55981,1.14221769820233e-08)(63474,7.69714336712468e-09)(71797,5.01628427684864e-09)(81232,2.8924045381018e-09)(91649,1.29169297480303e-09)};
\addplot [thick, blue]
coordinates {(10165,9.54086799787035e-06)(11562,4.3662471487238e-06)(13211,2.75713283848233e-06)(15081,1.58620937718013e-06)(17218,1.05718724263681e-06)(19582,6.96810655398394e-07)(22235,4.99345087945358e-07)(25279,3.56702072323856e-07)(28743,2.59127245882951e-07)(32597,1.89976493203403e-07)(37042,1.42515242274044e-07)(42139,1.06283225775883e-07)(47810,7.64528103047724e-08)(54269,5.44636371557772e-08)(61490,3.75526525431269e-08)(69429,2.5450278329231e-08)(78398,1.64942015601355e-08)(88501,9.53313161744518e-09)(99956,4.10108391690756e-09)};
\addplot [thick, red]
coordinates {(10369,0.000509226878156355)(11792,0.000431127777695828)(13353,0.000364035536191132)(15198,0.000308548728202407)(17290,0.000260739530994147)(19606,0.000223361926426691)(22263,0.000193924731041406)(25345,0.000169166450509461)(28778,0.000147853544724352)(32620,0.000130465845950223)(37000,0.0001143125292562)(41890,9.97719123296825e-05)(47347,8.69605991020705e-05)(53513,7.51057184933401e-05)(60491,6.42994275295017e-05)(68204,5.51771447216964e-05)(76894,4.74688552016822e-05)(86630,4.11449861341673e-05)(97651,3.56496577045284e-05)(109791,3.06343487750382e-05)(123386,2.63683527608194e-05)(138698,2.25659624355501e-05)(155746,1.89816244318131e-05)(174488,1.58351125891798e-05)(195482,1.2957766454047e-05)(218761,1.025884614414e-05)(245022,7.85497044830663e-06)(273965,5.74873080871896e-06)(306270,3.97633741799908e-06)(342218,2.47931232211229e-06)(382418,1.20032346728749e-06)};
\addplot [thick, green!50.0!black]
coordinates {(10189,0.00477045010908128)(11632,0.00378345222882004)(13217,0.00311949980218085)(15029,0.00262670020488542)(17091,0.00223004286793405)(19397,0.00187426617056108)(22033,0.00159719218140353)(25005,0.00137024371317995)(28360,0.00115521274433661)(32280,0.000979996588738707)(36611,0.000843200692677293)(41490,0.000731410331186844)(46978,0.000638636711102869)(53138,0.0005587187913072)(60097,0.000480802419279946)(67846,0.000415537684958167)(76632,0.000358349475940933)(86536,0.000309287409191228)(97506,0.000263158244204753)(109967,0.000219389434217732)(124030,0.000185465180803313)(139570,0.000156157541749025)(157039,0.000131792345729131)(176715,0.000110651576443743)(198631,9.23523892772948e-05)(223103,7.43485456853321e-05)(250415,5.81177191021354e-05)(281063,4.41659012961892e-05)(315298,3.17809267817992e-05)(353038,2.05096810441674e-05)(395555,9.75074954023825e-06)};
\addplot [thick, blue]
coordinates {(10147,0.0134896779761924)(11500,0.0108482591146277)(13118,0.00884714947567744)(14869,0.00723636645576953)(16895,0.0059716943109418)(19206,0.00501740978174148)(21813,0.0042085911462042)(24746,0.00359262878856015)(28069,0.0031117019131166)(31844,0.00268527225637172)(36101,0.00231255233120109)(40867,0.0019795070661619)(46292,0.00169294205684345)(52430,0.00144845380211223)(59312,0.00124363720384579)(67080,0.00106172256645465)(75754,0.000907961424831072)(85506,0.000783915230462284)(96309,0.000678069839901596)(108697,0.000582436647851736)(122526,0.000497036327590905)(138058,0.000418774257004983)(155330,0.000347659813888512)(174672,0.000285715834790423)(196448,0.000231155141181105)(220869,0.000182537487889434)(248364,0.00014043983460077)(279071,0.000103210724033076)(313217,7.18975427815849e-05)(351068,4.49754222042742e-05)(393246,2.16621926547589e-05)};

\end{loglogaxis}
\end{tikzpicture} \hspace{-0.2cm}} 
    \subfloat[Vase: residuals]{   
\begin{tikzpicture}

\begin{loglogaxis}[small,residuals]
\addplot [thick, red]
coordinates {(10790,4.84566754086121e-05)(12260,2.76984764302789e-05)(13930,1.60991697765911e-05)(15900,9.28708248686879e-06)(18173,5.47948672576073e-06)(20773,3.17984221361348e-06)(23760,1.85571430286167e-06)(27183,1.07001439857999e-06)(31133,5.53230855444237e-07)(35590,3.32301267303106e-07)(40702,1.85746716086039e-07)(46537,1.22562633219489e-07)(53194,6.63533305530794e-08)(60835,4.37232863885455e-08)(69507,2.74179630084864e-08)(79305,1.80579212035886e-08)(90517,1.25312474431997e-08)};
\addplot [thick, green!50.0!black]
coordinates {(10790,1.38837531783033e-05)(12300,8.54897078067649e-06)(14014,5.41111700137505e-06)(15916,3.42898201708025e-06)(18098,2.26087002281017e-06)(20645,1.48258183638118e-06)(23484,9.82343715829263e-07)(26751,6.61283381577462e-07)(30411,4.83265451433631e-07)(34680,3.3377320627669e-07)(39468,2.40054392470809e-07)(44904,1.79064582141844e-07)(51031,1.34592977325046e-07)(57972,1.01592666309401e-07)(65775,7.67417996791624e-08)(74525,5.87123637907314e-08)(84479,4.55301981379036e-08)(95754,3.53548930156612e-08)};
\addplot [thick, blue]
coordinates {(10662,5.92353809690449e-05)(12118,3.86538663476961e-05)(13801,2.52885629536648e-05)(15658,1.72203576498667e-05)(17789,1.17384074128244e-05)(20169,8.47840927060502e-06)(22904,5.74029447211565e-06)(25976,3.96339959115001e-06)(29468,2.8851320395051e-06)(33425,2.16391023626606e-06)(38027,1.60830710687964e-06)(43068,1.19898896652191e-06)(48807,9.07284211601249e-07)(55345,6.91034397962938e-07)(62584,5.26930570581885e-07)(70741,4.06103955166112e-07)(80005,3.11440621456654e-07)(90385,2.39999997635264e-07)};
\addplot [thick, red]
coordinates {(11242,0.000420940536659128)(12944,0.000340129584346895)(14886,0.000281347841282403)(17105,0.000234772221169177)(19689,0.000198622948094008)(22675,0.000169030820694356)(26066,0.000145443289596518)(29999,0.000124940376798607)(34536,0.0001079259532749)(39735,9.34360994197018e-05)(45636,8.13369978277517e-05)(52367,7.07350196346757e-05)(59983,6.14668359271341e-05)(68802,5.34756514259228e-05)(78732,4.67005945633557e-05)(90015,4.07957185128714e-05)(102818,3.56850274462171e-05)(117384,3.11993386890013e-05)(133928,2.73294919317595e-05)(152577,2.40220881641024e-05)(173664,2.1173404134659e-05)(197474,1.86313005564529e-05)(224148,1.64128628708835e-05)(254432,1.44361219913721e-05)(288464,1.271402677117e-05)(326651,1.12314794294847e-05)(369612,9.91957071205727e-06)};

\addplot [thick, green!50.0!black]
coordinates {(10630,0.0176274204245699)(12087,0.0145785327273215)(13740,0.0122864463099967)(15635,0.0103749442822781)(17815,0.00884042083276578)(20253,0.00755297805088961)(23013,0.00644311772969426)(26232,0.00549990936525513)(29869,0.00473501987632347)(33897,0.00412610829817971)(38393,0.00362418198130565)(43603,0.0031801492069458)(49451,0.00279605853710395)(55951,0.00246539770111051)(63477,0.00216589975844869)(71850,0.0018976366800117)(81296,0.00166125843560119)(92104,0.00145484609716537)(104141,0.00127866264357921)(117559,0.0011309380799934)(132502,0.00100673177734907)(149253,0.000896339605928935)(168117,0.000799425049851008)(189303,0.000710983919244156)(212839,0.000633978395602559)(239482,0.000564436424997214)(269172,0.000500484359546182)(302390,0.000443310329426654)(339876,0.000392485012037301)(381993,0.000347304306395994)};

\addplot [thick, blue]
coordinates {(10778,0.0644885277811755)(12303,0.0525887815141283)(14058,0.0432386559145153)(16064,0.0359968830357958)(18328,0.0301725865711141)(20978,0.0254394245787709)(23926,0.0216125892586442)(27331,0.0185151532298691)(31107,0.0160019559018686)(35474,0.0138310232712702)(40362,0.0119806902821224)(46014,0.0103910447554103)(52268,0.00903126423558243)(59392,0.00785816906429489)(67540,0.00684955843193027)(76636,0.00598808989182616)(86944,0.00524735626578482)(98689,0.00460530854842945)(111885,0.00405077807285398)(126606,0.00357271439198462)(143181,0.00314928534855892)(162020,0.00277983569191785)(182931,0.00245832270560694)(206463,0.00216964697496139)(233187,0.00191500925378121)(263072,0.00169015668180086)(296480,0.00149414248237804)(334256,0.001325312849454)(376782,0.00117517098805342)};
\end{loglogaxis}
\end{tikzpicture} \hspace{-0.2cm}}   
  \subfloat[Vase: differences]{
\begin{tikzpicture}

\begin{loglogaxis}[small,diffs]
\addplot [thick, red]
coordinates {(11002,2.12447816028138e-06)(12613,1.2767897578092e-06)(14498,7.13070038016461e-07)(16618,4.28429095644667e-07)(19128,2.39706062554745e-07)(21972,1.44021657821902e-07)(25233,8.07497989585571e-08)(28969,4.85169426855947e-08)(33302,2.72407422174936e-08)(38362,1.63358948590542e-08)(44106,9.14772135818254e-09)(50623,5.44105305255727e-09)(58036,3.00041658185535e-09)(66479,1.72882330673474e-09)(76084,8.91092644117464e-10)(87065,4.48462156299456e-10)(99371,1.56315960175846e-10)};
\addplot [thick, green!50.0!black]
coordinates {(10861,4.13192201342838e-07)(12356,2.52126787003704e-07)(14052,1.54619562042768e-07)(15953,9.59963775137851e-08)(18152,6.02128054083551e-08)(20664,3.82271738708084e-08)(23474,2.45535560772225e-08)(26673,1.59111492870068e-08)(30279,1.04562496439087e-08)(34348,6.9098264887657e-09)(39008,4.62771510001403e-09)(44171,3.1430089642015e-09)(50039,2.10985806603503e-09)(56652,1.42128753211068e-09)(63989,9.36775546023227e-10)(72440,5.81795944754049e-10)(81855,3.2849811759661e-10)(92278,1.36177735754472e-10)};
\addplot [thick, blue]
coordinates {(10822,1.36001197859059e-06)(12365,8.72047379196772e-07)(14074,5.3567019886458e-07)(16040,3.43038800920681e-07)(18194,2.19079510088704e-07)(20734,1.48302055968941e-07)(23536,1.00138506375913e-07)(26689,6.98192872405912e-08)(30278,4.87767568557729e-08)(34332,3.5218646665669e-08)(38933,2.51439766785211e-08)(43945,1.80136527916375e-08)(49733,1.2308756680568e-08)(56116,8.33552071810573e-09)(63305,5.5459672410052e-09)(71526,3.51745388371683e-09)(80586,2.04274464010723e-09)(90672,8.99174956714432e-10)};
\addplot [thick, red]
coordinates {(11077,3.39000913550658e-05)(12620,2.56282767532312e-05)(14398,1.95154203124215e-05)(16429,1.57043176578497e-05)(18693,1.26522021806075e-05)(21311,1.05305111741671e-05)(24337,8.8580643401448e-06)(27727,7.59258698079801e-06)(31540,6.53948923456937e-06)(35856,5.63755386162601e-06)(40753,4.84547266155211e-06)(46295,4.18003177893311e-06)(52556,3.61250087521636e-06)(59628,3.11878729919446e-06)(67468,2.67819925181545e-06)(76270,2.30385971333735e-06)(86299,1.98091319947302e-06)(97563,1.70159846435425e-06)(110256,1.46209310603052e-06)(124357,1.24602819329134e-06)(140149,1.04668387113094e-06)(158021,8.72538529872458e-07)(178007,7.2002592432785e-07)(200088,5.86436520011446e-07)(224883,4.69805253033861e-07)(252552,3.62701633060958e-07)(283116,2.68535794267777e-07)(317303,1.88036822224014e-07)(355614,1.16655414994504e-07)(398262,5.51606966769924e-08)};
\addplot [thick, green!50.0!black]
coordinates {(10492,0.00100982418692785)(11895,0.000831211444963564)(13500,0.000702080995252263)(15243,0.000603415921972772)(17236,0.000519474944081555)(19468,0.000445315354774767)(21919,0.000375278658727218)(24674,0.000319452414264898)(27797,0.000273334045032048)(31428,0.00023477866295174)(35443,0.000204382866571429)(39985,0.000179506505329829)(45039,0.000157832034716421)(50662,0.000139336839721338)(56939,0.000122579767219744)(63855,0.000107535486824073)(71625,9.26658301536598e-05)(80347,7.91342302548514e-05)(89967,6.70587085411789e-05)(100691,5.7313410571691e-05)(112850,4.85302829940082e-05)(126399,4.16081607532703e-05)(141569,3.56749924215194e-05)(158289,3.0439043730901e-05)(177150,2.57561003556006e-05)(197943,2.15651331916256e-05)(221097,1.77225829851579e-05)(246373,1.42559717617452e-05)(274718,1.08064153105936e-05)(306367,7.56564518766822e-06)(341093,4.6285830346271e-06)(379935,2.14585882041263e-06)};
\addplot [thick, blue]
coordinates {(10582,0.00441874252238783)(12011,0.00359047850033889)(13599,0.00296550387861672)(15392,0.00247219635952511)(17392,0.00206287654142212)(19732,0.00172584908273787)(22382,0.00147557470233295)(25389,0.00126660703293613)(28703,0.00109880985016808)(32432,0.000945231285554016)(36620,0.000818943935365191)(41243,0.000703901259884177)(46466,0.000610701664109037)(52336,0.000525332868058115)(58976,0.000445657226888052)(66473,0.000382749404504779)(74861,0.000329931257406102)(84292,0.000284447213875438)(94847,0.000247899200745749)(106463,0.000211977347622572)(119492,0.000180395007093992)(133907,0.000153448902113773)(149918,0.000127521413848974)(167851,0.000106339270325329)(188088,8.7124741910749e-05)(210680,6.82609749418361e-05)(235906,5.18833728166967e-05)(263894,3.82928897826318e-05)(295050,2.64827962732639e-05)(330030,1.6534341007457e-05)(368647,8.42498602526831e-06)};
\end{loglogaxis}

\end{tikzpicture} \hspace{-0.1cm}}     
  \end{center}
  \caption{Convergence for the three lowest eigenvalues of the mode $m=1$ (linear and quadratic methods). Axis ranges for all plots are the same as for cylindrical tanks, \autoref{fig:cyl_conv}.}
  \label{fig:silo_vase}
\end{figure}
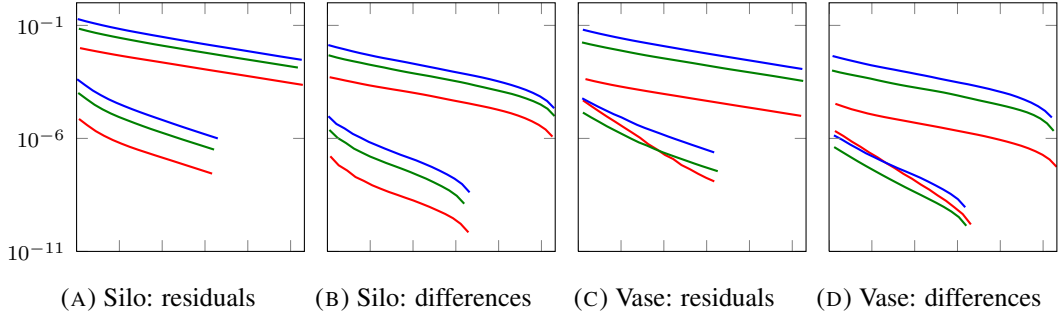

\autoref{fig:silo_vase} shows the residual errors and the differences to the best found approximation for adaptive loops on the silo (\autoref{fig:silo}) and the vase (\autoref{fig:vase}). The scales on the plots are identical to the cylindrical data from \autoref{fig:cyl_conv}. The lines for the silo are nearly identical to the ones for the cylindrical tank, with generally slightly higher errors. The vase also shows similar performance, but with the lowest eigenvalue showing higher initial residual errors. This seems to indicate that the reentrant corners (over 180 degrees) inside of the vase are a significant challenge for the numerical method. This is not unexpected, since smoothness of the theoretical solutions of any Laplace-type eigenvalue problem deteriorates significantly near reentrant corners. See Fox-Henrici-Moler \cite{FHM} and Betcke-Trefethen \cite{BT} for the significance of the reentrant corners and a different numerical method of handling these corners.

Note that the plots of the differences are almost straight, as expected. Except for the ends of the lines which show unwarranted improved convergence. The difference to the best approximation is nearly the same as the difference to the exact value when the errors are large. But once we reach the limits of the approximation scheme, we can see false convergence to overestimated best approximation. This problem does not apply to the residual errors, which exhibit consistent decay rates throughout the whole adaptive loop.

\subsubsection{A shape with the axisymmetric fundamental sloshing mode.}
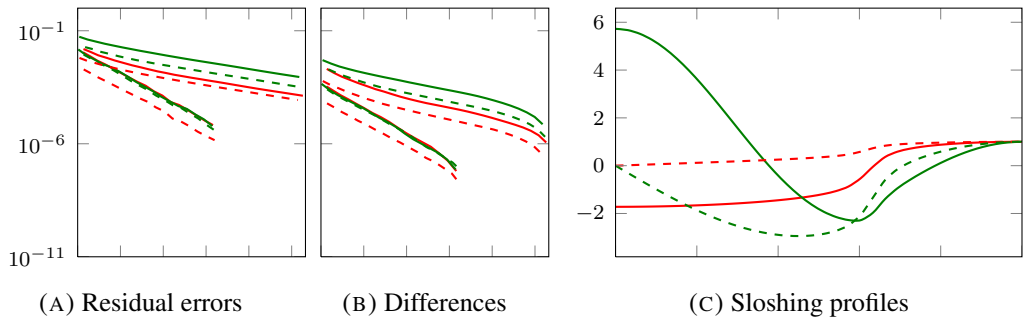
\begin{figure}[t]
  \begin{center}
    \subfloat[Residual errors\label{fig:symres}]{   
\begin{tikzpicture}
\begin{loglogaxis}[smally,residuals]
\addplot [thick, red]
coordinates {(10916,0.0111267911630643)(12229,0.00742520445093017)(13747,0.00474940304365768)(15554,0.00329868125773329)(17668,0.00215019245988458)(20157,0.00135311792968312)(23007,0.000856641225165901)(26313,0.000547942272727502)(30106,0.000368174225238385)(34451,0.000206849022100458)(39411,0.000135881480153923)(45083,6.99047223664283e-05)(51650,5.20902830631337e-05)(59285,3.13223609775733e-05)(68105,1.81813005986258e-05)(78103,1.0737491584707e-05)(89489,6.70714762923148e-06)};
\addplot [thick, green!50.0!black]
coordinates {(10106,0.0148768928213184)(11516,0.00815573441134882)(13109,0.00535061907294564)(14958,0.00337956263437551)(17098,0.00245362739171195)(19616,0.00156810092727585)(22484,0.00094031498263365)(25765,0.00054422035237939)(29483,0.000339336828609197)(33810,0.00021281130700145)(38781,0.000119986222872809)(44484,7.49326856791467e-05)(51082,4.64710231459149e-05)(58607,3.48082800282642e-05)(67270,1.97511221895772e-05)(77115,1.21081582158289e-05)(88534,6.07272248593958e-06)};

\addplot [thick,dashed,red]
coordinates {(11032,0.00193931608885726)(12402,0.00130569943255551)(14037,0.000841880482797416)(15935,0.000570369505225651)(18137,0.000376150301883246)(20741,0.000243408769910652)(23696,0.000161381349300628)(27058,0.000104681629591778)(30983,7.25001114572897e-05)(35474,4.07458576418595e-05)(40572,2.73649120064045e-05)(46440,1.30497719569603e-05)(53185,9.68459261162171e-06)(60989,6.01361342834012e-06)(69884,3.42058597278106e-06)(80121,2.18538152162675e-06)(91837,1.44337206909289e-06)};
\addplot [thick,dashed, green!50.0!black]
coordinates {(11013,0.00884040438059237)(12419,0.00582866627560091)(14032,0.00368690653044438)(15941,0.00264591521655352)(18110,0.00169835499171547)(20664,0.00102991236113568)(23624,0.000608388472891568)(27038,0.000380959179631672)(30993,0.000242946680517728)(35527,0.00013667245583344)(40764,8.62252405594562e-05)(46713,5.07011787971434e-05)(53538,3.80560949489767e-05)(61467,2.18121044550244e-05)(70482,1.3138018450985e-05)(80787,6.81762794533247e-06)(92541,3.81228396596518e-06)};

\addplot [thick, red]
coordinates {(10989,0.0154985513448922)(12300,0.0116222597336473)(13961,0.0081812650575615)(15975,0.00633592230992052)(18318,0.00479350541976287)(21043,0.00376216333368875)(24137,0.00296119757084044)(27709,0.0024159372440015)(31717,0.00196154020217633)(36382,0.0016000218928582)(41780,0.00135201389132823)(47895,0.00112735304345541)(54963,0.000972143982501756)(63020,0.000836066668582644)(72138,0.000729953407587751)(82577,0.000633808027325017)(94424,0.000553109012317577)(107623,0.000482119185246031)(122544,0.00041926358940442)(139647,0.000364509751618225)(158926,0.000317555456543103)(180675,0.00027777296852731)(205325,0.000244308479154922)(233152,0.000215978796471982)(264432,0.000191436651129373)(299676,0.000169711432137351)(339374,0.000150324462632513)(383700,0.000133206682062421)};
\addplot [thick, green!50.0!black]
coordinates {(10262,0.0542181234768271)(11759,0.0408065401780107)(13463,0.0330658099542006)(15520,0.0265863835907112)(17877,0.0220682961267767)(20605,0.0183637745378398)(23772,0.0154597374451783)(27412,0.0130121818407021)(31636,0.0110792406251474)(36447,0.00943610445794547)(41809,0.00807801895548063)(48170,0.0069337593041311)(55427,0.00596354471647349)(63787,0.00514685319005305)(73366,0.00447172262139892)(84224,0.00389811373786305)(96588,0.0033996097385802)(110873,0.00296142915241627)(127057,0.0025804724136525)(144940,0.00225165323662936)(165483,0.0019630539663888)(188941,0.0017125223020787)(215592,0.0014964111813015)(245776,0.00131154810297959)(280031,0.00115243470278516)(318539,0.00101773973777437)(362061,0.000898655566990786)};
\addplot [thick,dashed, red]
coordinates {(10302,0.00626471596539102)(11752,0.0045824725252659)(13456,0.00365323195900464)(15445,0.00284749607508625)(17760,0.00231222244308149)(20455,0.00186648446449865)(23584,0.00153809287166824)(27204,0.00127669198197402)(31302,0.00108092316172312)(36077,0.000918200851389388)(41582,0.000779360751557631)(47811,0.000669481785763031)(54987,0.000572717986774758)(63289,0.000495159505583054)(72724,0.000426672460642245)(83580,0.000369441721856244)(95894,0.000320764340993771)(109926,0.000279307501033944)(125775,0.000244451630021653)(143851,0.000214091374993578)(164299,0.000187571318484361)(187213,0.000164229051442769)(213194,0.00014395470975511)(242521,0.000126465966427887)(275493,0.00011120434231232)(312536,9.78331769488917e-05)(354129,8.6170669676716e-05)};
\addplot [thick,dashed, green!50.0!black]
coordinates {(11278,0.0188396308374681)(12854,0.0152347987971325)(14684,0.0114879428000261)(16857,0.00938647294908202)(19370,0.0074678562993875)(22264,0.00615874033474369)(25644,0.00507985815878116)(29522,0.00431994564575131)(34032,0.00365684116101774)(39159,0.00313298790936183)(45097,0.00271055121347837)(51779,0.00233813627153838)(59509,0.00202875803515177)(68291,0.00175442713822412)(78288,0.00152261014257012)(89677,0.00132456525099905)(102651,0.00115992940679609)(117508,0.00101544318338593)(134384,0.000889078934789738)(153440,0.000779995538284174)(174949,0.000684431979124528)(199120,0.000600437760490911)(226742,0.000526278675955773)(257695,0.000461405578042902)(292685,0.000404576362926373)(332144,0.000356254736624035)(376728,0.000314389589459397)};
\end{loglogaxis}
\end{tikzpicture}\hspace{-0.3cm} }   
\subfloat[Differences\label{fig:symdif}]{   
\begin{tikzpicture}
\begin{loglogaxis}[small,diffs]
\addplot [thick, red]
coordinates {(10916,0.000345555597273894)(12229,0.000221573941792452)(13747,0.000140369384786232)(15554,9.30625599517843e-05)(17668,5.91145132085025e-05)(20157,3.75740403973923e-05)(23007,2.3049699673372e-05)(26313,1.52093696242916e-05)(30106,9.35954611891443e-06)(34451,5.28885826056591e-06)(39411,3.05785043408502e-06)(45083,1.86543423541519e-06)(51650,1.17200941651685e-06)(59285,7.3439740155834e-07)(68105,3.66473982493787e-07)(78103,1.75296776205869e-07)(89489,6.22763588475905e-08)};
\addplot [thick, green!50.0!black]
coordinates {(10106,0.000445306767042908)(11516,0.000238944239475458)(13109,0.000143541791882917)(14958,9.26886776397851e-05)(17098,6.07126669933677e-05)(19616,3.92538908653428e-05)(22484,2.35073801091445e-05)(25765,1.39774308025586e-05)(29483,9.13312537065281e-06)(33810,5.44248488809629e-06)(38781,2.85549595169243e-06)(44484,1.5900463421481e-06)(51082,9.92617425144005e-07)(58607,6.21049337734547e-07)(67270,3.85775817512979e-07)(77115,2.01577109226747e-07)(88534,7.27014275447857e-08)};

\addplot [thick,dashed,red]
coordinates {(11037,6.1448457674862e-05)(12405,4.0618819481586e-05)(14037,2.51998777205387e-05)(15947,1.68139856816518e-05)(18218,1.04551537476461e-05)(20872,6.95103769310368e-06)(23899,4.32991694532259e-06)(27365,2.81464941276344e-06)(31419,1.73135107728317e-06)(36062,1.13838559068213e-06)(41435,6.95936797412422e-07)(47616,4.29809558055716e-07)(54705,2.37620419119011e-07)(62906,1.57995877492212e-07)(72430,8.7203142151715e-08)(83261,4.83352659008673e-08)(95747,1.57022012059471e-08)};
\addplot [thick,dashed, green!50.0!black]
coordinates {(11057,0.000255122318544299)(12447,0.000156171526207505)(14131,0.000100236930086872)(16106,6.53708988862522e-05)(18357,4.18918875635388e-05)(20992,2.74764634460301e-05)(24081,1.75869646699134e-05)(27653,1.05440910276311e-05)(31738,6.32685501189556e-06)(36467,4.124498376612e-06)(41860,2.43025208379599e-06)(48137,1.29898894041247e-06)(55275,7.1059604711543e-07)(63651,4.49531444512985e-07)(73293,2.57136320680473e-07)(84256,1.48131352295522e-07)(96981,6.2204958961587e-08)};

\addplot [thick, red]
coordinates {(10989,0.00205063379092107)(12300,0.00138234578470242)(13961,0.00093632100390495)(15975,0.000672209300118798)(18318,0.000480990783921276)(21043,0.000355195749062953)(24137,0.000261224378349079)(27709,0.000203563330368528)(31717,0.000155444847715236)(36382,0.000119619004519289)(41780,9.49924165732918e-05)(47895,7.69800035035351e-05)(54963,6.30921390434303e-05)(63020,5.32889666545744e-05)(72138,4.50274900948289e-05)(82577,3.79073869832824e-05)(94424,3.18852502257938e-05)(107623,2.65288663049423e-05)(122544,2.1718105009505e-05)(139647,1.75862180143982e-05)(158926,1.40716846566136e-05)(180675,1.11205763939637e-05)(205325,8.76968699659031e-06)(233152,6.83327391681754e-06)(264432,5.22023502114166e-06)(299676,3.74196023389395e-06)(339374,2.37710579265027e-06)(383700,1.1446218052269e-06)};
\addplot [thick, green!50.0!black]
coordinates {(10262,0.0050253665997797)(11759,0.00353814450324563)(13463,0.00262247176678176)(15520,0.00204652096726621)(17877,0.00162735983812601)(20605,0.001323297809396)(23772,0.00108688864390594)(27412,0.000897000029219441)(31636,0.000749267526944486)(36447,0.000624931719860555)(41809,0.000519649076073048)(48170,0.000434947230178118)(55427,0.000364370100664146)(63787,0.000305680411417431)(73366,0.000257678701554021)(84224,0.000218358302295218)(96588,0.00018419829363614)(110873,0.000153879813603819)(127057,0.000126897950107452)(144940,0.000102847301377551)(165483,8.16823023779278e-05)(188941,6.38187868942097e-05)(215592,4.85603521358513e-05)(245776,3.55950674513217e-05)(280031,2.44548208678808e-05)(318539,1.533095181383e-05)(362061,7.30337045107277e-06)};
\addplot [thick, dashed, red]
coordinates {(10302,0.000595197361498712)(11752,0.000388605368884631)(13456,0.000278247261349951)(15445,0.000191604482266627)(17760,0.000141278529543154)(20455,0.000101996559093043)(23584,7.77796073236914e-05)(27204,5.83951091069901e-05)(31302,4.65311296169446e-05)(36077,3.70182299291866e-05)(41582,2.98522636283938e-05)(47811,2.41165293520806e-05)(54987,1.97745878561806e-05)(63289,1.62115245837535e-05)(72724,1.33875047407805e-05)(83580,1.10327611014815e-05)(95894,9.12950375642207e-06)(109926,7.49431700597381e-06)(125775,6.20401890127908e-06)(143851,5.16007441020427e-06)(164299,4.23164710561252e-06)(187213,3.31934260988653e-06)(213194,2.52341923767663e-06)(242521,1.85308667266515e-06)(275493,1.27591650489478e-06)(312536,7.98656447753565e-07)(354129,3.7372175132333e-07)};
\addplot [thick, dashed, green!50.0!black]
coordinates {(11278,0.00195622444671817)(12854,0.00136806966445979)(14684,0.00098394351644826)(16857,0.000734982396125794)(19370,0.000553084210565435)(22264,0.000423765358303463)(25644,0.000330432386353552)(29522,0.000270128171555584)(34032,0.000220235051936912)(39159,0.000181301639170739)(45097,0.000151714322609076)(51779,0.000127017882543212)(59509,0.000106609026244264)(68291,8.92181656975133e-05)(78288,7.44734122868351e-05)(89677,6.22496334246048e-05)(102651,5.24843069118486e-05)(117508,4.41383062954603e-05)(134384,3.67939484762214e-05)(153440,3.03243085046923e-05)(174949,2.4519977616988e-05)(199120,1.93271733806011e-05)(226742,1.48282059311278e-05)(257695,1.08742268869477e-05)(292685,7.40550713462795e-06)(332144,4.51989867022995e-06)(376728,2.03776268659084e-06)};
\end{loglogaxis}
\end{tikzpicture} }   
\subfloat[Sloshing profiles\label{fig:symprof}]{
\begin{tikzpicture}

\begin{axis}[
xmin=0, xmax=1,
axis on top,
xticklabels={},
height=0.35\textwidth,
width=0.50\textwidth,
smooth,
]
\addplot [thick, red]
coordinates {(0,-1.72112851074149)(0.0301507537688442,-1.71977117551513)(0.0603015075376885,-1.71569170932143)(0.0904522613065326,-1.70885957544158)(0.120603015075377,-1.69922423174229)(0.150753768844221,-1.68671041332692)(0.180904522613065,-1.67121432727332)(0.21105527638191,-1.65259669465991)(0.241206030150754,-1.63067276769796)(0.271356783919598,-1.60519897368979)(0.301507537688442,-1.57585047978561)(0.331658291457286,-1.54218880642239)(0.361809045226131,-1.50360888803636)(0.391959798994975,-1.4592514994243)(0.422110552763819,-1.40784811854987)(0.452261306532663,-1.34742792117118)(0.482412060301508,-1.2747036105486)(0.512562814070352,-1.18362021319986)(0.542713567839196,-1.06126894800199)(0.57286432160804,-0.873621064604099)(0.603015075376884,-0.529648746495825)(0.633165829145729,-0.0463688676848797)(0.663316582914573,0.364421393438975)(0.693467336683417,0.585180526169833)(0.723618090452261,0.709297891001176)(0.753768844221106,0.791784169222853)(0.78391959798995,0.851452816247289)(0.814070351758794,0.896351156775161)(0.844221105527638,0.930592382212438)(0.874371859296482,0.956537244515441)(0.904522613065327,0.975647528535695)(0.934673366834171,0.988870751863642)(0.964824120603015,0.996832292665937)(0.994974874371859,0.999936160625229)(1,1)};
\addplot [thick, green!50!black]
coordinates {(0,5.7241983296977)(0.0301507537688442,5.67155372808136)(0.0603015075376885,5.51508392282185)(0.0904522613065326,5.2590905178501)(0.120603015075377,4.91060576223945)(0.150753768844221,4.47916915298712)(0.180904522613065,3.97654016433694)(0.21105527638191,3.41633793241469)(0.241206030150754,2.81362779356267)(0.271356783919598,2.18446359275902)(0.301507537688442,1.54540742394373)(0.331658291457286,0.913031046871441)(0.361809045226131,0.303420919129754)(0.391959798994975,-0.268293757853742)(0.422110552763819,-0.788383445587427)(0.452261306532663,-1.24491596974246)(0.482412060301508,-1.62809456947037)(0.512562814070352,-1.93051570856139)(0.542713567839196,-2.14726121390142)(0.57286432160804,-2.27459952498028)(0.603015075376884,-2.2815599437441)(0.633165829145729,-1.9707281679278)(0.663316582914573,-1.39272602287442)(0.693467336683417,-0.946386382914164)(0.723618090452261,-0.600865071020044)(0.753768844221106,-0.297678603684928)(0.78391959798995,-0.0220243567088687)(0.814070351758794,0.227589930130414)(0.844221105527638,0.448442920951203)(0.874371859296482,0.636643041874006)(0.904522613065327,0.788392861964522)(0.934673366834171,0.900604850733281)(0.964824120603015,0.971228647851426)(0.994974874371859,0.999416766471049)(1,1)};
\addplot [thick,dashed, red]
coordinates {(0,-1.09672484427392e-08)(0.0301507537688442,0.0178560190924424)(0.0603015075376885,0.0357383097193749)(0.0904522613065326,0.0536757335015214)(0.120603015075377,0.0716981793299647)(0.150753768844221,0.0898418067893347)(0.180904522613065,0.108148628971409)(0.21105527638191,0.126670811679043)(0.241206030150754,0.145472826544292)(0.271356783919598,0.164639640301842)(0.301507537688442,0.184278723461048)(0.331658291457286,0.204538359888438)(0.361809045226131,0.225620526223318)(0.391959798994975,0.247812680326352)(0.422110552763819,0.271541272244618)(0.452261306532663,0.29746782696603)(0.482412060301508,0.326701754395034)(0.512562814070352,0.36128433288229)(0.542713567839196,0.405566136587048)(0.57286432160804,0.47092204087152)(0.603015075376884,0.585948176820234)(0.633165829145729,0.730997857604898)(0.663316582914573,0.835969809687008)(0.693467336683417,0.889290685726148)(0.723618090452261,0.9198133758177)(0.753768844221106,0.940796373891502)(0.78391959798995,0.956572727879021)(0.814070351758794,0.968910615733118)(0.844221105527638,0.978689310752248)(0.874371859296482,0.986367211557031)(0.904522613065327,0.992211716132704)(0.934673366834171,0.9963804881761)(0.964824120603015,0.998954044574697)(0.994974874371859,0.999978467725144)(1,1)};
\addplot [thick,dashed , green!50!black]
coordinates {(0,2.89403938147359e-07)(0.0301507537688442,-0.30254610571426)(0.0603015075376885,-0.601977115296158)(0.0904522613065326,-0.895207051986949)(0.120603015075377,-1.17920826651013)(0.150753768844221,-1.45103835923296)(0.180904522613065,-1.70786614418391)(0.21105527638191,-1.94697911303101)(0.241206030150754,-2.16583272731855)(0.271356783919598,-2.36202731709518)(0.301507537688442,-2.53331594292915)(0.331658291457286,-2.67758050138674)(0.361809045226131,-2.79287348205559)(0.391959798994975,-2.87725712572576)(0.422110552763819,-2.92857073146177)(0.452261306532663,-2.94434361149348)(0.482412060301508,-2.92093442929863)(0.512562814070352,-2.85203079060082)(0.542713567839196,-2.72400982647629)(0.57286432160804,-2.49924286225264)(0.603015075376884,-2.06146153813331)(0.633165829145729,-1.33733688316003)(0.663316582914573,-0.598621013134138)(0.693467336683417,-0.150150566337937)(0.723618090452261,0.133697967389187)(0.753768844221106,0.343637878440933)(0.78391959798995,0.510087841539701)(0.814070351758794,0.645368951383698)(0.844221105527638,0.755344876399699)(0.874371859296482,0.843112093701875)(0.904522613065327,0.910464989391088)(0.934673366834171,0.958532028898197)(0.964824120603015,0.98809522509582)(0.994974874371859,0.999759130837872)(1,1)};
\end{axis}

\end{tikzpicture}

}
  \end{center}
  \caption{The two lowest eigenvalues of the modes $m=0$ (solid lines) and $m=1$ (dashed lines) found using linear and quadratic approximations, with separate adaptive loops. Note that the lowest eigenvalue belongs to $m=0$, solid red line on the rightmost plot.}
  \label{fig:sym}
\end{figure}

The last shape (from \autoref{fig:shapesym}) is the most interesting, both numerically and theoretically. Due to very narrow passage and reentrant corners we might expect slower convergence (compared to a cylindrical tank). Instead, \autoref{fig:symres} and \autoref{fig:symdif} show unchanged convergence rates, but with much higher initial errors (especially in the quadratic method). Notice also that the errors for the quadratic method are almost independent of the eigenvalue. This shows that the adaptive loop tries to refine the narrow passage (and reentrant corners) before tackling the Steklov boundary $F$. We still see that the residual error estimators correlate almost perfectly with the differences.

The profiles of the sloshing fluid for the last shape (\autoref{fig:symprof}) are the most interesting aspect of the tank. The lowest eigenvalue belongs to the mode $m=0$, not to $m=1$, as was the case in all other shapes (see \autoref{tab:symeig}). Therefore, we found an example for which $\nu_1\ne \lambda_{1,1}$ and the results of \cite{KK2012} do not apply (see \autoref{rigor}). However, our example is rather strange, and almost all practical shapes should satisfy the conditions imposed in \cite{KK2012}. 

We see that the eigenfunction for the lowest eigenvalue attains its maximum at the center of the tank, and the global high spot is almost twice the height of the local boundary maximum.

\begin{table}
  \centering
  \begin{tabular}{ccc}
    \toprule        
    & linear P1 & quadratic P2 \\
    \midrule
    $m=0$ & 1.046054, 6.366773  & 1.046038, 6.366661\\
    $m=1$ & 1.238800, 3.782383  & 1.238794, 3.782350\\
    \bottomrule
  \end{tabular}
  \caption{Approximations of the eigenvalues of the shape from \autoref{fig:symprof}. The smallest eigenvalue belongs to $m=0$!}
  \label{tab:symeig}
\end{table}

\subsubsection{Sizes of triangles created by the adaptive loop.}
\begin{figure}[t]
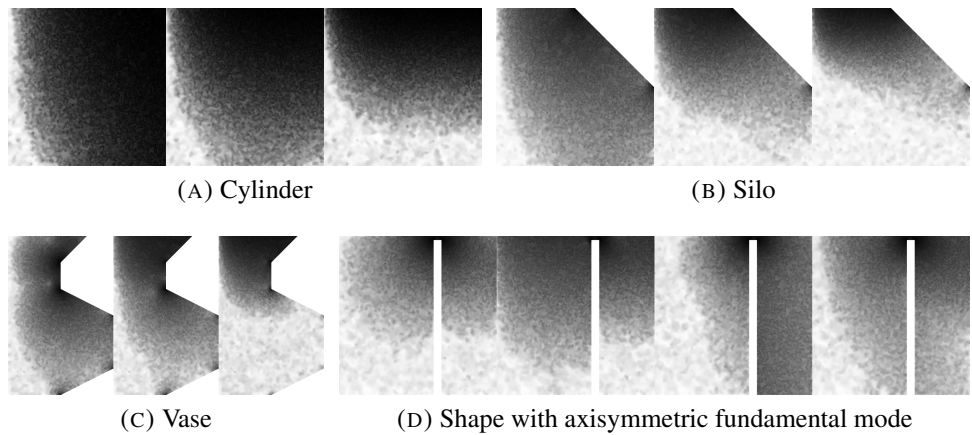

  \begin{center}
\subfloat[Cylinder]{
\includegraphics[height=0.15\textwidth]{{{cylindrical1.0}}}
}
\subfloat[Silo]{
\includegraphics[height=0.15\textwidth]{{{silo}}}
}

\subfloat[Vase\label{fig:denvase}]{
\includegraphics[height=0.15\textwidth]{{{vase}}}
}
\subfloat[Shape with axisymmetric fundamental mode\label{fig:densym}]{
\includegraphics[height=0.15\textwidth]{{{symmetric}}}
}
  \end{center}
  \caption{Sizes of the mesh triangles (darker color indicates smaller triangles). Adaptive loop refines mostly triangles near any corner and near $F$. For each domain we show the adaptively refined mesh for the smallest three eigenvalues of the mode $m=1$ (except for the last shape with two lowest eigenvalues for $m=0, 1$).}
  \label{fig:density}
\end{figure}
We end our investigations with triangle density plots, \autoref{fig:density}. The adaptive loop creates highly nonuniform meshes, with many small triangles in most important parts of the domain. It is clear that the free surface $F$ is the key to good approximation. However, any corners in the boundary of the domain are also heavily refined by the adaptive loop. This is especially noticable on the last two shapes (\autoref{fig:denvase} and \autoref{fig:densym}).

Note also, that the higher the eigenvalue, the higher the concentration of triangles near the free surface $F$. This can be observed on any of the subplots of \autoref{fig:density}. As the eigenfunctions become more complicated (change sign more often), the shape of the domain (corners) become less important for the residual error estimators.

\FloatBarrier
\subsection{Spherical tank and shape approximation.}\label{sec:sphres}
\begin{table}
  \catcode`+=\active\def+{\phantom0}
  \centering
  \begin{tabular}{clllll}
    \toprule        
    		    & $d=0.2$ & $d=0.6$ & $d=1.0$ & $d=1.4$ & $d=1.8$ \\
    \midrule
    $m=0$ 	& $+3.8261^{18}_2$ & $+3.6501^{44}_4$ & $+3.7451^{68}_7$ & $+4.3010^{22}_2$ & $+6.7641^{80}_8$ \\
    		& $+9.2561^{27}_3$ & $+7.2659^{63}_6$ & $+6.9763^{60}_6$ & $+7.8005^{48}_5$ & $12.113^{930}_9$ \\
		& $14.7556^{07}_1$ & $10.7449^{76}_8$ & $10.1474^{73}_8$ & $11.2558^{77}_9$ & $17.39^{5915}_6$ \\    
		& $20.118^{787}_8$ & $14.196^{333}_4$ & $13.304^{170}_2$ & $14.698^{295}_4$ & $22.65^{6823}_7$ \\
    \midrule
    $m=1$	& $+1.0723^{24}_2$ & $+1.2625^{06}_1$ & $+1.5601^{57}_6$ & $+2.1232^{02}_0$ & $+3.9593^{02}_0$ \\
    		& $+6.2008^{06}_1$ & $+5.3683^{22}_2$ & $+5.2755^{44}_5$ & $+5.9728^{26}_3$ & $+9.4534^{76}_8$ \\
		& $11.8821^{18}_2$ & $+8.9418^{05}_1$ & $+8.5044^{40}_4$ & $+9.4762^{22}_2$ & $14.7548^{34}_4$ \\
		& $17.358^{856}_9$ & $12.423^{266}_3$ & $11.683^{441}_5$ & $12.938^{003}_0$ & $20.022^{340}_4$ \\
    \midrule
    $m=2$	& $+2.1079^{16}_2$    & $+2.3876^{70}_7$ & $+2.8196^{93}_9$ & $+3.6335^{73}_8$    & $+6.3154^{63}_7$ \\
    		& $+8.3952^{30}_3$    & $+6.8866^{86}_9$ & $+6.6594^{10}_1$ & $+7.5087^{09}_1$    & $11.8582^{19}_2$ \\
		& $14.2944^{11}_4$    & $10.5081^{76}_8$ & $+9.9412^{89}_9$ & $11.058^{298}_{30}$ & $17.2101^{17}_3$ \\
		& $19.80^{8901}_{90}$ & $14.021^{632}_7$ & $13.149^{899}_9$ & $14.547^{385}_4$    & $22.509^{191}_3$ \\
    \midrule
    $m=3$	& $+3.1294^{80}_9$ & $+3.4664^{16}_2$ & $+3.9941^{61}_6$ & $+5.0122^{47}_5$ & $+8.4631^{03}_0$\\
    		& $10.4883^{17}_2$ & $+8.3121^{40}_4$ & $+7.9728^{06}_1$ & $+8.9708^{90}_9$ & $14.1373^{89}_9$\\
		& $16.5802^{04}_1$ & $12.0063^{64}_7$ & $11.3199^{57}_6$ & $12.578^{861}_9$ & $19.565^{290}_3$\\
		& $22.158^{469}_5$ & $15.563^{479}_6$ & $14.566^{780}_9$ & $16.10^{4206}_4$ & $24.91^{2201}_2$\\
    \bottomrule
  \end{tabular}
  \caption{Spherical tanks filled to the depth $d$. Numerical results for the lowest four eigenvalues of modes $m=0,1,2,3$, compared to McIver's analytic approximations. Using quadratic approximations with separate adaptive loops for each eigenvalue. Differences showed as stacked numbers, with McIver's results below. In most cases rounding gives exactly the same results.}
  \label{tab:sph}
\end{table}

Now we turn to the shapes which cannot be exactly triangulated. A perfect example is a spherical tank, which has rounded profile. We will use mesh snapping technique described in \autoref{snapping} to improve our approximations. For a spherical container that is filled to an arbitrary level, the profile is a semicircle with a piece cut out at the top. Note that by approximating that shape with an inscribed polygon, we shrink the domain, hence lower the eigenvalue (this follows from the somewhat counterintuitive domain monotonicity \cite[Proposition 3.2.1]{BKPS10}, that works backwards compared to other Laplace eigenvalue problems). At the same time, finite element methods generally overestimate the eigenvalues. Therefore we get additional weak error cancellation. Similar cancellation is possible for any convex profile.

We consider a sphere of radius 1, filled to a depth $d<2$. Note that the same parametrization was already considered by McIver \cite{McI89} and Faltinsen-Timokha \cite{FT2012}. In both papers authors use the geometry of the sphere to either numerically or analytically approximate the sloshing eigenvalues of the spherical tank. Our finite element approach gives nearly the same values as the other authors. In \autoref{tab:sph} we compare our approximations to McIver's Table 2.

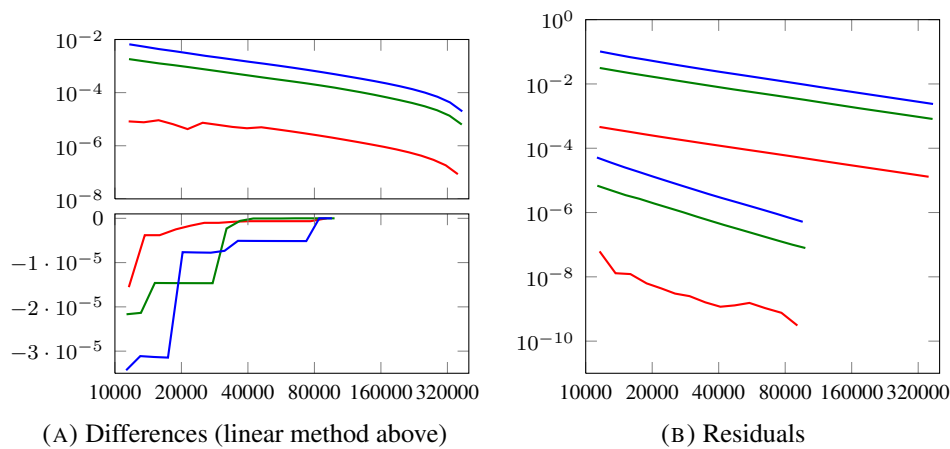
\begin{figure}[t]
  \begin{center}
    \subfloat[Differences (linear method above)\label{fig:sphdif}]{
\begin{tikzpicture}
\begin{loglogaxis}[large,diffs, ymin=1E-8,ymax=1E-2,
height=0.265\textwidth,
xticklabels={},
]

\addplot [thick, red]
coordinates {(11541,8.26455948965688e-06)(13517,7.64677100106859e-06)(15766,9.13773355426528e-06)(18313,6.48836337258096e-06)(21329,4.23432873164487e-06)(24955,7.33473826142195e-06)(29184,6.13071087740558e-06)(34024,5.11538981307069e-06)(39607,4.55498550211431e-06)(45999,4.96271237171442e-06)(53105,4.22634027508195e-06)(61767,3.5497431132292e-06)(71667,2.96281687184319e-06)(83099,2.45757499617305e-06)(96067,2.02675099725269e-06)(109999,1.67150702101715e-06)(125975,1.37455526982855e-06)(144126,1.1228333123281e-06)(164797,9.11460539265718e-07)(187971,7.27439248171535e-07)(214121,5.64848168460585e-07)(243639,4.22363218177679e-07)(276925,2.93773181336832e-07)(314477,1.83496191574761e-07)(356821,8.50748014258329e-08)};
\addplot [thick, green!50.0!black]
coordinates {(11581,0.00182824375947632)(13652,0.00150055717691089)(15871,0.00125672708149427)(18543,0.00107647840926184)(21785,0.00090138754867386)(25506,0.00075176626831297)(29809,0.000627462235573972)(34759,0.000524387762050438)(40366,0.000441725457635656)(46790,0.000371529602686493)(54564,0.000312641661665225)(63564,0.000264031796972652)(73915,0.000221934140538949)(85534,0.000185448506737629)(98911,0.000153561194283114)(113159,0.00012622241833693)(129607,0.000103425189035278)(148420,8.35586636886987e-05)(169641,6.64913637908882e-05)(194013,5.2387289924738e-05)(221351,4.08079847851184e-05)(252332,3.06511370089524e-05)(287362,2.16902526561924e-05)(327065,1.35208151990085e-05)(371798,6.28783017564416e-06)};
\addplot [thick, blue]
coordinates {(11616,0.00658654270377923)(13569,0.00537275845801943)(15829,0.00435559983501577)(18698,0.00360527863522897)(21924,0.0029621438575731)(25717,0.00241918086666182)(30046,0.00204223771955903)(34900,0.00172059456740215)(40482,0.00145130724374276)(47342,0.0012262841585855)(55193,0.0010302907131976)(64262,0.000859974910600414)(74496,0.000716002551678585)(86211,0.00059256018051812)(99396,0.000489378378198069)(113868,0.000403565595771482)(130509,0.00033255198036386)(149436,0.000271579668368105)(170693,0.000219432956283683)(194890,0.000174057480350598)(222196,0.000134821746707203)(253240,9.9925100069953e-05)(288438,6.97512799359146e-05)(328406,4.29463843829581e-05)(373648,1.97133863970578e-05)};
\end{loglogaxis}
\begin{semilogxaxis}[
ymin=-3.5E-5,
ymax=1E-6,
height=0.265\textwidth,
anchor=north west,
yshift=-0.2cm,
scaled y ticks = false,
xmin=10000, xmax=400000,
axis on top,
xtick={10000,20000,40000,80000,160000,320000},
xticklabels={10000,20000,40000,80000,160000,320000},
width=0.45\textwidth,
tick label style = {font=\tiny},
]
\addplot [thick, red]
coordinates {(11559,-1.55455020975914e-05)(13650,-3.80912567266023e-06)(15943,-3.80944376865422e-06)(18867,-2.49048095346538e-06)(21974,-1.6663672905981e-06)(25387,-1.01913819272603e-06)(29523,-1.01916699057902e-06)(34793,-7.65684396331068e-07)(40704,-6.44932295346123e-07)(47278,-6.44956112516581e-07)(55097,-6.44989587073042e-07)(64841,-6.45001175136883e-07)(76938,-6.45006164035067e-07)(90844,-2.69584354839481e-12)};
\addplot [thick, green!50.0!black]
coordinates {(11287,-2.16734542801689e-05)(13117,-2.13517580522904e-05)(15162,-1.46035583865611e-05)(17502,-1.462536228658e-05)(20468,-1.46385545169991e-05)(23826,-1.46484230620914e-05)(27691,-1.465506564724e-05)(32013,-2.30395512357262e-06)(36828,-5.80452562459755e-07)(42317,-3.76801718715569e-08)(49019,-3.77455444677821e-08)(56551,-3.61519010283473e-08)(65235,2.25125251773761e-09)(75168,1.48033230118472e-09)(86205,8.56409165805871e-10)(98703,3.41484174271045e-10)};
\addplot [thick, blue]
coordinates {(11244,-3.43181981854457e-05)(13026,-3.11220100375209e-05)(15010,-3.13199149921672e-05)(17387,-3.14616188532568e-05)(20252,-7.64672657993515e-06)(23506,-7.7148984978237e-06)(27179,-7.76082979925263e-06)(31355,-7.34353532472198e-06)(36038,-5.08397072351841e-06)(41337,-5.10320943725162e-06)(47956,-5.11659695590083e-06)(55338,-5.12646438899367e-06)(63852,-5.13378813238319e-06)(73502,-5.13950742941915e-06)(84149,5.40082645272832e-09)(96051,2.27112195716472e-09)};
\end{semilogxaxis}
\end{tikzpicture}
}
\subfloat[Residuals\label{fig:sphres}]{
\begin{tikzpicture}
\begin{loglogaxis}[large,residuals]
\addplot [thick, red]
coordinates {(11559,6.19542621305204e-08)(13650,1.29462000225398e-08)(15943,1.21834507828761e-08)(18867,6.2641183035979e-09)(21974,4.37080814246099e-09)(25387,3.01187445692035e-09)(29523,2.50911045362344e-09)(34793,1.6078936432692e-09)(40704,1.17494790547464e-09)(47278,1.2946380910475e-09)(55097,1.54454435642909e-09)(64841,1.06709304879706e-09)(76938,7.6059729643478e-10)(90844,3.05911269414475e-10)};
\addplot [thick, green!50.0!black]
coordinates {(11287,6.83588936317423e-06)(13117,4.82860376037415e-06)(15162,3.47259105592916e-06)(17502,2.67628943249267e-06)(20468,1.89984493931677e-06)(23826,1.38709324849936e-06)(27691,1.01455977005974e-06)(32013,7.36540045468933e-07)(36828,5.48117297541416e-07)(42317,4.10060044608416e-07)(49019,3.07092801344954e-07)(56551,2.30848607155922e-07)(65235,1.73262420867134e-07)(75168,1.3055593002458e-07)(86205,9.99623502451797e-08)(98703,7.88136087101538e-08)};
\addplot [thick, blue]
coordinates {(11244,5.14688548138958e-05)(13026,3.56774179426024e-05)(15010,2.54160606880061e-05)(17387,1.83736040988438e-05)(20252,1.30200343353832e-05)(23506,9.40063558191816e-06)(27179,6.86138016151098e-06)(31355,5.05674164846538e-06)(36038,3.74820023212518e-06)(41337,2.83311649950391e-06)(47956,2.11927565629181e-06)(55338,1.59561385911175e-06)(63852,1.19889285657835e-06)(73502,9.00797594851119e-07)(84149,6.76461052267912e-07)(96051,5.15868199015518e-07)};
\addplot [thick, red]
coordinates {(11541,0.00046029532898736)(13517,0.000385057650325587)(15766,0.000325069116023721)(18313,0.000274429696707717)(21329,0.000232472668650494)(24955,0.000197066761302107)(29184,0.000167658281318187)(34024,0.000142757072683451)(39607,0.000122374083651372)(45999,0.000105187456133467)(53105,9.08085444733378e-05)(61767,7.81105805023464e-05)(71667,6.72210850309165e-05)(83099,5.78570160810946e-05)(96067,4.98729872396559e-05)(109999,4.31689621728324e-05)(125975,3.74600251520609e-05)(144126,3.2605393202112e-05)(164797,2.84657643932469e-05)(187971,2.49126037129913e-05)(214121,2.18338854802594e-05)(243639,1.91706854598296e-05)(276925,1.68187594867616e-05)(314477,1.47769560601773e-05)(356821,1.29795609427642e-05)};
\addplot [thick, green!50.0!black]
coordinates {(11581,0.0314365776032413)(13652,0.0258378005541269)(15871,0.0216844079348461)(18543,0.0182546792446543)(21785,0.0153479185189974)(25506,0.0129239473988409)(29809,0.0109141052466498)(34759,0.00925744724236345)(40366,0.00789493958664225)(46790,0.00675405933498187)(54564,0.00579199662924668)(63564,0.00497826705807078)(73915,0.00428090033207312)(85534,0.00368814497293275)(98911,0.00317870459773898)(113159,0.00274919945846342)(129607,0.00238392991303014)(148420,0.00206523610988893)(169641,0.00179336556395991)(194013,0.00156467949640333)(221351,0.00137225975623999)(252332,0.00120403934693078)(287362,0.0010570066881718)(327065,0.000927243185052514)(371798,0.00081417660769792)};
\addplot [thick, blue]
coordinates {(11616,0.102272817278456)(13569,0.0834116643089668)(15829,0.068437758682171)(18698,0.0567468268505453)(21924,0.0470825544626591)(25717,0.0391480938916784)(30046,0.0329041140365687)(34900,0.027935470646652)(40482,0.0238340099325547)(47342,0.0203220456979839)(55193,0.01735893988444)(64262,0.0148173753413441)(74496,0.0127055480177891)(86211,0.0108865988539354)(99396,0.0093719594466434)(113868,0.00808879742152024)(130509,0.00701353157850711)(149436,0.00609866932355486)(170693,0.00531857192798408)(194890,0.00464409336804597)(222196,0.00406113483282337)(253240,0.00354872842951413)(288438,0.0031076326879905)(328406,0.00272057836700087)(373648,0.00238178770980585)};
\end{loglogaxis}
\end{tikzpicture}

}
  \end{center}
  \caption{Convergence of the adaptive loop for a half-full spherical tank and $m=1$. Using linear and quadratic methods, but with one adaptive loop per eigenvalue. Note that the differences to best result are now negative for the quadratic method, with sharp jumps indicating successful perimeter refinement. At the same time residual norms behave almost exactly as in earlier, exact boundary cases. In the linear method, perimeter refinement seems insignificant. }
  \label{fig:sphconv}
\end{figure}

The decay of the differences of the approximations is no longer easy to check, due to perimeter approximation having significant impact on eigenvalues (especially in the quadratic method). We illustrate this on \autoref{fig:sphdif} for a half-full tank. The large jumps on the lower subplot (quadratic method) indicate steps with successful perimeter refinement. The almost flat parts of the graphs (in fact slightly decreasing) show steps without perimeter refinement, only with the residual norm based mesh refinement. As mentioned before, the perimeter snapping enlarges the domain, hence increases the eigenvalue. This effect dwarfs any decreases caused by finer mesh eigenvalue calculations onthe quadratic method. 

Crucially, the residual norms still converge in a predictable manner (see \autoref{fig:sphres}). Only the lowest eigenvalue, with the simplest eigenfunction is slightly influenced by the changing shape of the domain. This confirmes our observation from previous section, that the lower the eigenvalue, the more import the boundary for the adaptive loop. The linear approach seems unaffected by the perimeter approximation, with positive differences to the best result. 

One might try to further improve the quadratic approach by trying to balance the perimeter and the eigenvalue approximations. However, extreme density of triangles near the curved boundary will eventually lead to significant rounding errors, singular matrices. 

\begin{figure}[t]
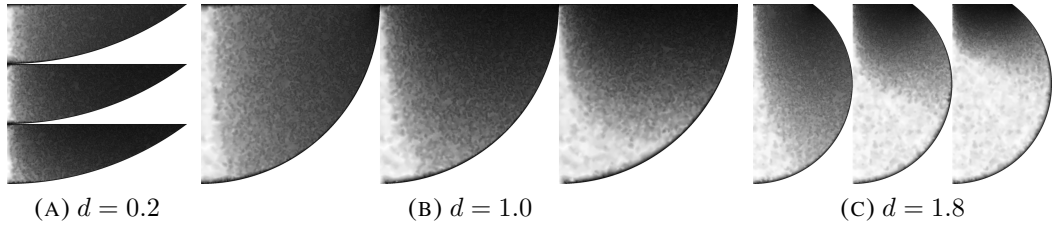

  \begin{center}
\subfloat[$d=0.2$]{
\includegraphics[height=0.17\textwidth]{{{spherical0.2}}}
}
\subfloat[$d=1.0$]{
\includegraphics[height=0.17\textwidth]{{{spherical1.0}}}
}
\subfloat[$d=1.8$]{
\includegraphics[height=0.17\textwidth]{{{spherical1.8}}}
}
  \end{center}
  \caption{Sizes of the mesh triangles (darker color indicates smaller triangles) for the mode $m=1$ of a half-full spherical tank. Note very small triangle on the curved boundary.}
  \label{fig:sphden}
\end{figure}

For our adaptive loop we obtained the triangle densities from \autoref{fig:sphden} (fill levels $d=0.2$, $1.0$ and $1.8$). Note that the deeper the tank, the less relevant its bottom, and the higher the eigenvalue, the more triangles near the free boundary $F$. At the same time, the curved boundary is very heavily refined due to perimeter snapping.

\begin{figure}
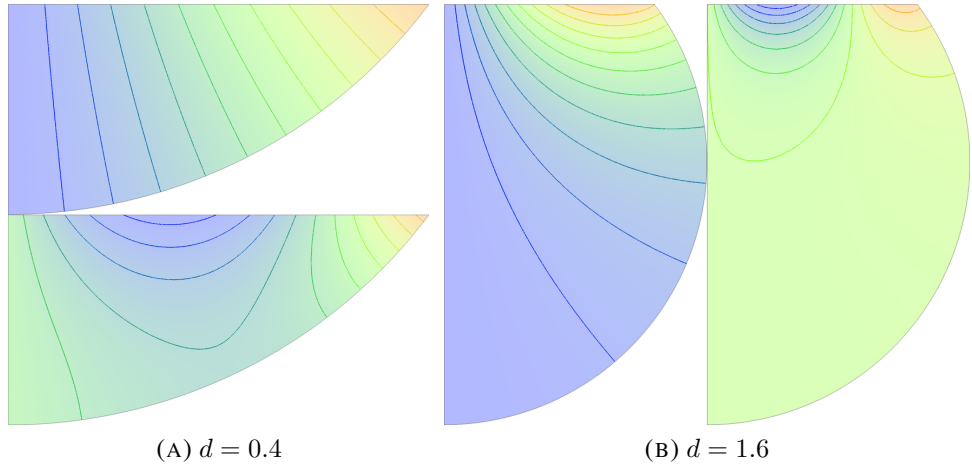

  \begin{center}
\subfloat[$d=0.4$]{
\includegraphics[height=0.4\textwidth]{{{spherical0.4plots}}}
}
\subfloat[$d=1.6$]{
\includegraphics[height=0.4\textwidth]{{{spherical1.6plots}}}
}

  \end{center}
  \caption{Plots of the first two eigenfunctions of the mode $m=1$ for spherical tanks.}
  \label{fig:sphplots}
\end{figure}

Finally we make density plots (with a few level curves) of two eigenfunctions for the spherical tanks filled to the levels $d=0.4$ and $d=1.6$ on \autoref{fig:sphplots}. We again notice that the higher eigenfunctions fluctuate more near $F$, and that the deep parts of he containers are nearly insignificant. 

\begin{figure}[t]
  \begin{center}

\begin{tikzpicture}

\begin{axis}[
xmin=0, xmax=1,
ymin=0, ymax=1.7,
axis on top,
xtick = {0,0.5,1},
ytick = {0},
width=0.8\textwidth,
height=0.7\textwidth,
extra x ticks={0.64824,0.73869,0.80402,0.86432,0.93467,0.686},
extra y ticks={1.64948,1.22728,1.10704,1.04258,1.00965,1.42041},
extra tick style={grid=major},
extra x tick style={tick label style={font=\tiny,rotate=45, anchor=north east, inner sep=0mm}},
extra y tick style={tick label style={font=\tiny,scale=0.8}},
smooth,
cycle list name=color list,
legend pos = outer north east,
]
\addplot+[very thick]
coordinates {(0,1.28471233651339e-09)(0.0552763819095478,0.0591527063421573)(0.110552763819095,0.118234581168559)(0.165829145728643,0.177173874176923)(0.221105527638191,0.235902202532114)(0.276381909547739,0.294343929955767)(0.331658291457286,0.352432918827825)(0.386934673366834,0.410093489004416)(0.442211055276382,0.46725833084183)(0.49748743718593,0.523854867299153)(0.552763819095477,0.579811374048791)(0.608040201005025,0.635058723786493)(0.663316582914573,0.689518591651246)(0.718592964824121,0.74312321241776)(0.773869346733668,0.795799821772983)(0.829145728643216,0.847471676593978)(0.884422110552764,0.898062954808726)(0.939698492462312,0.947494681488172)(0.994974874371859,0.995686299067956)(1,1)};
\addplot+[very thick]
coordinates {(0,-2.78467982006703e-09)(0.0552763819095478,0.0685750423640017)(0.110552763819095,0.136901879185455)(0.165829145728643,0.204733048317133)(0.221105527638191,0.271821696584162)(0.276381909547739,0.33792577554339)(0.331658291457286,0.402797806847702)(0.386934673366834,0.466197863685437)(0.442211055276382,0.527884982861759)(0.49748743718593,0.587622461004)(0.552763819095477,0.645171290872169)(0.608040201005025,0.700303689439877)(0.663316582914573,0.752773491088544)(0.718592964824121,0.802349407464824)(0.773869346733668,0.848785375901556)(0.829145728643216,0.891824867244683)(0.884422110552764,0.931195707985509)(0.939698492462312,0.966549504394348)(0.994974874371859,0.997441667460953)(1,1)};
\addplot+[very thick]
coordinates {(0,-3.36334382566861e-09)(0.0552763819095478,0.0816789357042746)(0.110552763819095,0.162846991173673)(0.165829145728643,0.242996768978078)(0.221105527638191,0.321624894026119)(0.276381909547739,0.398233185573552)(0.331658291457286,0.472335461992044)(0.386934673366834,0.543454595893737)(0.442211055276382,0.611125653467105)(0.49748743718593,0.674895695867401)(0.552763819095477,0.734326160483831)(0.608040201005025,0.788990829585913)(0.663316582914573,0.838467812128258)(0.718592964824121,0.882337672803094)(0.773869346733668,0.92018485052496)(0.829145728643216,0.951549305190492)(0.884422110552764,0.975897624502842)(0.939698492462312,0.992497349198902)(0.994974874371859,0.999918918902407)(1,1)};
\addplot+[very thick]
coordinates {(0,-8.89247209201471e-09)(0.0552763819095478,0.0905864165025276)(0.110552763819095,0.180473706156145)(0.165829145728643,0.268965013821128)(0.221105527638191,0.355376799248153)(0.276381909547739,0.439032534648439)(0.331658291457286,0.519277422617135)(0.386934673366834,0.595475733339885)(0.442211055276382,0.667016361323646)(0.49748743718593,0.733302074697168)(0.552763819095477,0.793781483051841)(0.608040201005025,0.847912555334823)(0.663316582914573,0.895178492743078)(0.718592964824121,0.935074600324982)(0.773869346733668,0.96709544393197)(0.829145728643216,0.990688333620009)(0.884422110552764,1.0051873100964)(0.939698492462312,1.00959994737446)(0.994974874371859,1.00162812824891)(1,1)};
\addplot+[very thick]
coordinates {(0,-4.3989665243771e-09)(0.0552763819095478,0.102128203197196)(0.110552763819095,0.203300419738273)(0.165829145728643,0.302570346437031)(0.221105527638191,0.399005015806348)(0.276381909547739,0.491697219803731)(0.331658291457286,0.579765307464148)(0.386934673366834,0.662371453690631)(0.442211055276382,0.73871802284041)(0.49748743718593,0.808039569808679)(0.552763819095477,0.869641155121782)(0.608040201005025,0.922857640261724)(0.663316582914573,0.967077513586277)(0.718592964824121,1.00171267127486)(0.773869346733668,1.0261768252495)(0.829145728643216,1.0398527067236)(0.884422110552764,1.04194013184846)(0.939698492462312,1.03113992553663)(0.994974874371859,1.0038576115827)(1,1)};
\addplot+[very thick]
coordinates {(0,-3.35633343340191e-09)(0.0552763819095478,0.118060667846222)(0.110552763819095,0.234795881383892)(0.165829145728643,0.348895481287531)(0.221105527638191,0.459074483862872)(0.276381909547739,0.564091043873417)(0.331658291457286,0.662755204041521)(0.386934673366834,0.753947905535934)(0.442211055276382,0.836610284467964)(0.49748743718593,0.909781553867749)(0.552763819095477,0.972574801414869)(0.608040201005025,1.02419490779933)(0.663316582914573,1.06393359983396)(0.718592964824121,1.09112360821833)(0.773869346733668,1.10515568908461)(0.829145728643216,1.10535719816285)(0.884422110552764,1.0908373896637)(0.939698492462312,1.05990603817549)(0.994974874371859,1.00698788405325)(1,1)};
\addplot+[very thick]
coordinates {(0,1.13472262697201e-08)(0.0552763819095478,0.142877791652753)(0.110552763819095,0.283825292826232)(0.165829145728643,0.420939041126973)(0.221105527638191,0.552365751953799)(0.276381909547739,0.676325719069518)(0.331658291457286,0.791138733611962)(0.386934673366834,0.895248722258057)(0.442211055276382,0.987226490822118)(0.49748743718593,1.06579757805453)(0.552763819095477,1.12984214688849)(0.608040201005025,1.17840659611094)(0.663316582914573,1.21069577483801)(0.718592964824121,1.22603590852283)(0.773869346733668,1.22383047451233)(0.829145728643216,1.20344952048164)(0.884422110552764,1.16393334210485)(0.939698492462312,1.10313326658871)(0.994974874371859,1.01202161007362)(1,1)};
\addplot+[very thick]
coordinates {(0,2.48436902161586e-08)(0.05,0.161951714854667)(0.1,0.321792536294208)(0.15,0.477436220744647)(0.2,0.626856302466695)(0.25,0.768115573779695)(0.3,0.899365300617768)(0.35,1.01890348110718)(0.4,1.12517637811128)(0.45,1.21679918632081)(0.5,1.29256116365302)(0.55,1.35145517778732)(0.6,1.39265691547394)(0.65,1.41554062834584)(0.7,1.41963282094984)(0.75,1.40459696129769)(0.8,1.37013009511943)(0.85,1.31582802030354)(0.9,1.24066739241936)(0.95,1.14162742630606)(1,1)};
\addplot+[very thick]
coordinates {(0,5.49715073461243e-08)(0.0552763819095478,0.219538494941033)(0.110552763819095,0.435159212968955)(0.165829145728643,0.643021399471351)(0.221105527638191,0.839408060490942)(0.276381909547739,1.02085503971553)(0.331658291457286,1.18413350633618)(0.386934673366834,1.32635519507355)(0.442211055276382,1.44501560586844)(0.49748743718593,1.53801442329137)(0.552763819095477,1.60369680961375)(0.608040201005025,1.6408478814766)(0.663316582914573,1.64868032424625)(0.718592964824121,1.62678057640103)(0.773869346733668,1.57499515233448)(0.829145728643216,1.49317282932686)(0.884422110552764,1.38053821461665)(0.939698492462312,1.23360142134812)(0.994974874371859,1.02983210740764)(1,1)};

\legend{$d=0.2$,$d=0.6$,$d=1.0$,$d=1.2$,$d=1.4$,$d=1.6$,$d=1.8$,$d=1.95$,ice-fishing}
\end{axis}

\end{tikzpicture}
  \end{center}
  \caption{Sloshing profiles for the lowest eigenvalues of spherical tanks, with high-spots marked. The ice-fishing problem is equivalent to $d=2.0$.}
  \label{fig:sphprof}
\end{figure}
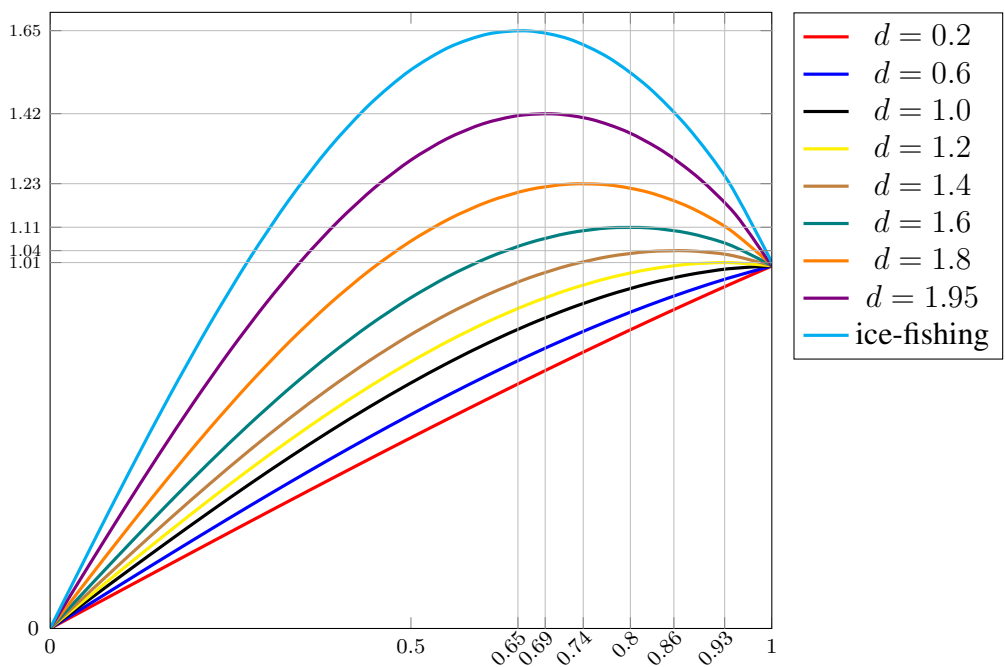

We also provide sloshing profiles for the fundamental modes for various depths on \autoref{fig:sphprof}, where we also mark the high-spots (cf. Figure 2 in Faltinsen-Timokha \cite{FT2012}). We include the ice-fishing problem (discussed and solved numerically in \autoref{sec:ice}), since it can be understood as the limiting case $d=2$ (full tank), see McIver \cite{McI89}. Note that numerical approximation of the ice-fishing using cylindricals tank is much easier than using spherical tanks.

\FloatBarrier

\section{Images of high spots}\label{photo}

\begin{figure}[H]
  \begin{center}
    \subfloat[Small fish tank\label{smallfishsize}]{
\begin{tikzpicture}[scale=1.3] 
  \draw[->] (-0.5,0) -- (1.5,0) node[above] {\footnotesize $r$};
  \draw[->] (0,-2.5) -- (0,0.5) node[right] {\footnotesize $z$};
  \begin{scope}
  \clip (0,0) rectangle (2,-2);
  \fill[fill=black!15!white,draw=black,thick] (0,-0.9) ellipse (1.4 and 1.3);
  \end{scope}
  \draw[thick] (0,0) |-  (0.76,-2) node [pos=0.25,left=-3pt] {\tiny $180$} node [pos=0.75,below=-3pt] {\tiny $60$};
  \draw (0.5,0) node [above=-3pt] {\tiny $72.5$};
  \draw[dotted] (0,-0.9) -- ++(1.4,0) node [above=-3pt,pos=0.5] {\tiny $115$};
\end{tikzpicture} 
    }
    \subfloat[Coctail glass\label{coctailsize}]{
\begin{tikzpicture}[scale=1.3] 
  \draw[->] (-0.5,0) -- (2,0) node[above] {\footnotesize $r$};
  \draw[->] (0,-2.5) -- (0,0.5) node[right] {\footnotesize $z$};
  \fill[fill=black!15!white,draw=black,thick]
  (0,0) |- (0.2,-2) node [left=-3pt,pos=0.2] {\tiny $70$} node [below=-3pt,pos=0.75] {\tiny $5$} -- (1.5,0) ;
  \draw (0.75,0) node [above=-3pt] {\tiny $56$};
\end{tikzpicture}
    }
    \subfloat[Large fish tank\label{largefishsize}]{
    \includegraphics[width=0.3\textwidth,]{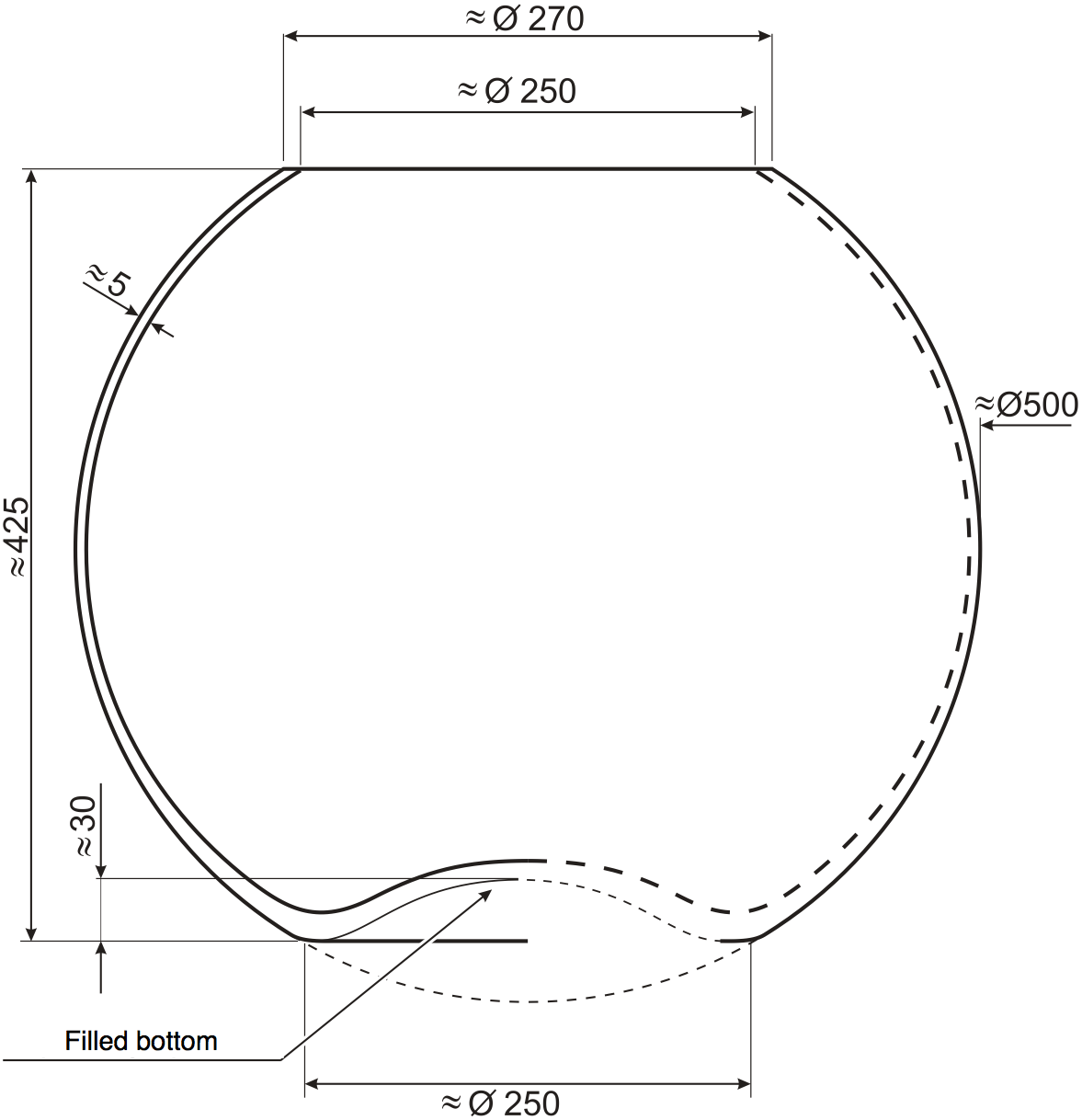}}
  \end{center}
  \caption{Shapes of the tanks we used to take photographs (lengths in millimeters). The first two profiles were also used in Kuznetsov et. al. \cite{Notices2014}.}
  \label{fig:expshapes}
\end{figure}
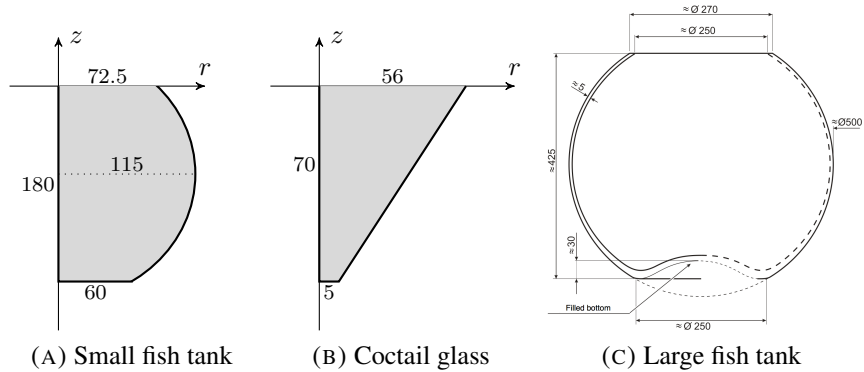

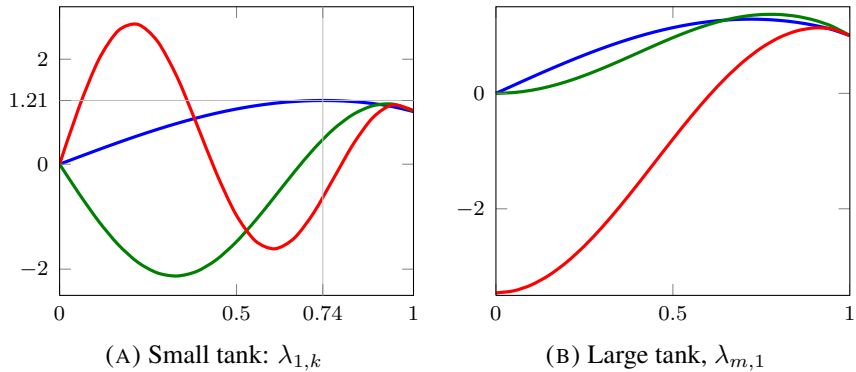
\begin{figure}[H]
  \begin{center}
    \subfloat[Small tank: $\lambda_{1,k}$]{
\begin{tikzpicture}

\begin{axis}[
xmin=0, xmax=1,
ymin=-2.5, ymax=3,
axis on top,
xtick = {0,0.5,1},
extra x ticks={0.743718592964824},
extra y ticks={1.21377852855208},
extra tick style={grid=major},
smooth,
width=0.45\textwidth
]
\addplot [very thick, blue]
coordinates {(0,0)(0.0552763819095478,0.140110116240619)(0.110552763819095,0.278362710867925)(0.165829145728643,0.412921350856326)(0.221105527638191,0.542001752633126)(0.276381909547739,0.663887037280533)(0.331658291457286,0.776949517151918)(0.386934673366834,0.879693938847255)(0.442211055276382,0.97071417977208)(0.49748743718593,1.0487784620665)(0.552763819095477,1.11279304589157)(0.608040201005025,1.16180541323794)(0.663316582914573,1.19502273283702)(0.718592964824121,1.21176839830663)(0.773869346733668,1.21142190931772)(0.829145728643216,1.19333421270727)(0.884422110552764,1.15652479958056)(0.939698492462312,1.098842931318)(0.994974874371859,1.01151666300101)(1,1)};
\addplot [very thick, green!50.0!black]
coordinates {(0,0)(0.0552763819095478,-0.564882288029247)(0.110552763819095,-1.08875943962865)(0.165829145728643,-1.53394741690729)(0.221105527638191,-1.86913383672201)(0.276381909547739,-2.07188689158948)(0.331658291457286,-2.13036849821893)(0.386934673366834,-2.04422972391595)(0.442211055276382,-1.82451806333066)(0.49748743718593,-1.49256991152514)(0.552763819095477,-1.07816981567326)(0.608040201005025,-0.616967961062183)(0.663316582914573,-0.147479450606334)(0.718592964824121,0.292069628316365)(0.773869346733668,0.666750113806462)(0.829145728643216,0.947432195847706)(0.884422110552764,1.11230661911993)(0.939698492462312,1.14606837991392)(0.994974874371859,1.02480642458426)(1,1)};
\addplot [very thick, red]
coordinates {(0,0)(0.0552763819095478,1.09188692340864)(0.110552763819095,1.99164725028501)(0.165829145728643,2.54465489945306)(0.221105527638191,2.6636379993712)(0.276381909547739,2.34472560466864)(0.331658291457286,1.66655996677391)(0.386934673366834,0.772861814169826)(0.442211055276382,-0.157774643035768)(0.49748743718593,-0.948362490080155)(0.552763819095477,-1.4587528471429)(0.608040201005025,-1.61225239598279)(0.663316582914573,-1.40844364222628)(0.718592964824121,-0.919887539993659)(0.773869346733668,-0.273941764028172)(0.829145728643216,0.375872090817509)(0.884422110552764,0.882158889696616)(0.939698492462312,1.13065116383451)(0.994974874371859,1.03617155660439)(1,1)};

\end{axis}

\end{tikzpicture}
}
\subfloat[Large tank, $\lambda_{m,1}$]{
\begin{tikzpicture}

\begin{axis}[
xmin=0, xmax=1,
ymin=-3.5, ymax=1.5,
axis on top,
xtick = {0,0.5,1},
smooth,
width=0.45\textwidth
]
\addplot [very thick, blue]
coordinates {(0,0)(0.05,0.139320717862126)(0.1,0.277005691686926)(0.15,0.411438986232737)(0.2,0.541042347031343)(0.25,0.664292030574903)(0.3,0.779744333958055)(0.35,0.886037271960334)(0.4,0.98191827388963)(0.45,1.06625041438857)(0.5,1.13801956205148)(0.55,1.19634796118188)(0.6,1.24048077062205)(0.65,1.26980920911418)(0.7,1.28381819029032)(0.75,1.28208861904257)(0.8,1.26421246266855)(0.85,1.22967420703572)(0.9,1.17751109187454)(0.95,1.10526844803029)(1,1)};
\addplot [very thick, green!50.0!black]
coordinates {(0,0)(0.05,0.0133494860820809)(0.1,0.0528965005198328)(0.15,0.117154629121828)(0.2,0.203703747856972)(0.25,0.309272360714743)(0.3,0.429852110518765)(0.35,0.560832527138465)(0.4,0.697163761306362)(0.45,0.833524058649839)(0.5,0.964500642316976)(0.55,1.08477233331783)(0.6,1.18928357300839)(0.65,1.27339696881833)(0.7,1.33303505278542)(0.75,1.36476116042917)(0.8,1.36581041154442)(0.85,1.33398465884209)(0.9,1.2672390720308)(0.95,1.16214449482417)(1,1)};
\addplot [very thick, red]
coordinates {(0,-3.45623849370994)(0.05,-3.4216043384947)(0.1,-3.31878531648917)(0.15,-3.150988781264)(0.2,-2.92343953065024)(0.25,-2.64320068074249)(0.3,-2.31893320933906)(0.35,-1.96060273482784)(0.4,-1.57914356032954)(0.45,-1.18609322090862)(0.5,-0.793210745825188)(0.55,-0.412094684400381)(0.6,-0.0538149494936846)(0.65,0.271423138213673)(0.7,0.55456543269978)(0.75,0.787920445961671)(0.8,0.965258596762769)(0.85,1.08171365257627)(0.9,1.13323528921524)(0.95,1.11448035183512)(1,1)};

\end{axis}

\end{tikzpicture}
}

  \end{center}
  \caption{Profiles of the sloshing water in a fish tank.}
  \label{fig:smallprof}
\end{figure}

To further check our numerical methods, and the mathematical model itself, we attempted to photograph a sloshing liquid in various containers (tank profiles shown on \autoref{fig:expshapes}, numerical sloshing profiles on \autoref{fig:smallprof}). Note that throughout this paper we used the simplified linear model of sloshing. Clearly, it cannot be applied to strong turbulent sloshing. But we visually check that it is very accurate for infinitesimal sloshing. To reiterate, we did not attempt to eliminate any nonlinearities intrinsic to strong sloshing from our experiment, opting instead for capturing long time behavior. After a long time, all nonlinear phenomena (with higher order of decay) are insignificant in comparison to linear wave modes we study.

In order to get accurate experimental results for infinitesimal sloshing we needed a nearly still water surface. We achieve this by first nudging a container several times in roughly equal intervals, corresponding to theoretical sloshing frequency. After that, we allow the water waves to dampen to a point of being barely visible. 

\begin{figure}[t]
  \centering
  \subfloat[Long exposure image of the reflection in a fish bowl. Blurred lines show gradient of the sloshing mode, with high spots at points with vanishing gradient.\label{fishbowl}]{
\begin{tikzpicture} 

  \node[anchor=south west,inner sep=0] (image) at (0,0) {\includegraphics[width=0.35\textwidth]{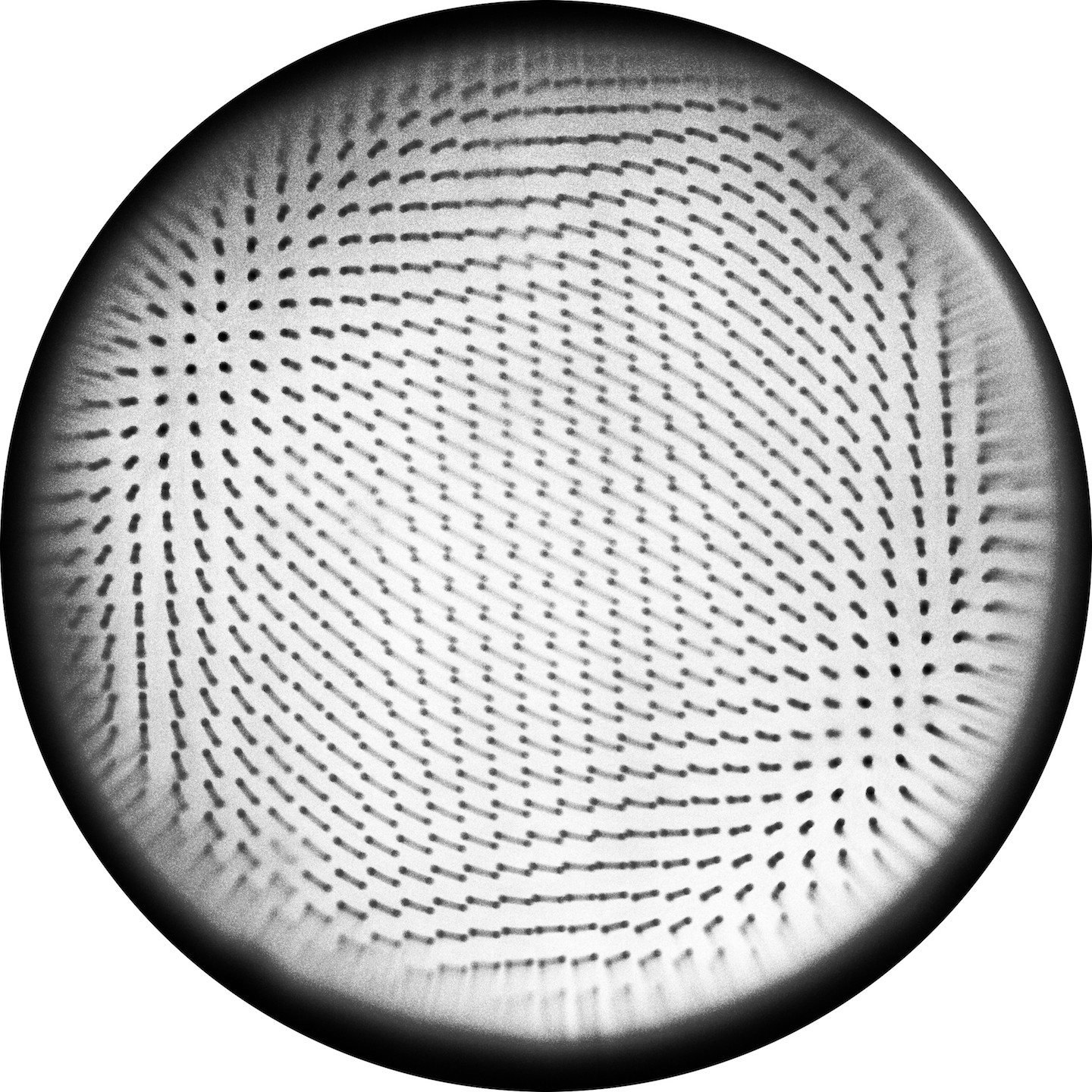}};
    \begin{scope}[x={(image.south east)},y={(image.north west)}]
      \draw[red,very thick] (0.5,0.5) circle (0.48);
      \draw[dotted,red,very thick] (0.5,0.5) circle ({0.743*0.48});
    \end{scope}
  \end{tikzpicture}}
  \subfloat[Whirling effect in a fish bowl. High spots stay a certain distance from the center. Points closer to the center trace elliptical paths. \label{obrot}]{
\begin{tikzpicture}

    \node[anchor=south west,inner sep=0] (image) at (0,0) {\includegraphics[width=0.35\textwidth]{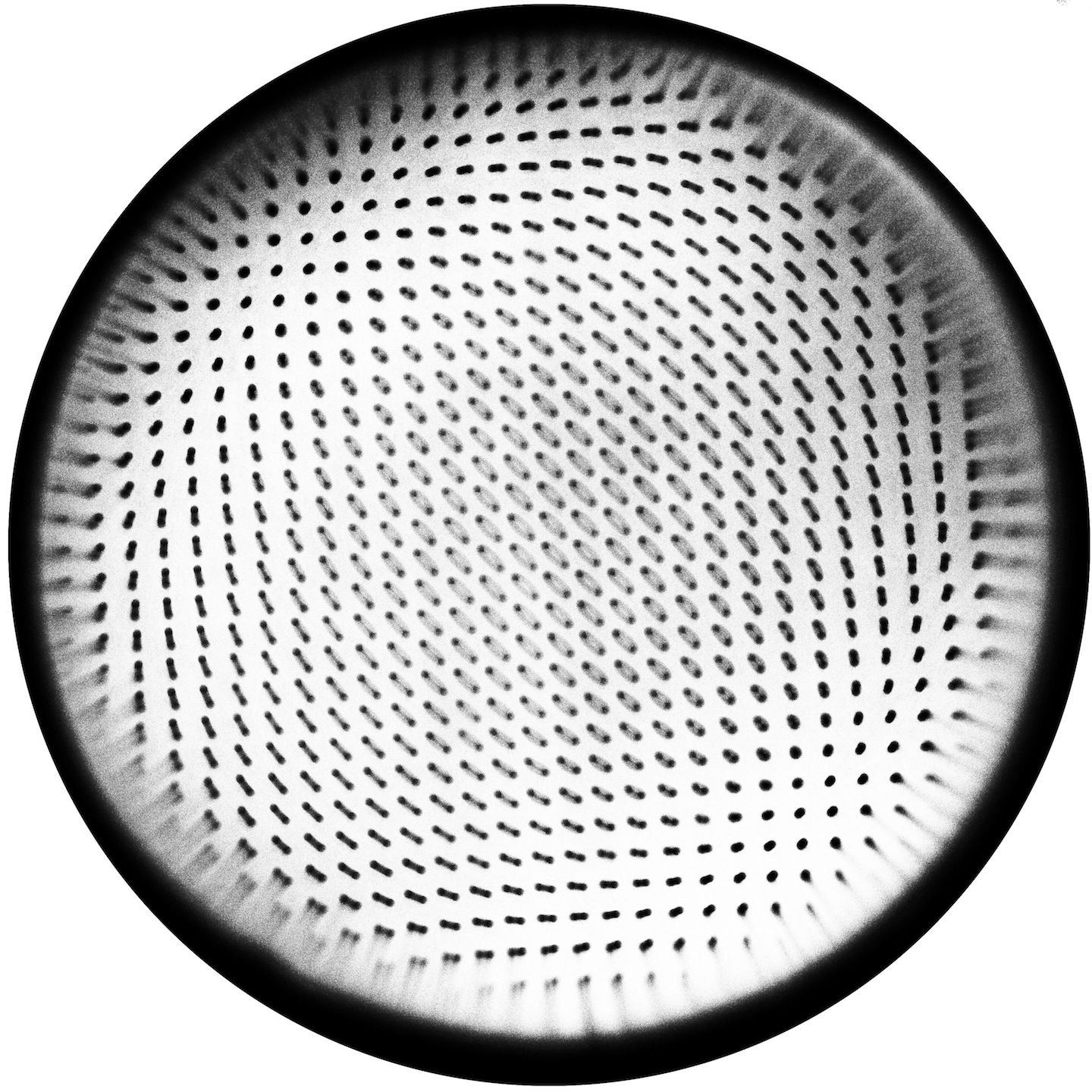}};
    \begin{scope}[x={(image.south east)},y={(image.north west)}]
      \draw[red,very thick] (0.5,0.5) circle (0.48);
      \draw[dotted,red,very thick] (0.5,0.5) circle ({0.743*0.48});
    \end{scope}
  \end{tikzpicture}}

  \subfloat[An image for a cocktail glass, showing no points with vanishing gradient.\label{cocktail}]{
  \includegraphics[width=0.35\textwidth]{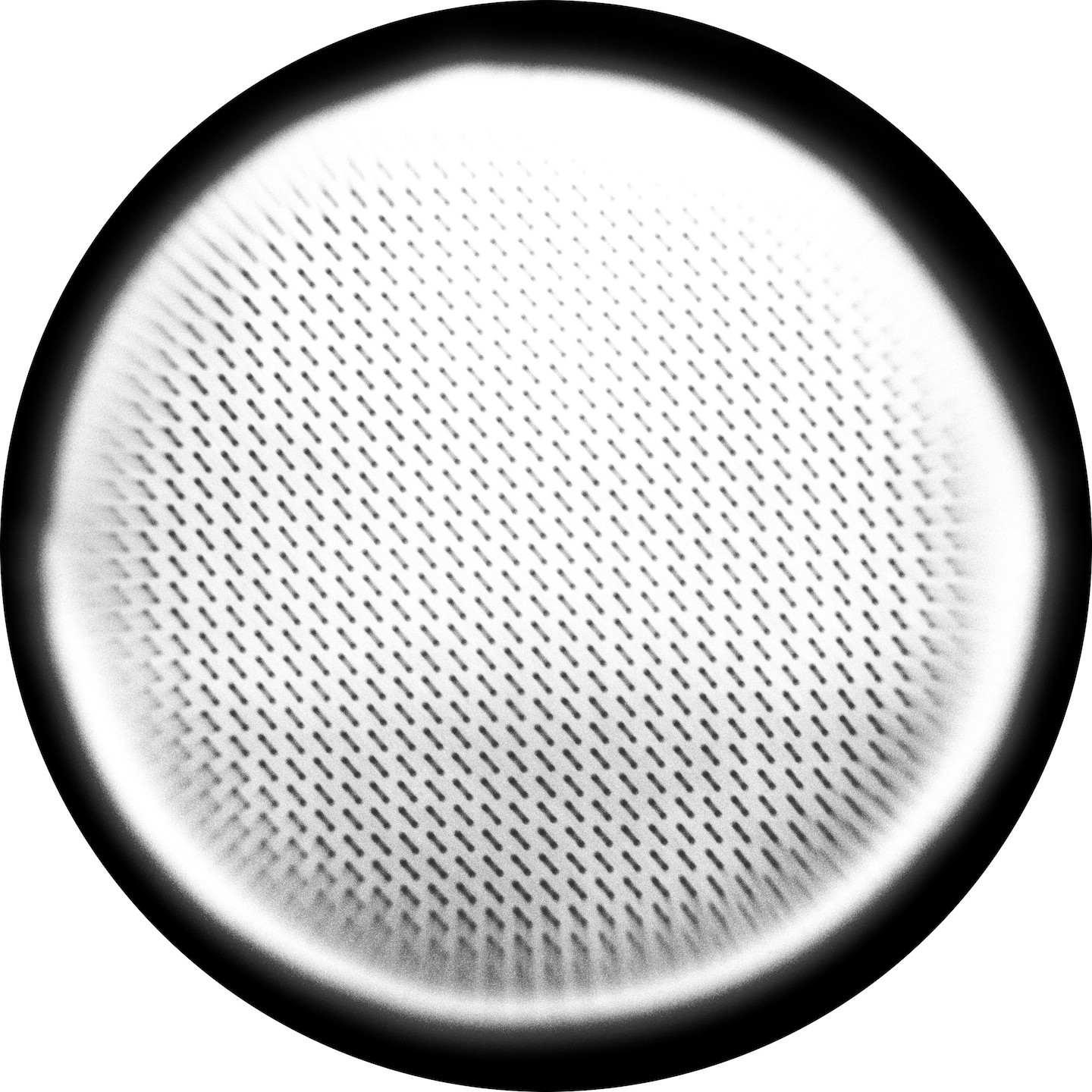}}
  \subfloat[Experimental setup\label{experiment}]{
\begin{tikzpicture}[scale=1.6]

    \draw[thick] (-2,0) -- (2,0); 
    \filldraw[fill=black!10!white,thick] (-0.35355,0.85355) arc (135:405:0.5); 
    \draw (0.5,0.7) node[right]{{\footnotesize sloshing tank}};
    \begin{scope}[shift={(0.55,3)}, rotate=250] 
      \fill (0,0) rectangle (0.2,0.5);
      \fill (0.22,0.09) rectangle (0.25,0.41);
      \fill (0.27,0.11) rectangle (0.30,0.39);
      \fill (0.32,0.13) rectangle (0.35,0.37);
      \fill (0.37,0.15) rectangle (0.40,0.35);
      \fill (0.42,0.17) rectangle (0.48,0.33);
    \end{scope}
    \draw (1,2.6) node[right]{{\footnotesize camera}};
    \draw[very thick] (-1.03539,2) -- (0.196246,2); 
    \draw (-0.9,1.95) node[above]{{\footnotesize dotted paper}};
    \draw[dashed] (0.6,2.4) -- (0.30355,0.90355) -- (0.096246,1.95); 
    \draw[dashed] (0.6,2.4) -- (-0.30355,0.90355) -- (-0.93539,1.95); 
  \end{tikzpicture}}
  \caption{Photographic experiment}
  \label{fig:expnotices}
\end{figure}
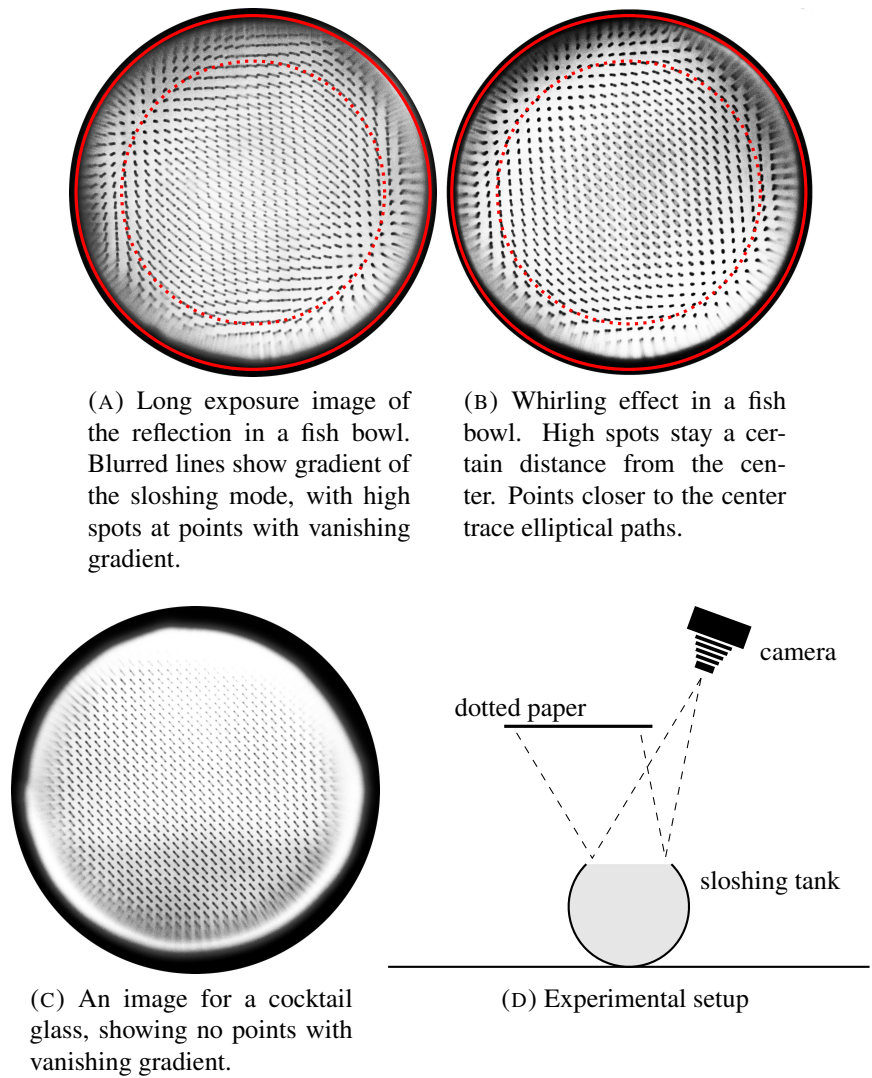

Next, we photograph a reflection of a dotted piece of paper on the nearly still surface of the water, using a few seconds long exposure to blur the dots. The setup of the experiment is shown on \autoref{fig:expnotices}, together with the results on the first two containers from \autoref{fig:expshapes}. This figure is reproduced from \cite{Notices2014}, but we overlay the numerical results on the boulbous container images \autoref{fishbowl} and \autoref{obrot}. The dotted red line shows the numerically calculated distance from the center of the free surface to the high spot, assuming the solid line is the boundary of the surface. We used our finite element method to find the radius of the dotted circle, using container measurements from \autoref{smallfishsize}.

The results of the experiment on the small fishbowl (profile from \autoref{smallfishsize}) agree with the infinitesimal, linear model. Most of the dots trace a path caused by the changing tangent planes of the moving surface. Only the dots corresponding to the local extrema of the sloshing amplitude remain still, hence sharp on our photographs. Note that the lowest sloshing eigenvalue has multiplicity two, hence it is possible to get phase-shifted combination of the two corresponding modes, giving an illusion of whirling from \autoref{obrot}. The sloshing is irrotational, so the wave is a standing wave, but the extrema travel around the red dotted circle.  

The coctail glass (profile from \autoref{coctailsize}) also gives the results (\autoref{cocktail}), that are consistent with the model. The high spot seems to be on the boundary and all dots trace roughly the same path. This suggests that the sloshing profile is nearly a flat plane. In fact, certain cones do have exactly planar sloshing profile (see Troesch \cite{T60} or a brief discussion in \autoref{sec:harmonic}).

\subsection{Larger tank and higher modes}
\begin{figure}[t]
  \centering
  \subfloat[high spots for $m=1$\label{largehighspot}]{
\begin{tikzpicture}
    \node[anchor=south west,inner sep=0] (image) at (0,0) {\includegraphics[width=0.49\textwidth]{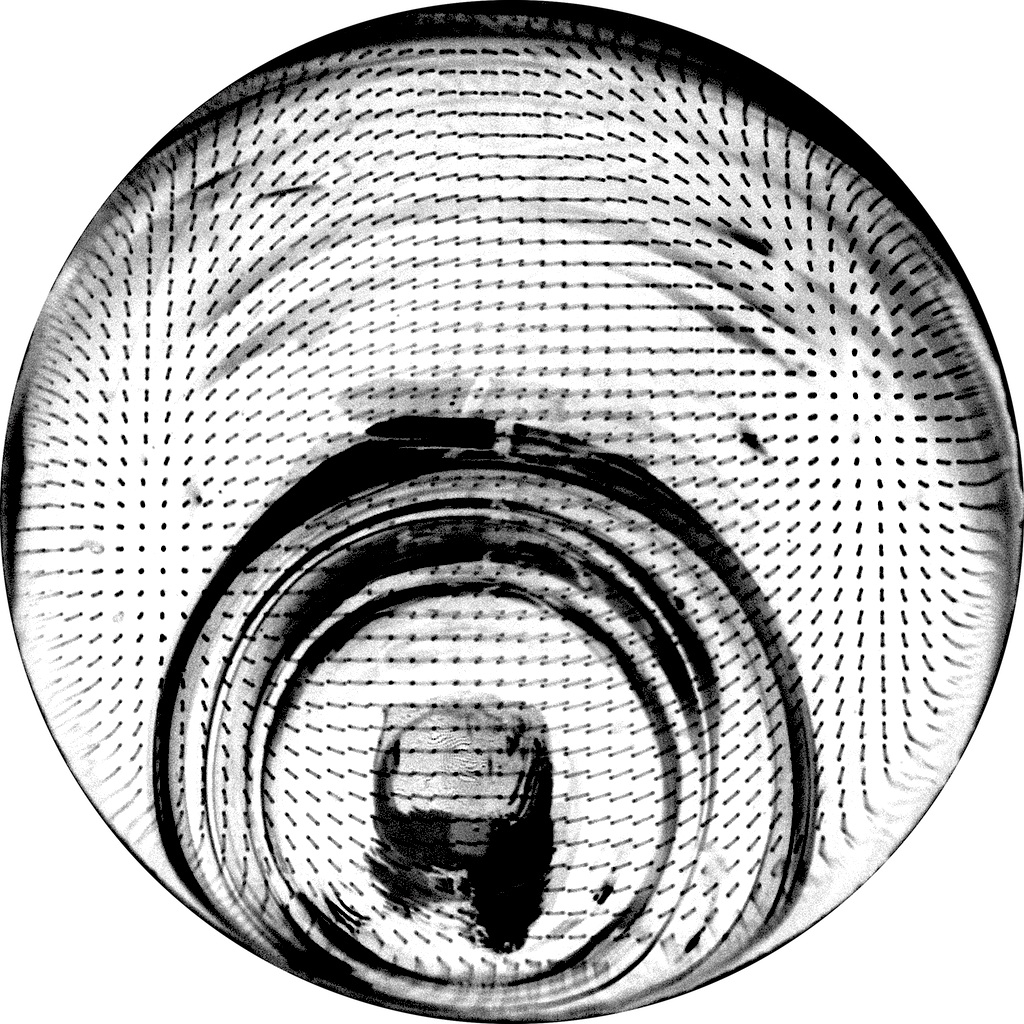}};
    \begin{scope}[x={(image.south east)},y={(image.north west)}]
      \draw[red,very thick] (0.5,0.5) circle (0.5);
      \draw[dotted,red,very thick] (0.5,0.5) circle ({0.719*0.5});
    \end{scope}
  \end{tikzpicture}}
  \subfloat[rotating mix of $m=1$ modes \label{obrot2}]{
\begin{tikzpicture}

    \node[anchor=south west,inner sep=0] (image) at (0,0) {\includegraphics[width=0.49\textwidth]{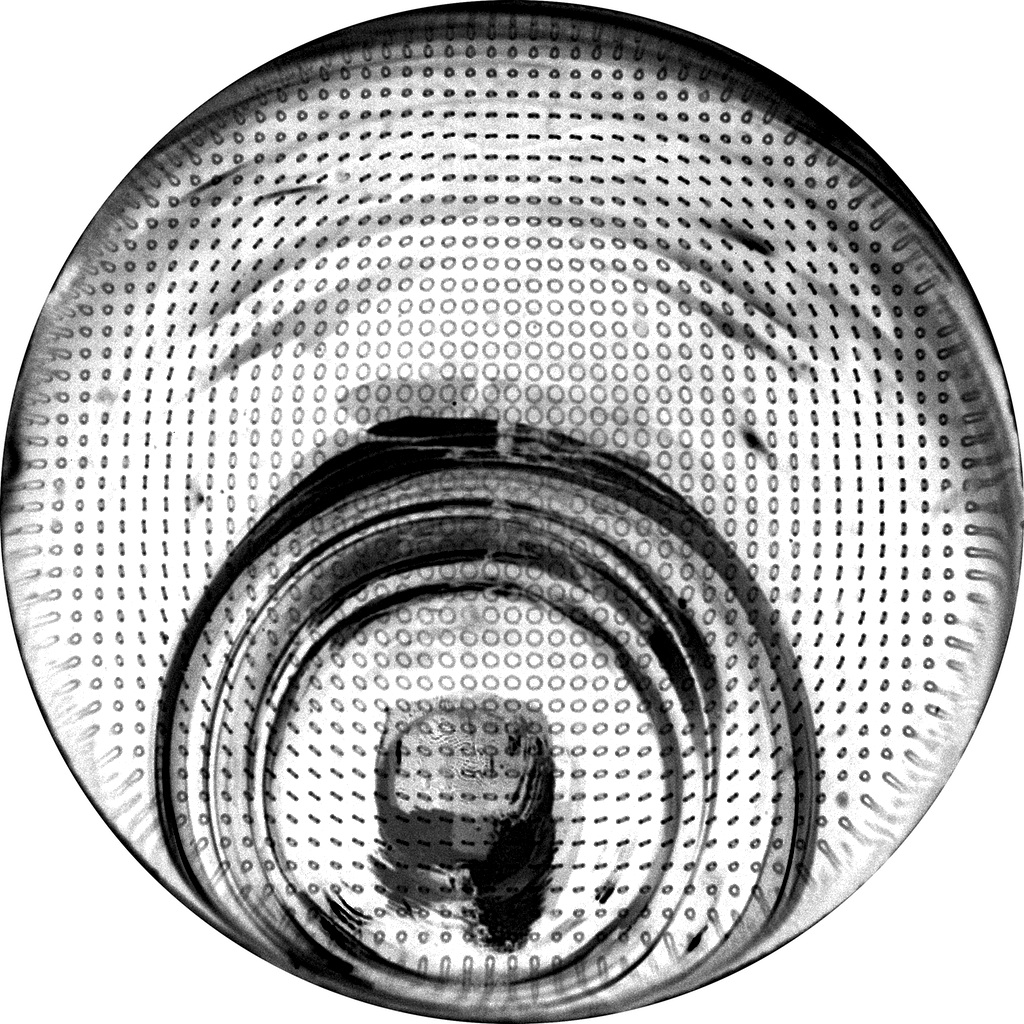}};
    \begin{scope}[x={(image.south east)},y={(image.north west)}]
      \draw[red,very thick] (0.5,0.5) circle (0.5);
      \draw[dotted,red,very thick] (0.5,0.5) circle ({0.719*0.5});
    \end{scope}
  \end{tikzpicture}}\\

  \vspace{0.5cm}
  \subfloat[lowest $m=2$ mode \label{mode2}]{
\begin{tikzpicture}
    \node[anchor=south west,inner sep=0] (image) at (0,0) {\includegraphics[width=0.49\textwidth]{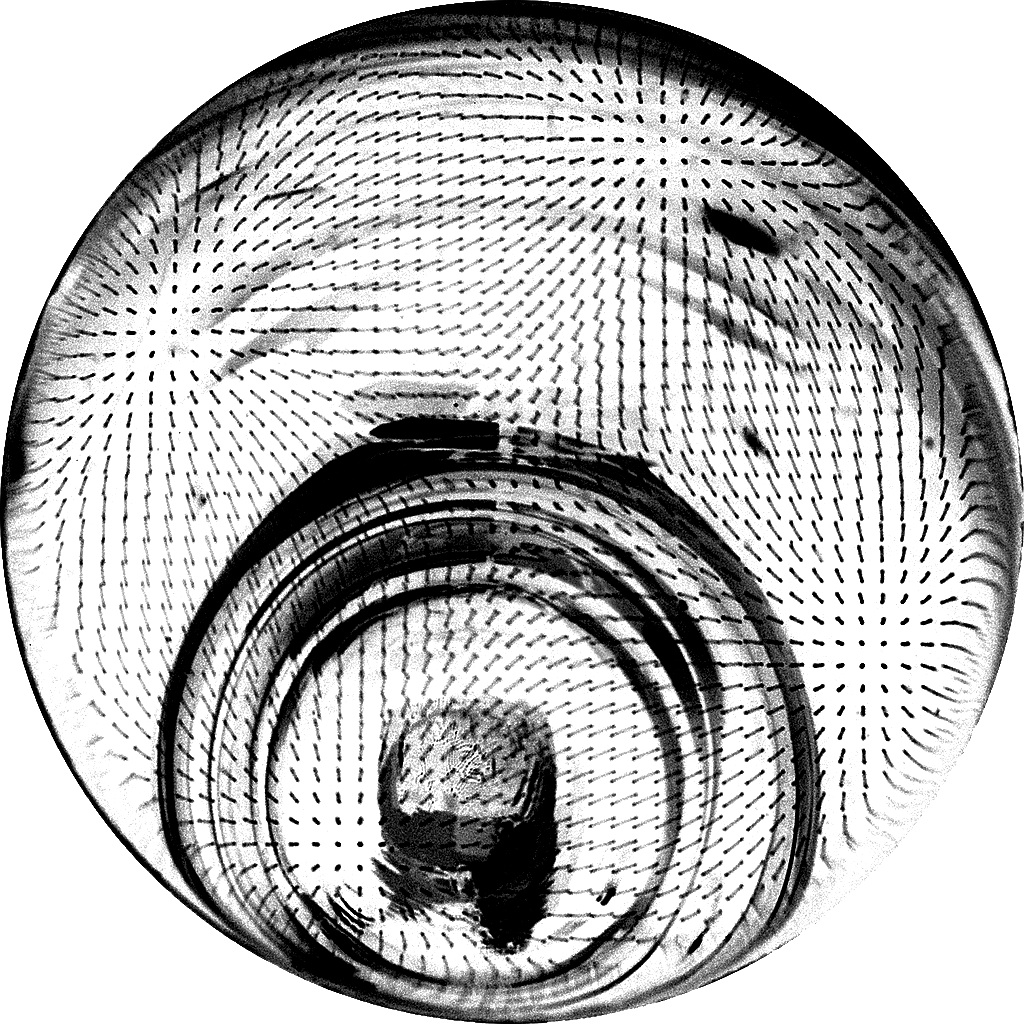}};
    \begin{scope}[x={(image.south east)},y={(image.north west)}]
      \draw[red,very thick] (0.5,0.5) circle (0.5);
      \draw[dotted,red,very thick] (0.5,0.5) circle ({0.777*0.5});
    \end{scope}
  \end{tikzpicture}}
  \subfloat[lowest $m=0$ mode \label{mode0}]{
\begin{tikzpicture}
    \node[anchor=south west,inner sep=0] (image) at (0,0) {\includegraphics[width=0.49\textwidth]{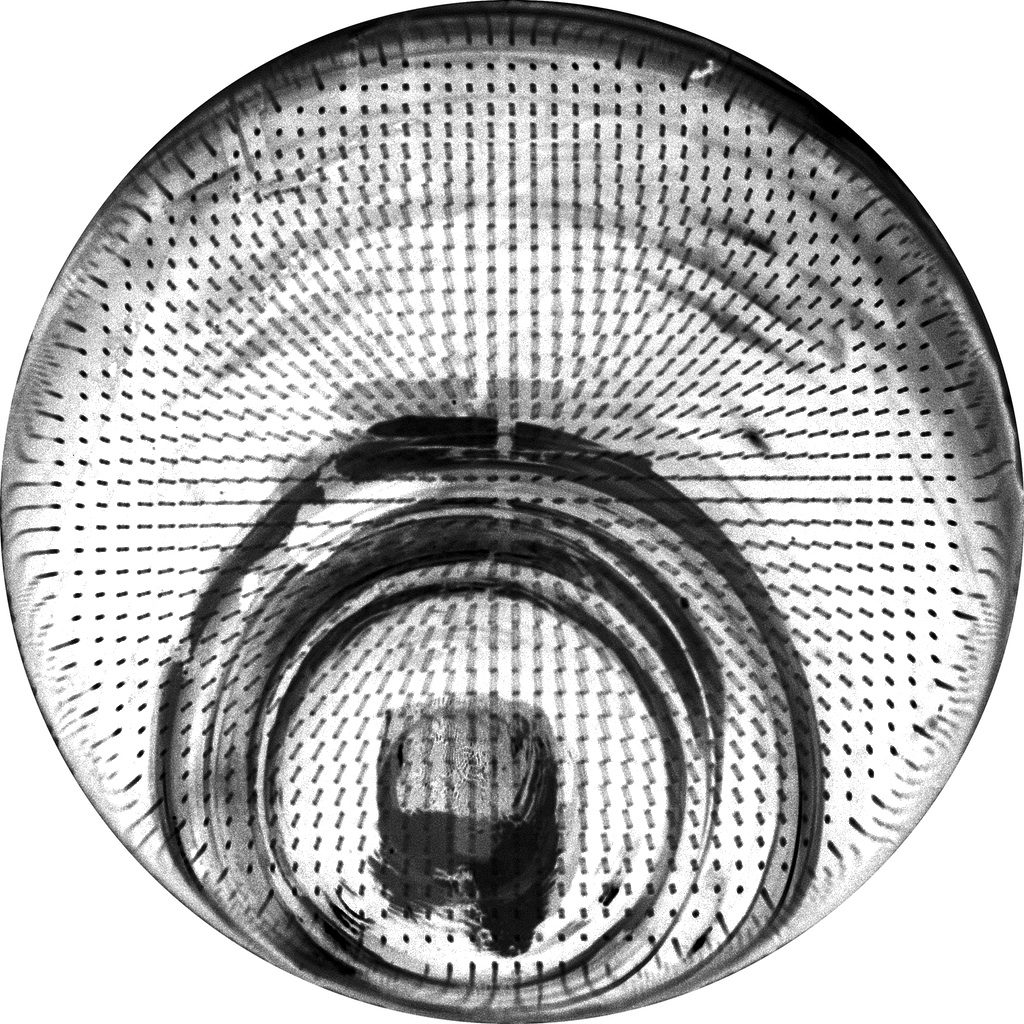}};
    \begin{scope}[x={(image.south east)},y={(image.north west)}]
      \draw[red,very thick] (0.5,0.5) circle (0.5);
      \draw[dotted,red,very thick] (0.5,0.5) circle ({0.912*0.45});
    \end{scope}
  \end{tikzpicture}}
  \vspace{0.5cm}
  \caption{Large fishbowl. Pictures of the eigenfunctions for the lowest three modes. Note the reflection caused by the protruding, convex bottom of the tank, showing a slightly off-center perspective of our camera setup from see \autoref{experiment}. 
  }
  \label{explarge}
\end{figure}
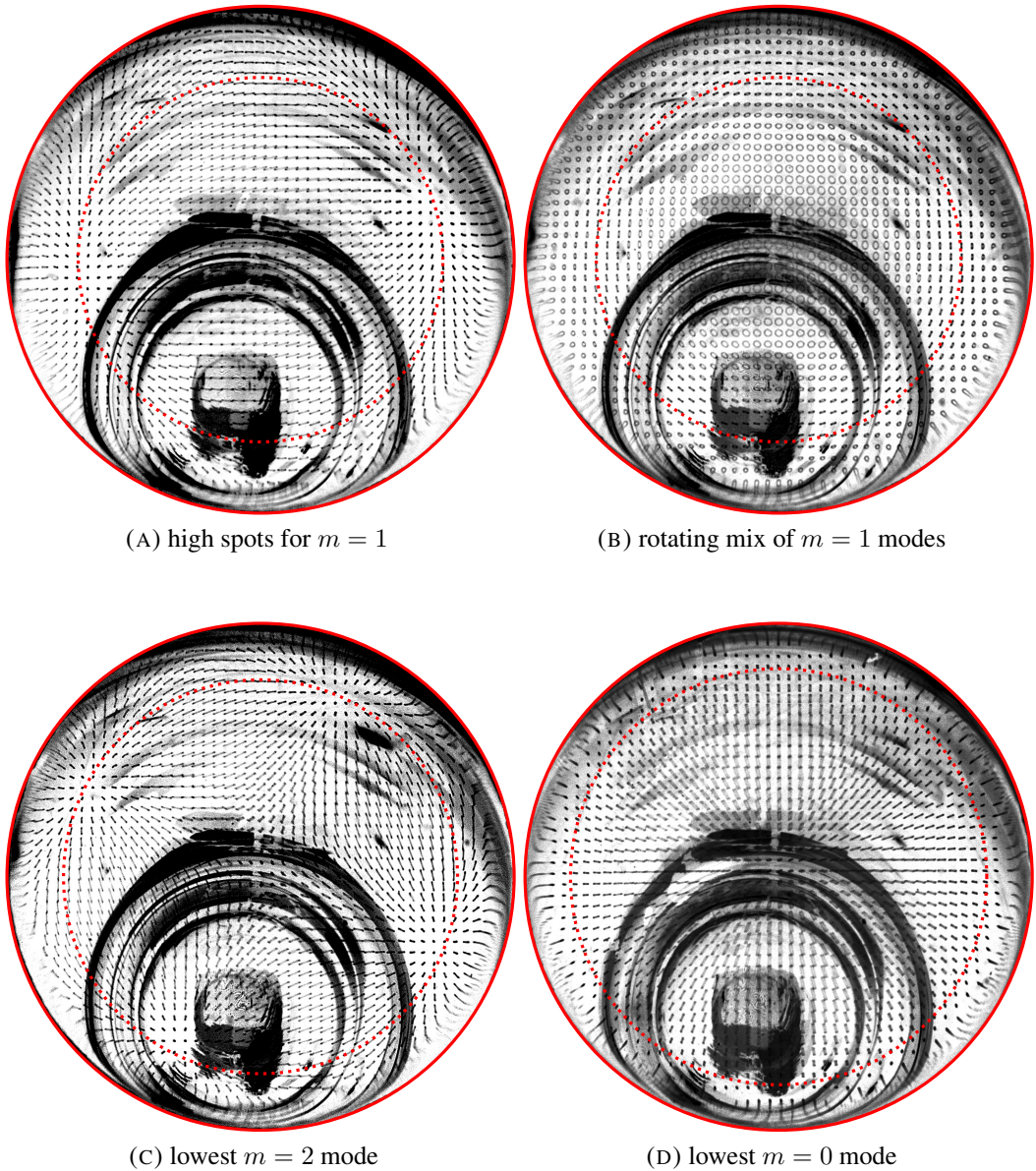

Our numerical results seem to almost exactly match the experiments. However, alert reader may notice very significant blur (on all photographs) very near the edges of the free surface. This is most likely caused by the interaction of the water surface and the container. As noted earlier, we did not try to eliminate such nonlinear phenomena. Instead, we performed the same experiment in a larger fishbowl with the shape from \autoref{largefishsize}. Much larger free surface should yield less significant boundary interaction. Indeed, \autoref{explarge} still shows great accuracy. \autoref{largehighspot} and \autoref{obrot2} show the modes for the lowest eigenvalue (cf. \autoref{fishbowl} and \autoref{obrot}). The latter emphasizes the whirling effect we first noticed on the smaller tank.

The larger tank allowed us to capture the eigenfunctions for the lowest eigenvalues of higher modes of oscillations: $m=2$ on \autoref{mode2} and $m=0$ on \autoref{mode0}. To capture the second mode we changed the frequency of nudging to match the second natural frequency. As a result we have four local maximal positioned on the numerically calculated dotted circle (consistent with $\cos(2\theta)$ in the eigenfunction decomposition \eqref{rep}). To capture the lowest radial mode $m=0$ we repeatedly dipped a tip of a pencil (with appropriate frequency). The extrema form a circle that is not far from the numerically predicted position. Nevertheless, that last circle is already quite close to the boundary, so the interactions of the water and the container seem to influence accuracy of the experiment.

A protruding bottom of the large fish tank (see \autoref{largefishsize}) allows us to also test the dependence of the eigenvalues on the shape of the bottom of the tank. Clearly, in a deep container the dependence should be nearly insignificant. The eigenvalue of the tank (assuming it is perfectly spherical) is contained between the eigenvalues of the spherical tanks with 125mm radius cut at the top and either 125mm or 155mm straight cuts at the bottom (the shape from \autoref{smallfishsize} with slightly different proportions). We can use domain monotonicity argument (see \cite[Proposition 3.2.1]{BKPS10}) to give upper and lower bounds for the eigenvalues of the original large tank (note that somewhat counterintuitively the smaller the tank, the smaller the eigenvalue). \autoref{tab:largefish} shows the bounds for the eigenvalues as well as the high spots for the lowest three eigenfunctions (corresponding to the lowest eigenvalues of modes $m=1$, $2$ and $0$). Note that the high spot positions and heights are not bounds, but merely approximations. 

\begin{table}
  \centering
  \catcode`+=\active\def+{\phantom0}
  \begin{tabular}{clll}
    \toprule        
    		    & eigenvalue & high spot position & high spot height \\
    \midrule
    $m=1$	& $2.453^{5073}_{0649}$ & $0.7^{19}_{20}$ & $1.28^{504}_{486}$\\
    $m=2$	& $3.86169^{67}_{23}$   & $0.777$ & $1.3693$ \\
    $m=0$	& $4.07333^{33}_{29}$   & $0$     & $3.4562$ \\
    		&			& $0.912$ & $1.1354$ \\
    \bottomrule
  \end{tabular}
  \caption{Upper and lower bounds for the eigenvalues of the large fish tank from \autoref{largefishsize} as well as positions of the high spots (extrema) and their relative height compared to boundary value. The bounds result from replacing the curved bottom with flat one, so that the new tank is either contained (lower bound) or contains (upper bound) the original tank. The accuracy of the position of the high spot is limited by the sampling of the numerical solution on the free surface. Heights do differ but on too high digits to be of any significance, especially given low position accuracy. }
  \label{tab:largefish}
\end{table}

\section{Relation to the classical spectral problems}\label{sec:relation}
Traditionally, Laplace eigenvalues are intuitively understood via heat flow. The Dirichlet eigenvalues of a domain $F$ govern the heat on a plate $F$ with zero temperature on the boundary, while the Neumann eigenvalues are related to the insulated plate. This explains the name ``hot spots conjecture'', for one the famous open problems (discussed below). The famous paper ``Can one hear the shape of drum?'' by Kac \cite{Kac} emphasizes another practical interpretation for Dirichlet spectrum. The corresponding eigenfunctions give the shapes of the drum membrane vibrating in one of the characteristic frequencies. The Neumann spectrum is harder to explain this way, as one would need a membrane that is free to float up and down while vibrating, but is horizontal near the boundary. This behavior can however be attributed to the free surface of an ideal liquid in a container. Therefore, sloshing not only generalizes the Neumann spectral problem, but also provides a clear physical intuition. In particular the hot spots problem (see below) might as well be called the high spot problem for a container with vertical walls.

Roughly speaking, the hot spots conjecture states that in a thermally insulated domain, for ``typical'' initial conditions, the hottest point will move towards the boundary of the domain, as time passes. The mathematical formulation of the hot spots conjecture is the following: the extrema of every fundamental eigenfunction (eigenvalue may have multiplicity) of the Neumann problem for a domain $F$ are located on the boundary of $F$. The conjecture was proved for certain sufficiently symmetric planar domains by Ba\~nuelos-Burdzy \cite{BB1999} and Jerison-Nadirashvili \cite{JN2000}, but disproved for some strange domains with holes by Burdzy and Werner \cite{BW99}, and just one hole by Burdzy \cite{B2005}. See \emph{Nature} article by Stewart~\cite{S1999} for more information, and Terence Tao's Polymath project~\cite{Tpoly} for current developments. This problem can be embedded in the studied sloshing problem as follows:

Function $\varphi$ is a sloshing mode in the cylinder
\[
 W = \{ (x, y, z) : (x, y) \in F, \, z \in (-h, 0) \}
\]
if and only if $\psi(x, y) = \varphi(x, y, 0)$ is a Neumann eigenfunction on $F$. Furthermore
\[
  \varphi(x, y, z) = \psi(x, y) \cosh(\sqrt{\mu} (z + h))
\]
and $\nu = \sqrt{\mu} \tanh(\sqrt{\mu}h)$, where $\mu$ is the Neumann eigenvalue corresponding to $\psi$ and $\nu$ is the sloshing eigenvalue corresponding to $\varphi$ by \cite[Proposition 3.1]{KK2009}.  Due to the very explicit connection between these problems, many results and conjectures for Neumann spectrum can be quickly translated to the sloshing language. In particular, the high spots problem~(\ref{slosh1}--\ref{slosh4}) generalizes the hot spots conjecture. 

By the above identification and the results of~\cite{BB1999, JN2000}, for any convex domain $F$ with two orthogonal axes of symmetry (e.g. ellipses), the high spots of the fundamental sloshing modes for cylindrical tanks with cross-section $F$ are located on the boundary. Note that the counterexample for the hot spots conjecture given by Burdzy and Werner \cite{BW99} is conceptually rather similar to the shape from \autoref{fig8c} or \autoref{fig:shapesym}. Indeed, their domain consists of a disk and a larger annulus, connected using spokes. The profile from \autoref{fig:shapesym} shows two deep containers with a disk and an annulus as free surfaces, connected using extremely shallow portion. In both cases one would like to have two disjoint subdomains, but a minimal connection must be made. In the high spot problem there are more ways to make the domain connected due to one extra dimension. For that reason it is much easier to construct intuitive counterexamples for high spots problem. Nevertheless, we expect the high spot conjecture to hold for convex containers.

The connection between Neumann problem and sloshing was also used to conjecture bounds for the eigenvalues. In particular the Dirichlet to Neumann eigenvalue comparisons \cite{LW86,Fr95,Fi04} was already extended to mixed Steklov eigenvalues \cite{BKPS10}. Furthermore, bounds for spectral functionals for Neumann eigenvalues naturally extend to the cylindrical sloshing case (see \cite[Section 13]{LS14}), opening a whole new area of research for general containers. Finally, sloshing problem can be used to tackle, seemingly unrelated, spectral problems for nonlocal fractional Laplacian  \cite{BK04,KKMS2010, K2012}.

\section*{Acknowledgements} Bart{\l}omiej Siudeja was partially supported by the Polish National Science Centre grant 2012/07/B/ST1/03356. T. Kulczycki and M. Kwaśnicki were supported in part by Project S30011/I-18 of the Institute of Mathematics and Computer Science of Wrocław University of Technology.

The authors would like to thank N. Kuznetsov for many useful discussions on the subject of the paper and for sharing his knowledge about the sloshing problem.


\begin{thebibliography}{99}


\bibitem{AO00}\textbf{M.~Ainsworth and J.~T. Oden}, \textit{A posteriori error estimation in finite element analysis}, Pure and Applied Mathematics (New York), Wiley-Interscience [John Wiley \& Sons], New York, 2000. \mref{MR1885308}

\bibitem{AB11}
  \textbf{H.~A. Ardakani and T.~J. Bridges}, \textit{Shallow-water sloshing in vessels undergoing prescribed rigid-body motion in three dimensions}, J. Fluid Mech. \textbf{667} (2011), 474--519. 

\bibitem{AP08}\textbf{M.~G. Armentano and C.~Padra}, \textit{A posteriori error estimates for  the {S}teklov eigenvalue problem},  Appl. Numer. Math. \textbf{58} (2008),  \textit{no.~5}, 593--601. \mref{MR2407734}

\bibitem{BO91}\textbf{I.~Babu{\v{s}}ka and J.~Osborn}, \textit{Eigenvalue problems}, Handbook  of numerical analysis, {V}ol.\ {II}, Handb. Numer. Anal., II, North-Holland,  Amsterdam, 1991, pp.~641--787. \mref{MR1115240}


\bibitem{BB1999}\textbf{R.~Ba{\~n}uelos and K.~Burdzy}, \textit{On the ``hot spots'' conjecture  of {J}.\ {R}auch},  J. Funct. Anal. \textbf{164} (1999), \textit{no.~1},  1--33. \mref{MR1694534}

\bibitem{BKPS10}\textbf{R.~Ba{\~n}uelos, T.~Kulczycki, I.~Polterovich, and B.~Siudeja},  \textit{Eigenvalue inequalities for mixed {S}teklov problems}, Operator  theory and its applications, Amer. Math. Soc. Transl. Ser. 2, vol. 231, Amer.  Math. Soc., Providence, RI, 2010, pp.~19--34. \mref{MR2758960}

\bibitem{BK04}\textbf{R.~Ba{\~n}uelos and T.~Kulczycki}, \textit{The {C}auchy process and the  {S}teklov problem},  J. Funct. Anal. \textbf{211} (2004), \textit{no.~2},  355--423. \mref{MR2056835}

\bibitem{BDM99}\textbf{C.~Bernardi, M.~Dauge, and Y.~Maday}, \textit{Spectral methods for  axisymmetric domains}, Series in Applied Mathematics (Paris), vol.~3,  Gauthier-Villars, \'Editions Scientifiques et M\'edicales Elsevier, Paris;  North-Holland, Amsterdam, 1999, Numerical algorithms and tests due to Mejdi  Aza{\"{\i}}ez. \mref{MR1693480}

\bibitem{BT}\textbf{T.~Betcke and L.~N. Trefethen}, \textit{Reviving the method of particular solutions}, SIAM Rev. \textbf{47} (2005), \textit{no.~3}, 469--491 (electronic). \mref{MR2178637}

\bibitem{BB92}\textbf{P.~Blanchard and E.~Br{\"u}ning}, \textit{Variational methods in  mathematical physics}, Texts and Monographs in Physics, Springer-Verlag,  Berlin, 1992, A unified approach, Translated from the German by Gillian M.  Hayes. \mref{MR1230382}

\bibitem{B2005}\textbf{K.~Burdzy}, \textit{The hot spots problem in planar domains with one  hole},  Duke Math. J. \textbf{129} (2005), \textit{no.~3}, 481--502.  \mref{MR2169871}

\bibitem{BW99}\textbf{K.~Burdzy and W.~Werner}, \textit{A counterexample to the ``hot spots''  conjecture},  Ann. of Math. (2) \textbf{149} (1999), \textit{no.~1},  309--317. \mref{MR1680567}


\bibitem{FT2012}\textbf{O.~M. Faltinsen and A.~N. Timokha}, \textit{Analytically approximate  natural sloshing modes for a spherical tank shape},  J. Fluid Mech.  \textbf{703} (2012), 391--301. \mref{MR2949919}

\bibitem{FT2014} \textbf{O.~M. Faltinsen and A.~N. Timokha}, \textit{Analytically approximate natural sloshing modes and frequencies in two-dimensional tanks}, European Journal of Mechanics B/Fluids \textbf{47} (2014), 176-187.

\bibitem{Fi04}\textbf{N.~Filonov}, \textit{On an inequality for the eigenvalues of the {D}irichlet and {N}eumann problems for the {L}aplace operator}, Algebra i Analiz \textbf{16} (2004), \textit{no.~2}, 172--176. \mref{MR2068346}

\bibitem{FHM}\textbf{L.~Fox, P.~Henrici, and C.~Moler}, \textit{Approximations and bounds  for eigenvalues of elliptic operators},  SIAM J. Numer. Anal. \textbf{4}  (1967), 89--102. \mref{MR0215542}

\bibitem{FK83}\textbf{D.~W. Fox and J.~R. Kuttler}, \textit{Sloshing frequencies},  Z. Angew.  Math. Phys. \textbf{34} (1983), \textit{no.~5}, 668--696. \mref{MR723140}

\bibitem{Fr95}\textbf{L.~Friedlander}, \textit{Remarks on {D}irichlet and {N}eumann eigenvalues}, Amer. J. Math. \textbf{117} (1995), \textit{no.~1}, 257--262. \mref{MR1314466}

\bibitem{GM11}\textbf{E.~M. Garau and P.~Morin}, \textit{Convergence and quasi-optimality of  adaptive {FEM} for {S}teklov eigenvalue problems},  IMA J. Numer. Anal.  \textbf{31} (2011), \textit{no.~3}, 914--946. \mref{MR2832785}

\bibitem{GHLST} \textbf{I. Gavrilyuk, M. Hermann, I. Lukovsky, O. Solodun and A. Timokha}, \textit{Natural sloshing frequencies
  in rigid truncated conical tanks}, Eng. Comput. \textbf{25} (2008), 518–-540.

\bibitem{HW12} 
  \textbf{A. Herczy\'nski and P. D. Weidman}, \textit{Experiments on the Periodic Oscillations of Free Containers Driven by Liquid Sloshing}, J. Fluid Mech. \textbf{693} (2012).

\bibitem{Ibr} \textbf{R. Ibrahim}, \textit{Liquid sloshing dynamics}, Theory and applications (Cambridge University Press), Cambridge,
UK, 2005.

\bibitem{JN2000}\textbf{D.~Jerison and N.~Nadirashvili}, \textit{The ``hot spots'' conjecture  for domains with two axes of symmetry},  J. Amer. Math. Soc. \textbf{13}  (2000), \textit{no.~4}, 741--772. \mref{MR1775736}

\bibitem{Kac}\textbf{M.~Kac}, \textit{Can one hear the shape of a drum?},  Amer. Math.  Monthly \textbf{73} (1966), \textit{no.~4, part II}, 1--23. \mref{MR0201237}

\bibitem{KK01} 
\textbf{N.~D. Kopachevsky and S.~G. Krein}, \textit{Operator approach to linear  problems of hydrodynamics. {V}ol. 1}, Operator Theory: Advances and  Applications, vol. 128, Birkh\"auser Verlag, Basel, 2001, Self-adjoint  problems for an ideal fluid. \mref{MR1860016}

\bibitem{KK2004}\textbf{V.~Kozlov and N.~Kuznetsov}, \textit{The ice-fishing problem: the  fundamental sloshing frequency versus geometry of holes},  Math. Methods  Appl. Sci. \textbf{27} (2004), \textit{no.~3}, 289--312. \mref{MR2023011}

\bibitem{K80}\textbf{A.~Kufner}, \textit{Weighted {S}obolev spaces}, Teubner-Texte zur  Mathematik [Teubner Texts in Mathematics], vol.~31, BSB B. G. Teubner  Verlagsgesellschaft, Leipzig, 1980, With German, French and Russian  summaries. \mref{MR664599}

\bibitem{KK2009}\textbf{T.~Kulczycki and N.~Kuznetsov}, \textit{`{H}igh spots' theorems for  sloshing problems},  Bull. Lond. Math. Soc. \textbf{41} (2009),  \textit{no.~3}, 495--505. \mref{MR2506833}

\bibitem{KK2011}\textbf{T.~Kulczycki and N.~Kuznetsov}, \textit{On the `high spots' of  fundamental sloshing modes in a trough},  Proc. R. Soc. Lond. Ser. A Math.  Phys. Eng. Sci. \textbf{467} (2011), \textit{no.~2129}, 1491--1502.  \mref{MR2782167}

\bibitem{KK2012}\textbf{T.~Kulczycki and M.~Kwa{\'s}nicki}, \textit{On high spots of the  fundamental sloshing eigenfunctions in axially symmetric domains},  Proc.  Lond. Math. Soc. (3) \textbf{105} (2012), \textit{no.~5}, 921--952.  \mref{MR2997042}


\bibitem{KKMS2010}\textbf{T.~Kulczycki, M.~Kwa{\'s}nicki, J.~Ma{\l}ecki, and A.~Stos},  \textit{Spectral properties of the {C}auchy process on half-line and  interval},  Proc. Lond. Math. Soc. (3) \textbf{101} (2010), \textit{no.~2},  589--622. \mref{MR2679702}


\bibitem{Notices2014} \textbf{N.~Kuznetsov, T.~Kulczycki, M.~Kwa{\'s}nicki,
A.~Nazarov, S.~Poborchi, I.~Polterovich
and B.~Siudeja}, \textit{The Legacy of Vladimir Andreevich Steklov}, Notices of AMS \textbf{61} (2014), \textit{no.~1},  9--22.

\bibitem{K2012}\textbf{M.~Kwa{\'s}nicki}, \textit{Eigenvalues of the fractional {L}aplace  operator in the interval},  J. Funct. Anal. \textbf{262} (2012),  \textit{no.~5}, 2379--2402. \mref{MR2876409}

\bibitem{L1932} \textbf{H.~Lamb}, \textit{Hydrodynamics}, Cambridge University Press (1932).

\bibitem{LS14}\textbf{R.~S. Laugesen and B.~A. Siudeja}, \textit{Sharp spectral bounds on  starlike domains},  J. Spectr. Theory \textbf{4} (2014), \textit{no.~2},  309--347. \mref{MR3232813}

\bibitem{LW86}\textbf{H.~A. Levine and H.~F. Weinberger}, \textit{Inequalities between  {D}irichlet and {N}eumann eigenvalues},  Arch. Rational Mech. Anal.  \textbf{94} (1986), \textit{no.~3}, 193--208. \mref{MR846060}

\bibitem{LMW} \textbf{A.~Logg, K.-A. Mardal, and G.~Wells}, \textit{Automated solution of  differential equations by the finite element method. The FEniCS book.},  {Lecture Notes in Computational Science and Engineering 84. Berlin: Springer.  xiii, 723~p.} (2012).

\bibitem{LBK}
\textbf{I.~A. Lukovski{\u\i}, M.~Y. Barnyak, and A.~N. Komarenko},  \textit{Approximate methods for solving problems of the dynamics of a bounded volume of fluid}, ``Naukova Dumka'', Kiev, 1984. \mref{MR780750}

\bibitem{McI89} \textbf{P.~McIver}, \textit{Sloshing frequencies for cylindrical and spherical  containers filled to an arbitrary depth},  J. Fluid Mech. \textbf{201}  (1989), 243--257. \mref{MR993197}

\bibitem{MR82}\textbf{B.~Mercier and G.~Raugel}, \textit{R\'esolution d'un probl\`eme aux  limites dans un ouvert axisym\'etrique par \'el\'ements finis en {$r$}, {$z$}  et s\'eries de {F}ourier en {$\theta $}},  RAIRO Anal. Num\'er. \textbf{16}  (1982), \textit{no.~4}, 405--461. \mref{MR684832}

\bibitem{M64}
\textbf{N.~N. Moiseev}, \textit{Introduction to the theory of oscillations of  liquid-containing bodies}, Advances in {A}pplied {M}echanics, {V}ol. 8,  Academic Press, New York, 1964, pp.~233--289. \mref{MR0167074}

\bibitem{RL10} \textbf{S. Rebouillat and D. Liksonov}, \textit{Fluid–structure interaction in partially filled liquid containers: A comparative review of numerical approaches}, Computers \& Fluids \textbf{39} (2010), 739--746.

\bibitem{S1999} 
  \textbf{I. Stewart}, \textit{Mathematics: Holes and hot spots}, Nature \textbf{401} (1999), 863--865.

\bibitem{Tpoly}
  \textbf{T.~Tao et al.}, \textit{Polymath7: The hot spots conjecture}, work in progress, \href{http://michaelnielsen.org/polymath1/index.php?title=The_hot_spots_conjecture}{michaelnielsen.org/polymath1/index.php?title=The\_hot\_spots\_conjecture}.

\bibitem{T60}\textbf{B.~A. Troesch}, \textit{Free oscillations of a fluid in a container},  Boundary problems in differential equations, Univ. of Wisconsin Press,  Madison, Wis., 1960, pp.~279--299. \mref{MR0114437}

\bibitem{WC09}\textbf{C-H. Wu and B-F. Chen}, \textit{ {Sloshing waves and resonance modes of fluid in a 3D tank by a time-independent finite difference method}}, Ocean Engineering \textbf{26} (2009), \textit{no.~6--7}, 500--510.
\end{thebibliography}
\end{document}